\documentclass[12pt,aps,pre,preprint,showpacs,showkeys]{revtex4}   

\usepackage{amsmath}    
\usepackage{amsfonts}   
\usepackage{amssymb}
\usepackage{graphicx}   
\usepackage{subfigure}

\DeclareMathOperator*{\argmin}{arg\,min}

\newcommand{\vrr}{\vec{r}}
\begin{document}

\title{Classification of missing values in spatial data using spin models}
\author{Milan \v{Z}ukovi\v{c}}
 \email{mzukovic@mred.tuc.gr}
 \thanks{Current address: Institute of Physics, Faculty of Science,
 Pavol Jozef \v{S}af\'{a}rik
University, Park Angelinum 9, 040 01 Ko\v{s}ice, Slovak Republic;
e-mail: milan.zukovic@upjs.sk.}  
 \affiliation{Technical University of Crete, Geostatistics Research Unit,
Chania 73100, Greece}
\author{Dionissios T. Hristopulos}
 \email{dionisi@mred.tuc.gr}   
 \homepage[web address: ]{http://www.mred.tuc.gr/home/hristopoulos/dionisi.htm}
 \affiliation{Technical University of Crete, Geostatistics Research Unit,
Chania 73100, Greece}

\date{\today}

\begin{abstract}
A problem of current interest is the estimation of spatially
distributed processes at locations where measurements are missing.
Linear interpolation methods rely on the Gaussian assumption, which
is often unrealistic in practice, or normalizing transformations,
which are successful only for mild deviations from the Gaussian
behavior. We propose to address the problem of missing values
estimation on two-dimensional grids by means of spatial
classification methods based on spin (Ising, Potts, clock) models.
The ``spin'' variables provide an interval discretization of the
process values, and the spatial correlations are captured in terms
of interactions between the spins. The spins at the unmeasured
locations are classified by means of the ``energy matching''
principle: the correlation energy of the entire grid (including
prediction sites) is estimated from the sample-based correlations.
We investigate the performance of the spin classifiers in terms of
computational speed, misclassification rate, class histogram and
spatial correlations reproduction using simulated realizations of
spatial random fields, real rainfall data, and a digital test image.
We also compare the spin-based methods with standard classifiers
such as the $k$-nearest neighbor, the fuzzy $k$-nearest neighbor,
and the Support Vector Machine. We find that the spin-based
classifiers provide competitive choices.
\end{abstract}

\pacs{02.50.-r, 02.50.Ey,  02.60.Ed, 75.10.Hk, 89.20.-a, 89.60.-k}

\keywords{stochastic estimation,  non-Gaussian, spatial
classification, Ising model,  Potts model, clock model, optimization}

\maketitle

\section{Introduction}
\label{sec:intro}

To date there is a growing demand for various types of spatial data,
including remote sensing images,
such as thematic maps representing geographical variables, natural
resources, land use, environmental indicators, or economic and
demographic information. At the
same time,  new methods are needed for the efficient and
reliable processing, analysis, and digital storage of this
information. Common issues that arise in the
processing phase include the heterogeneity of the data, i.e., the
fact that they come from different sources (sensors) operating at
different resolutions and often with different biases. In addition,
the data coverage is often incomplete due to  limited resources
(material, human, or technical), equipment limitations (detection
level, resolution), meteorological conditions (observations hindered
by clouds) and sensor malfunctions.

In the following, we will assume that an observed spatially
distributed process in two dimensions is a realization of a spatial
random field (SRF) \cite{yagl87}. An SRF, $Z(\vrr,\omega)$, where
$\vrr=(x,y) \in \mathbb{R}^{2}$ is the location and $\omega$ is the
state index, represents an ensemble of states (realizations), the
spatial correlations of which are determined by a joint probability
density function (p.d.f.). The state index will be suppressed in the
following to keep the notation concise. In order to use standard
tools for the analysis of space-time data, the observed process
needs to be estimated at the missing value locations. The spatial
estimation can be performed by means of established interpolation
and classification techniques~\citep{atk99}. However, considering
the ever increasing size of spatial data, classical methods, e.g.,
kriging~\citep{wack03} and minimum curvature estimation~\cite{min-curv},
become impractical due to high computational
complexity. Furthermore, the linear spatial interpolation methods
assume a jointly Gaussian p.d.f., which is often
not justified by data. In addition, the use of such methods
typically requires a considerable  degree of subjective
input~\citep{dig07}. The  nonlinear indicator kriging (IK) method is
based on a binary discretization of the data distribution with respect to
an arbitrary number of thresholds~\cite{journel83}. IK does not
require the Gaussian assumption, but the predicted indicator values
are not guaranteed to take $0$ or $1$ values, and may even lie
outside the $[0, 1]$ range.

Recent studies investigate applications  of statistical
physics concepts in various non-traditional fields, such as economy
and finance~\cite{man_stan99,bo_pot00,bouch00}, materials
science~\cite{kard95,wouts07,zhen06}, and biology~\cite{gran92}.
Most studies have focused on long-range correlations, while
short-range correlations that can be used for spatial or temporal
interpolation have received less attention. Nevertheless, within
the Gibbs-Markov Random Field framework, the Potts and Ising models
have been used in image analysis~\cite{besa86,gimel99,kato06,zhao07,tana95}.
The Potts model in the superparamagnetic regime has also been applied to the data
clustering problem~\cite{blatt96} and a Potts-model-based non-parametric approach 
employing simulated annealing to the oil reservoir characterization~\cite{farm92}.

Let us consider the Gibbs
probability density function $ f_{\rm Z} =
{e^{-H[Z(\vrr)]/k_{B}T}}/{\mathcal{Z}},$ where $H[Z(\vrr)]$ is an
interaction functional  that involves spatial operators acting on
the field $Z(\vrr)$, $k_{B}$ is the Boltzmann's constant, $T$ is the
temperature (we can set $k_{B}\, T=1 $ without loss of generality),
and ${\mathcal{Z}}$ is the partition function. Assume that $H[Z(\vrr)]$
measures the ``spatial disorder energy'' of the observed process.
Hence,  $H[Z(\vrr)]$ is minimized for a uniform state and increases
proportionally to the fluctuations of $Z$, as prescribed by the
interactions embodied in $H[Z(\vrr)]$. The family of Spartan Spatial
Random Fields (SSRF) is based on the idea that the spatial
correlations can be adequately determined from a general
$H[Z(\vrr)]$~\cite{dth03} with local interactions. However, the SSRF
models are still based on the restrictive Gaussian assumption.

Mild deviations from Gaussian dependence including lognormal dependence
can be treated by means of
normalizing non-linear transformations. Alternatively, one can try
to incorporate the  non-Gaussian terms directly into the statistical
model. However, it is not clear what type of interactions in
$H[Z(\vrr)]$ can generate the  non-Gaussian behavior encountered in
different data sets. Even if such interactions could be defined,
calculation of the
statistical moments would entail numerical integrations. The purpose
of this study is to present non-parametric
 and computationally efficient methods that do not assume a specific
 form for the p.d.f. of the observed process. The main idea is to employ
interaction models either of the form $H[S^{q}(Z(\vrr))],$ where
$S^{q}(Z(\vrr))$ is a binary discretization of the continuous SRF
$Z(\vrr)$ and $q$ is a level index (i.e., Ising model), or
of the form $H[S(Z(\vrr))]$, where $S(Z(\vrr))$ is a multilevel
discretization of $Z$ (i.e., Potts and clock models). For the interaction
functional we use spin (Ising, Potts and clock) Hamiltonians.
Parameter estimation in such systems is typically based on
computationally intensive Monte Carlo simulations or potentially
inaccurate approximation methods. To overcome these problems, we
develop a
 non-parametric, Monte Carlo based approach, which benefits
 from an informative assignment of
 the initial spin states.


The rest of the paper is organized as follows.  In Section
\ref{sec:spat_pred}, we discuss the problem of spatial
classification/prediction and review some established machine
learning classification algorithms. In Section~\ref{sec:spin_clas}
we develop our  spatial classification approach based on spin
models. Section~\ref{sec:class-simul} investigates the application
of the proposed models to synthetic realizations of spatial random
fields. Section~\ref{sec:real} focuses on the application of the
spin-based classifiers on real data. Finally, in
Section~\ref{sec:conclusions} we summarize our results and discuss
issues for future research.

\section{Spatial prediction and classification}
\label{sec:spat_pred}

We consider observations distributed on rectangular grids $\tilde{G}$ of size
$N_{G}=L_{x} \times L_{y}$, where $L_{x}$ and $L_{y}$ are
respectively the horizontal and vertical dimensions of the rectangle
(in terms of the unit length). We denote the set of sampling points
by $G_{s}=\{\vec{r}_{i}\} $, where $\vec{r}_{i}=(x_i, y_i) \in
{\mathbb{R}}^{2},$ $i=1,\ldots,N$ are points scattered  on the grid
and $N=(1-p)\, N_{G}<N_{G}$. Let $z_{i}$ be the observation at
$\vec{r}_{i}$ obtained from a joint p.d.f.
$f_{\rm Z}(z_{1}, \ldots,z_{N_G})$. The set $Z_{s}=\{z_{i} \in
{\mathbb{R}} \}$ represents the sample of the process. Let
$G_{p}=\{\vec{r}_{p}\},$ $p=1,\ldots, P=p\, N_{G}$, be the
prediction set
 so that $\tilde{G}=G_{s} \cup
G_{p}$. In the following we discretize the continuous distribution of
$Z$ using a number of classes (intervals), ${\mathcal C}_q, \, q=1,\ldots,N_c$.  The
classes are defined with respect to threshold levels $t_k,
\, k=1,\ldots,N_{c}+1$. If $Z_{\rm min} = \min(z_{1},\ldots,z_{N}),$ and
 $Z_{\rm max} = \max(z_{1},\ldots,z_{N})$ denote respectively the
minimum and maximum values of the data and $\epsilon$ is an
infinitesimal positive number, then
 $t_{1}= Z_{\rm min}-\epsilon$ and
$t_{N_{c}+1}=Z_{\rm max}$. The remaining thresholds are defined by means of
$t_{q}= (q-1)\,\left( Z_{\max}-Z_{\min}\right)/N_c + Z_{\min}$, for $q=2,\ldots,N_c.$
Each class ${\mathcal C}_q$
corresponds to an interval  ${\mathcal C}_q=( t_{q},
t_{q+1}]$ for $q=1,\ldots,N_c$.
  The \textit{class indicator field} $I_{Z}(\vec{r})$
 is defined so that $I_{Z}(\vec{r})= q$  if $z(\vec{r})
\in {\mathcal C}_q,$ for $q=1,\ldots,N_c$, i.e.,
\begin{equation}
\label{eq:indic}
 I_{Z}(\vec{r}_{i})= \sum_{q=1}^{N_c} q \, \left[ \theta\left( z_{i} - t_{q} \right)
 - \theta\left( z_{i} - t_{q+1} \right) \right], \quad \forall
 i=1,\ldots,N,
\end{equation}
where $\theta(x)=1$ for $x>0$ and  $\theta(x)=0$ for $x \le 0$ is
the unit step function.
 Prediction of the field values $Z\{G_{p}\}$ is
then mapped onto a classification problem, i.e., the estimation of
$\hat{I}_{Z}\{G_{p}\}$.
For environmental monitoring and risk
management applications useful answers can be obtained in terms of a small
number of levels (e.g., eight). For the analysis of gray-scale
digital images $256$ levels are typically required. As the number of
levels tends to infinity, the discretized interval representation
tends to the continuum, and spatial classification tends to spatial
interpolation.

 To test the performance of the spin-based classification methods,
 we use synthetic
 Gaussian (normal) and lognormal random field realizations
  on $L \times L$ grids (where $L=50, 100, 200$), continuous
 precipitation data sampled on a $50 \times 50$ grid, and a digital $512 \times 512$
 image of $256$ gray-scale values. Each data set is randomly
 and exhaustively partitioned into sampling and prediction sets.
  We use the points in the prediction set for validation,
  i.e., to compare the true value of the process  with the
  classifier predictions.
Next, we briefly review two
non-parametric,  machine learning classification algorithms,
which we use as benchmarks for the performance of the spin models.

\subsection{$k$-nearest neighbor models}
\label{knn_model} The $k$-nearest neighbor (KNN) model is probably  the simplest of all
machine learning algorithms \cite{dasa91}. The classified value is
assigned to the most populated class among the $k$ nearest neighbors
of the classification point. The distance metric used is typically
Euclidean, and the neighbors are selected from the points in $G_{s}$.
The optimal choice of the parameter $k$ is data dependent. Various
heuristic techniques can be used to determine the optimal $k$, e.g.
cross-validation. However, the method is sensitive to noise or the
presence of irrelevant features at the scales of interest.
Nevertheless, KNN typically outperforms many other flexible
nonlinear methods, particularly when the number of explanatory
variables is high~\cite{cher96}. It is also easy to implement, and
its classification error is asymptotically optimal.

An extension of KNN is the method of  fuzzy $k$-nearest neighbor (FKNN)
classification~\cite{kell85}.
In FKNN,  the sampling points $\vec{r}_{j}$ are first  assigned a membership to each class
${\mathcal C}_q, \, q=1,\ldots,N_c$
 by means of the function
$u_q(\vec{r}_{j}).$ Then, each prediction point $\vec{r}_{p}$  is
also assigned class membership  according to the function
$u_q(\vec{r}_{p})=\big[\sum_{j=1}^{k} u_q(\vec{r}_{j}) \,
||\vec{r}_{p}-\vec{r}_{j}||^{2/(1-m)} \big]
/\big(\sum_{j=1}^{k}||\vec{r}_{p}-\vec{r}_{j}||^{2/(1-m)}\big)$, for
$ q=1,\ldots,N_c$. The  parameter $m$ controls the influence of
distant samples. Following~\cite{kell85}  we set $m=2$. The prediction points are
classified according to the maxima of the membership functions.
The FKNN classifier statistically reduces the
effect of noisy samples and produces overall more accurate results
than the classical KNN classifier. To eliminate the impact of
an arbitrary $k$, we repeat the classification for
$k=1,\ldots,k_{\max}$, to determine a
$k_{\rm opt}$ that minimizes the misclassification rate. This
adaptive approach guarantees that the lowest misclassification rates
achievable by the KNN and the FKNN algorithms (based on the Euclidean distance
metric) are used in the comparisons.

\subsection{Support vector machines}
\label{svm_model} The support vector machines (SVM)  classifier  is
a supervised learning algorithm~\cite{vapnik63,aizerman64,vapn98}
which in several comparative studies outperformed other
methods~\cite{herm99,shari03,mull97}. The original SVM
algorithm~\cite{vapnik63} is a linear classifier that segregates
 data into  two classes using  maximum-margin hyperplanes.
The SVM algorithm has been extended to multi-class and non-linear
problems using the kernel function trick~\cite{aizerman64}, by means
of which nonlinear dependence is linearized in a higher-dimensional
``feature'' Hilbert space. The SVM method is robust and can handle
high-dimensional data. However, it is computationally intensive,
especially for large $N$ and $N_{c}$, because it requires the careful
tuning of hyperparameters for each binary classifier followed by the solution
of a quadratic problem with $N$ variables.


We solve the $N_c > 2$ classification problem  by means of $N_c$
binary classifiers operating in one-to-rest fashion
e.g.~\cite{hsulin02}. We use the software GeoSVM, which implements an
adaptation of SVM for spatial data classification~\cite{kane04}. The
code is run with radial basis function (RBF) kernels and involves
two tunable hyperparameters: the  kernel bandwidth $\sigma_k$ and
the regularization parameter $C$; the latter controls the trade-off
between the machine complexity and the number of linearly
non-separable points.
The hyperparameters are tuned to minimize the
misclassification rate. Due to the high computational cost of
tuning and training the SVM, it is only applied to
the rainfall data.

\section{Spatial classification based on spin models}
\label{sec:spin_clas}

Below we propose three non-parametric classifiers that use  spin model
Hamiltonians from statistical physics. In the following, the spins
correspond to discretized levels of the continuous variable $Z$ and
should not be confused with magnetic moments. The main idea is that
the spatial correlations of $Z$ can be captured by the spin
interactions. By focusing on the spins it is not necessary to assume
a specific form of the joint p.d.f. $f_{Z}.$
The non-parametric form
of the classifiers derives from the fact that the state of the spin
systems is constrained by the sample data instead of the interaction
couplings $J$ and the temperature $T$. This is convenient since $J$
and $T$ are unknown \textit{a priori}, their estimation from the
data can be computationally intensive due to the intractability of
the partition function, and their uniqueness is not guaranteed for
any given sample. To classify the values at the prediction points we
use the heuristic principle of energy matching: we calculate the
 correlation energy of the sample normalized by the number of interacting spin pairs,
 and then determine the spin values at the prediction sites so that
 the normalized energy
for the entire grid matches the respective sample value. Herein we
focus on nearest neighbor correlations, but this constraint can be
relaxed.

The idea of
correlation energy matching has been applied in the statistical
reconstruction of digitized (binary) random media from
limited morphological information
\cite{yeong98a,yeong98b}. The classifiers proposed here employ
 both \textit{local} (sample values) and \textit{global} (statistical)
 information. In particular, this is achieved by
 performing conditional
simulations, in which the sample values are respected locally and
the correlation energy globally. This means that while the
interactions are translation invariant, the state of the system is
not necessarily stationary (statistically homogeneous). The
correlation energy matching presupposes that the nearest-neighbor
separations in the sample capture the target scale of the prediction
set. For example, assume that  a sample is drawn
 from a square grid
with step $\alpha$ by removing $50\%$ of the points. The
energy matching will be more effective if the points are removed at random
 than if every second point is removed. In the first case, it is likely that
 contiguous groups of points will be removed, leaving
pairs of neighbors separated by $\alpha$, while in the second case
the minimum separation
between the sampling points will be $2\alpha.$

The Ising model~\cite{mccoy73} was introduced to describe the
energy states of magnetic materials and later found many
applications, e.g., as a model of neuron activity in the brain. It
involves binary variables $s_{i}$ (spins), which can take the values
$1$ or $-1$. In the absence of an external filed, the energy of the
spin system  can be expressed by the Hamiltonian $ H_{I} = -
\sum_{i,j}J_{ij}\, s_{i} \, s_{j}$. The coupling parameter $J_{ij}$
controls the type ($J_{ij}>0$ for ferromagnetic and $J_{ij}<0$ for
antiferromagnetic coupling) and strength of the pairwise
interactions. The model is usually defined on a regular grid, the
interactions are considered uniform, and their range is limited to
nearest neighbors. Generalizations to  irregular grids and
longer-range interactions are also possible. The Ising model has
been solved in one dimension and in $d=2$ without external field.
The Potts model is a generalization of  the Ising model~\cite{wu82}.
Each spin variable is assigned an integer value $s_{i} \in
\{1,\ldots,N_c\}$, where $N_c$ represents the total number of
states. The Hamiltonian of the Potts model is given by $H_{P} = -
\sum_{i,j}J_{ij}\delta_{(s_{i},s_{j})},$ where $\delta$ is the
Kronecker symbol. Hence,  nonzero contributions to the energy only
come from spins in the same state. For $N_c=2$ the Potts model is
equivalent to the 2D Ising model. The clock model, also known  as
the vector Potts model, assumes that the spins take one of $N_c$
possible values, which are distributed uniformly around a circle.
The clock Hamiltonian is given by $ H_{C} = - \sum_{i,j}J_{ij}
\cos\Big[\frac{2\pi}{N_c}( s_i- s_j)\Big].$ In contrast with the
Potts model,  in the clock model interactions between spins in
different states contribute to the interaction energy. The XY model
is obtained from the clock model at the continuum limit $N_c
\rightarrow \infty$. The presence of an external field $h_{i}$
implies an additional term
  $-\sum_{i}h_{i}s_{i}$ in the Hamiltonian. This term
 breaks the symmetry of the energy under spin reversals,
and  controls the total magnetization of the spins.

In typical applications of the spin models the parameters  $J_{ij}$
and $h_{i}$ are assumed to be known, and the problem of interest is
the estimation of various correlation functions. In the case of
spatial data analysis, the parameters are not known \textit{a
priori}. Hence, prediction of the spins at unmeasured locations
requires us to determine the model
parameters from the available spin configuration (sample). The
standard procedure for solving such an inverse problem is to infer the
parameters by means of the maximum likelihood estimation (MLE)
method. Then, the spin values at unsampled locations can be
predicted by maximizing the p.d.f. $f_{Z}$ (equivalently, by
minimizing $H$) under the data constraints. However,  the analytical
intractability of ${\mathcal{Z}}$ hampers the application of MLE.
Possible ways of circumventing this problem, such as the maximum
pseudo-likelihood~\cite{besa75} approach or Markov chain Monte Carlo
techniques~\cite{chen97} can be inaccurate or prohibitively slow.

To bypass the problems mentioned above, we propose a non-parametric
approach. For lattice data we assume that $J_{ij} = J$ for nearest
neighbors and $J_{ij} = 0$ otherwise. In addition, we consider a
zero external field. Nevertheless, as shown by the simulation
results below, the marginal class distributions of the data are
recovered based on the interaction energy of the discretized levels
and a judicious choice of the initial states (for each class). The
use of different spin models in the study allows us to investigate
the impact  on the misclassification rate and prediction errors of
different classification strategies (sequential vs. simultaneous),
different couplings (intra-level vs. inter-level interactions), as
well as the transition to the  continuum case (XY model).



\subsection{Classification based on the Ising Model}
Here we present a spatial classification approach based on the Ising
spin model with nearest neighbor correlations (INNC). The main idea
is to use a sequential discretization scheme to estimate the class
indicator field $\hat{I}_{Z}(G_{p}).$ In this scheme the sample
$G_{s}^{q}$ and prediction, $G_{p}^{q}$ grids are progressively
updated for increasing class index $q$. For the lowest class
$G_{s}^{1}=G_{s}, \, G_{p}^{1}=G_{p},$ where $G_{s}$ and $G_{p}$ are
the sampling and prediction grids respectively.  Let $N_q$ denote
the number of sites on $G_{s}^{q}$ that have  fixed class indicator
values at level $q$; at the first level $N_1=N$ and $N_{q+1} \ge
N_q$ for $q=1,\ldots, N_{c}-1$ since the sample at level $q+1$ is
augmented by the points that are assigned fixed indicator values at
level $q$. Let  $S^{q}_{s}=\{s^{q}_{i_q}\}$, $q=1,\ldots, N_c;$
$i_q=1,\ldots,N_q$ be the set that includes all the sample spin
values  with respect to level $q$, and
 $\tilde{S}^{q}=S^{q}_{s} \cup
S^{q}_{p}$ denote all the spin values on $\tilde{G}$. The Ising
model is used to represent interactions between the spins
$\tilde{S}^{q}$.

At each level, the discretization is binary with respect to the
upper threshold; i.e., $s^{q}_{i}=-1$ if $z_{i} \le t_{q+1}$ and
$s^{q}_{i}=1$ if $z_i
> t_{q+1}$, for all $\vec{r}_{i} \in \tilde{G}$.
The classification algorithm sweeps  sequentially through the $q$
values. All spin $-1$ assignments at level $q$  fix the class
indicator value for the respective sites;  i.e.,
$\hat{I}_{Z}(\vec{r}_i)=q.$ For $q>1$, the sample grid $S^{q}_{s}$
is augmented by the points ${\vec{r}_l}$ for which $s^{q-1}_{l}
=-1,$ while the prediction grid is accordingly diminished.  It
follows that  $N_{q>1} \ge N$. The ``gaps'' in the prediction grid
$G_{p}$ are gradually filled as the algorithm proceeds through
consecutive levels. The reduced prediction grid $G^{q}_{p}$ at
level $q$ contains $P^{q}$ points so that $P^{1}=P$ and  $P^{q} \le
P^{q'}$ for $q>q'$.

\begin{figure}[!t]
  \begin{center}
    \includegraphics[scale=0.5]{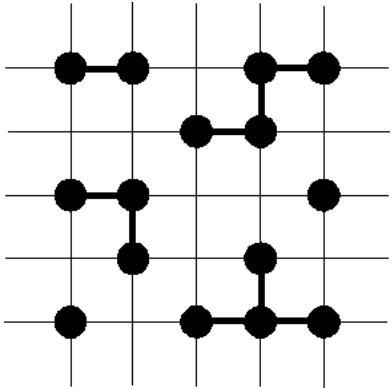}
  \end{center}
  \caption{Schematic depicting the sample sites (solid circles) and the nearest-neighbor
  pairs  (solid circles linked with bold lines) contributing to the sample
  correlation energy $C_{s}^{q}$.}
  \label{fig:schem}
\end{figure}

The spin assignments at each level are controlled by the
 cost function,
$U(S^{q}_{p}|S^{q}_{s}),$ which measures the deviation between the
correlation energy (per spin pair) of the simulated state,
$\tilde{C}^{q}$, and the respective sample energy $C_{s}^{q}$. This
cost function is given by
\begin{eqnarray}
\label{cost} U(S^{q}_{p}|S^{q}_{s}) = \left\{
\begin{array}{ll}
\Big[ 1  - \tilde{C}^{q}(S^{q}_{p},S^{q}_{s})/C_{s}^{q}(S^{q}_{s}) \Big]^2,
    & {\rm{for}}\ C_{s}^{q}(S^{q}_{p},S^{q}_{s}) \neq 0, \\ \\
\left[ \tilde{C}^{q}(S^{q}_{p},S^{q}_{s}) \right]^2,        & {\rm{for}}\
C_{s}^{q}(S^{q}_{p},S^{q}_{s}) = 0,
\end{array} \right.
\end{eqnarray}
\noindent where $C_{s}^{q}=\langle s^{q}_{i}s^{q}_{j} \rangle
_{G_{s}^{q}}$ is the spin two-point correlation of the $q-$level
sample (see Fig.~\ref{fig:schem}) and $\tilde{C}^{q}=\langle
s^{q}_{i}s^{q}_{j} \rangle _{\tilde{G}}$ is the respective
correlation over the entire grid. The $q-$level classification
problem is equivalent to finding the optimal configuration
$\hat{S}^{q}_{p}$ that minimizes~(\ref{cost}):
\begin{eqnarray}
\label{optim} \hat{S}^{q}_{p} = \argmin_{S^{q}_{p}}
U(S^{q}_{p}|S^{q}_{s}).
\end{eqnarray}

\subsubsection{Monte Carlo Relaxation (Greedy Scheme)}
Determining the optimum spin state is based on a Markov Chain Monte
Carlo (MC) walk over the ensemble of possible spin configurations at
sites that are not assigned a fixed class indicator value.  The
generation of new states  is realized using Metropolis updating at
$T=0.$
The zero-temperature condition means that there is no stochastic
selection of unfavorable spins; i.e., the spin tested  is flipped
unconditionally if it lowers the cost function. This so-called
``greedy'' Monte Carlo algorithm~\citep{pap82} guarantees
convergence, which  is typically very fast. In contrast, in
simulated annealing $T$ is slowly lowered from an initial
high-temperature state. This approach is much slower
computationally, but the resulting configuration is less sensitive
to the initial state. The sensitivity of the greedy algorithm is
 pronounced in high-dimensional spaces with
non-convex energies, since in such cases it is more
likely to be trapped in local minima. However, this is not a concern in the current problem.
In fact, targeting the global minimum of  $U(S^{q}_{p}|S^{q}_{s})$
 emphasizes exact matching of the sample correlation energy,
which is subject to  sample-to-sample
fluctuations.

\subsubsection{Initial State Selection}
To provide a fast and unsupervised automatic classification
mechanism, the initial configuration of the spins at the prediction
sites should require little or no user intervention and minimize the
relaxation path  to the equilibrium (in state space). Assuming a
degree of spatial continuity, which is common in geospatial data sets, we
propose to determine the initial states based on the sample values
in the immediate neighborhood of the individual prediction points
${\vec{r}_p}$. More specifically, the initial spin states at the
prediction sites are based on majority votes of the sample spins
within an adaptable $m \times m$ stencil (where $m=2l+1$) centered
at ${\vec{r}_p}$~\footnote{A circular stencil would correspond to
the Euclidean-distance-based KNN classification algorithm, $k$ being
the number of sampling points inside the stencil.}.
The stencil size $m \le m_{\max}$ is set
adaptively starting with $m=3$, reflecting
the local sampling density and the spin values distribution, to the
smallest size that contains a clear majority of spin values.
Imposing the arbitrary upper bound $m_{\rm max}$ on the stencil size
prevents oversmoothing and increasing the
 computational load (memory and CPU time).
Intuitively,  $m_{\rm max}$ should be higher for sparser sampling
patterns. If a majority is not achieved for $m \le m_{\rm max}$, the
initial spin value is assigned at random from the sample spin values
with the most votes. If a majority can not be achieved due to a lack of
sampling points inside the maximum stencil, the initial value is
drawn randomly from the entire range of spin values~\footnote{In the
INNC model there is practically no distinction between the two cases
of ambiguity, since at each level only two spin values are possible.
The distinction is meaningful for the vector-valued spin models
described below.}. We will refer to this procedure for determining
the initial state as the \textit{adaptable stencil size} (ASS)
procedure. The indeterminacy of the initial  state injects a
stochastic element in the classification. Hence, by performing
multiple initializations one can assess the classification
uncertainty due to the initial state ambiguity.

The parametrization required by the algorithm involves  only $m_{\max}$
and the definition of the class intervals. The number of
classes depends on the study objective: if the goal is to determine
areas where a specific level is exceeded, a binary classification is
sufficient. For environmental monitoring and decision making
purposes a moderate number of classes (e.g., eight) is often
sufficient. A larger number of classes is necessary for the
reconstruction of gray-scale images, and even larger numbers of
classes are used for spatial interpolation.

\subsubsection{Vectorized Algorithm for MC Simulation}
On commensurate lattices, the grid structure and the short range of
the interactions enable vectorization of the algorithm, thus
improving the computational efficiency of the Monte Carlo
relaxation. Vectorization  is achieved by partitioning  $\tilde{G}$
in two interpenetrating subgrids $\tilde{G}_k,\ k=1,2$, so that  the
spins on one subgrid interact only with spins on the other. Hence,
each subgrid can be updated simultaneously, while one sweep through
the entire grid involves just two subgrid updating steps.

 Starting from the initial state, Monte Carlo spin updating cycles
 are generated.
 Each cycle  consists  of two steps. Each step focuses on  one sublattice,
 simultaneously generating
 updates for all the prediction sites. The updates are drawn uniformly from
 the set of possible values (e.g., $\pm 1$ for the Ising model,
 $1,\ldots, N_c$ for the Potts
 and clock models). The updates are
 accepted or rejected locally for each site, depending on whether they raise or lower the
 energy of the specific spin's interactions.  The algorithm proceeds until the
 cost function becomes zero within a specified tolerance (termination criterion I) or
 an updating cycle ends in rejection of all the updates (termination criterion II).

Based on the above, the vectorized Monte Carlo (MC) algorithm for the INNC
classification model consists of the following steps:

\begin{center}\textbf{Algorithm for INNC model} \end{center}
\noindent {\bf (1)} {\em Initialization}
\newline \indent
{\rm (1.1)} {\em Define $N_c$ and $m_{\max}$}
\newline \indent
{\rm (1.2)} {\em Set the
indicator field on the entire grid by means of}
$\hat{I}_{Z}(\tilde{G})={\rm NaN}$
\newline
{\bf (2)} {\em Set the simulation level (class index) to $q=1$}
\newline
{\bf (3)} {\bf While [loop 1]} $q \le N_c -1$
\newline
\indent {\rm (3.1)} {\em Define $S_{s}^{q}$ by discretizing}
$Z\{G_{s}\}$ {\em with respect to} $t_{q+1}$
\newline \indent
{\rm (3.2)} {\em Assign spin values} $S^{q}_{s}$ to the points on $G^{q}_{s}$
\newline \indent
{\rm (3.3)} {\em Given the data} $S^{q}_{s}$,
{\em calculate the sample correlation energy} $C_{I;s}^{q}$
\newline \indent
{\rm (3.4)} {\em Initialize spin values on $G^{q}_{p},$
based on ASS, i.e., generate $\hat{S}^{q \,(0)}_{p}$}
\newline \indent
{\rm (3.5)} {\em Given} $S^{q}_{s}$ and $\hat{S}^{q \,(0)}_{p}$,
{\em calculate the initial simulated correlation energy} $\tilde{C}_{I}^{q \,(0)}$
\newline \indent
{\rm (3.6)} {\em Initialize the MC index} $i_{mc}=0$ {\em
and the rejected states index} $i_r =0$
\newline \indent
{\rm (3.7)}\;{\bf If} $\tilde{C}_{I}^{q \, (0)}<C_{I;s}^{q}$
\newline \indent \indent
{\rm (3.7.1)} {\bf While [loop 1.1] $(\tilde{C}_{I}^{q \, (i_{mc})}<C_{I;s}^{q}$) \& $(i_r < 2)$ update grid}
\newline \indent \indent \indent {\em Initialize  the rejected states index} $i_r =0$
\newline \indent \indent \indent {\bf For} $k=1,2$ [{\em subgrid
updating}]
\newline \indent \indent \indent {\rm (3.7.1.1)}{\em Generate a new state}
$\hat{S}^{q \, (i_{mc}+1)}_{p}$ {\em by perturbing} $\hat{S}^{q
\,(i_{mc})}_{p}$  {\em on subgrid} $\tilde{G}_k$
\newline \indent \indent \indent {\rm (3.7.1.2)}{\em Identify prediction sites}
$\{i_{k} \}= K_k \subset \tilde{G}_k$,
{\em such that} \newline
\indent \indent \indent \indent \indent \indent
$\{s_{i_k} \, s_{j}\}^{(i_{mc}+1)} > \{s_{i_k} \,
s_{j}\}^{(i_{mc})}$
\newline \indent \indent \indent {\rm (3.7.1.3)}\;{\bf If} $K_k \neq \emptyset$
{\em update the sites on} $K_k$
\newline \indent \indent \indent \indent {\bf else} {\em keep the ``old'' state} \;;
$\; i_r \rightarrow i_r+1$; {\bf endif}
\newline \indent \indent \indent
{\rm (3.7.1.4)} $i_{mc} \rightarrow i_{mc}+1$
\newline \indent \indent \indent {\bf end} [{\em subgrid
updating}]
\newline \indent \indent  {\bf end [loop 1.1]}
\newline \indent
{\bf elseif} $\tilde{C}_{I}^{q \, (0)}>C_{I;s}^{q}$
\newline \indent \indent
{\rm (3.7.2)} {\bf While [loop 1.2] $(\tilde{C}_{I}^{q \, (i_{mc})} > C_{I;s}^{q}$) \& $(i_r < 2)$ update grid}
\newline \indent \indent \indent {\em Initialize  the rejected states index} $i_r =0$
\newline \indent \indent \indent {\bf For} $k=1,2$ [{\em subgrid
updating}]
\newline \indent \indent \indent {\rm (3.7.2.1)}{\em Generate a new state}
$\hat{S}^{q \, (i_{mc}+1)}_{p}$ {\em by perturbing} $\hat{S}^{q
\,(i_{mc})}_{p}$  {\em on subgrid} $\tilde{G}_k$
\newline \indent \indent \indent {\rm (3.7.2.2)}{\em Identify prediction sites}
$\{i_{k} \}= K_k \subset \tilde{G}_k$,
{\em such that} \newline
\indent \indent \indent \indent \indent \indent
$\{s_{i_k} \, s_{j}\}^{(i_{mc}+1)}<\{s_{i_k} \,
s_{j}\}^{(i_{mc})}$
\newline \indent \indent \indent {\rm (3.7.2.3)}\;{\bf If} $K_k \neq \emptyset$
{\em update the sites on} $K_k$
\newline \indent \indent \indent \indent \; {\bf else} {\em keep the ``old'' state} \;;
$\; i_r \rightarrow i_r+1$; {\bf endif}
\newline \indent \indent \indent
{\rm (3.7.2.4)} $i_{mc} \rightarrow i_{mc}+1$
\newline \indent \indent \indent {\bf end} [{\em subgrid
updating}]
\newline \indent \indent {\bf end [loop 1.2]}
\newline \indent {\bf else}
\newline \indent \indent $i_{mc} \rightarrow i_{mc}+1$
\newline \indent {\bf endif}
\newline \indent {\rm (3.8)} {\em Assign the}  $-1$
{\em spins to the $q$-th class,  i.e.}
\newline \indent \indent $i_{mc} \rightarrow i_{mc}-1$;
\newline \indent \indent {\bf If}
$\hat{S}^{q(i_{mc})}(\{\vec{r}_{i}\})=-1$, $\{\vec{r}_{i}\} \in
\tilde{G},$  {\em set} $\hat{I}_{Z}(\{\vec{r}_{i}\})=q$
\newline \indent \indent
{\em where} $\hat{S}^{q(i_{mc})}(\{\vec{r}_{i}\} \in G_{s}) \equiv S^{q}_{s}$; {\bf endif}
\newline \indent
{\rm (3.9)} {\em Increase the class index,} $q \rightarrow q+1$,
{\em return to step} {\bf (3)}
\newline {\bf end [loop 1]}
\newline
{\bf (4)} {\em For} $q=N_c$, $\forall \vec{r}_{i} \;
(i=1,\ldots,N_{\tilde{G}})$ {\em such that}
$\hat{I}_{Z}(\{\vec{r}_{i}\})={\rm NaN}$, {\em set}
$\hat{I}_{Z}(\{\vec{r}_{i}\})=N_c \; $.

\vspace{12pt} In the above, the symbol NaN denotes
a non-numeric value used to initialize the class assignments.
Loop 1 sequentially assigns values to each  class.
In loop 1.1 the sample energy is approached from below, while in loop 1.2 from above.
In both cases the algorithm is vectorised by partitioning the lattice in two
subgrids.
The termination criterion  requires that either the spin correlation energy
match the sample energy (within machine precision) or that one sweep over entire grid not produce any successful updates. In steps 3.7.1.2 and
3.7.2.2 the terms $\{s_{i_k} \, s_{j}\}$ imply a summation over all the
neighbor spins $s_{j}$ on the complementary subgrid for each
$s_{i_k}$ on the perturbed subgrid. Step 3.8 assigns the $-1$ spins to the
current class and adds the respective sites to the sampling set for the next higher
level $q$.
In the end, all the remaining spins with NaN values are assigned to the highest class.

\subsection{Classification  based on Multivalued Spin Models}
This approach is based on models that involve multivalued  spin
variables with nearest neighbor correlations, such as the Potts,
clock and XY Hamiltonians. The same algorithm structure is used for
the nearest-neighbor correlation models based on the
Potts (PNNC), clock (CNNC), and XY (XYNNC) models. The PNNC and CNNC
models differ only in the form of the correlation energy, while the
XYNNC differs from the CNNC model in the number of discretization
levels.  In contrast with the INNC model, the classification for the
multi-valued models is performed simultaneously over all levels.
Accordingly,
 we drop the index $q$ which refers
to the current level, and the relevant quantities in the algorithm
are calculated from the spin values $S=\{s_{i}\}$, $i=1,\ldots,N_{\tilde{G}}$
and $s_i \in \{1,\ldots,N_c\}$. The normalized correlation energies over
the sample and entire grid are calculated by
$C_{s}=\langle \delta_{(s_{i},s_{j})} \rangle_{G_{s}}$ and
$\tilde{C}=\langle \delta_{(s_{i},s_{j})} \rangle _{\tilde{G}}$   for the PNNC model,
 and by $C_{s}=\left\langle
\cos\left[\frac{\pi ( s_i- s_j)}{2N_c}\right] \right\rangle_{G_{s}}$
and $\tilde{C}=\left\langle \cos\Big[\frac{\pi( s_i-
s_j)}{2N_c}\Big] \right\rangle _{\tilde{G}}$  respectively  for the
CNNC and XYNNC models.   Note that the prefactor in the
argument of the cosine function is changed from $2\pi$ to
$\pi/2$ to ensure that each energy level corresponds to a unique
value of $|s_{i}-s_{j}|$.
Based on the above, the MC algorithm for the
 PNNC models consists of the following steps:

 \begin{center}\textbf{Algorithm for PNNC model} \end{center}
\noindent {\bf (1)} {\em Initialization}
\newline \indent
{\rm (1.1)} {\em Define $N_c$ and $m_{\max}$}
\newline \indent
{\rm (1.2)}
{\em Define $S_{s}$ by discretizing} $Z\{G_{s}\}$ {\em with respect to} $t_{k},\
k=1,\ldots,N_{c}+1$
\newline \indent
{\rm (1.3)} {\em Assign spin values} $S_{s}$ to the points on $G_{s}$
\newline
{\bf (2)} {\em Given the data} $S_{s}$, {\em calculate the sample
correlation energy} $C_{P;s}$
\newline
{\bf (3)} {\em Initialize the spin values on $G_{p},$
based on ASS, i.e. generate} $\hat{S}^{(0)}_{p}$
\newline
{\bf (4)} {\em Calculate the initial correlation} $\tilde{C}_{P}^{(0)}$
\newline
{\bf (5)} {\em Initialize the MC index} $i_{mc}=0$ {\em and the rejected states index} $i_r =0$
\newline
{\bf (6)}\;{\bf If} $\tilde{C}_{P}^{(0)}<C_{P;s}$
\newline \indent
{\rm (6.1)} {\bf While [loop 1] $(\tilde{C}_{P}^{(i_{mc})}<C_{P;s}$)
\& $(i_r < 2)$  update grid}
\newline \indent \indent {\em Initialize the rejected states index} $i_r =0$
\newline \indent \indent {\bf For} $k=1,2$ [{\em subgrid updating}]
\newline \indent \indent {\rm (6.1.1)} {\em Generate a new state}
$\hat{S}^{(i_{mc}+1)}_{p}$ {\em by perturbing}
$\hat{S}^{(i_{mc})}_{p}$ {\em on subgrid} $\tilde{G}_k$
\newline \indent \indent {\rm (6.1.2)} {\em Identify prediction sites}
$\{i_{k} \}= K_k \subset \tilde{G}_k$,
{\em such that} \newline
\indent \indent \indent \indent \indent \indent
$\{\delta(s_{i_k},s_{j})\}^{(i_{mc}+1)}> \{\delta(s_{i_k},s_{j})\}^{(i_{mc})}$
\newline \indent \indent {\rm (6.1.3)}
{\bf If} $K_k \neq \emptyset$ {\em update the sites on} $K_k$
\newline \indent \indent \indent {\bf else} {\em keep the ``old'' state};
$\; i_r \rightarrow i_r+1$;
{\bf endif}
\newline \indent \indent {\rm (6.1.4)} $i_{mc} \rightarrow i_{mc}+1$
\newline \indent \indent {\bf end} [{\em subgrid
updating}]
\newline \indent  {\bf end [loop 1]}
\newline  {\bf elseif} $\tilde{C}_{P}^{(0)}>C_{P;s}$
\newline \indent
{\rm (6.2)} {\bf While [loop 2] $(\tilde{C}_{P}^{(i_{mc})}>C_{P;s}$)
\& $(i_r < 2)$  update grid}
\newline \indent \indent {\em Initialize  the rejected states index} $i_r =0$
\newline \indent \indent {\bf For} $k=1,2$ [{\em subgrid
updating}]
\newline \indent \indent {\rm (6.2.1)} {\em Generate a new state}
$\hat{S}^{(i_{mc}+1)}_{p}$ {\em by perturbing}
$\hat{S}^{(i_{mc})}_{p}$ {\em on subgrid} $\tilde{G}_k$
\newline \indent \indent {\rm (6.2.2)} {\em Identify prediction sites}
$\{i_{k} \}= K_k \subset \tilde{G}_k$,
{\em such that} \newline
\indent \indent \indent \indent \indent \indent
$\{\delta(s_{i_k},s_{j})\}^{(i_{mc}+1)}< \{\delta(s_{i_k},s_{j})\}^{(i_{mc})}$
\newline \indent \indent {\rm (6.2.3)}
{\bf If} $K_k \neq \emptyset$ {\em update the sites on} $K_k$
\newline \indent \indent \indent {\bf else} {\em keep the ``old'' state};
$\; i_r \rightarrow i_r+1$;
{\bf endif}
\newline \indent \indent {\rm (6.2.4)} $i_{mc} \rightarrow i_{mc}+1$
\newline \indent \indent {\bf end} [{\em subgrid
updating}]
\newline \indent  {\bf end [loop 2]}
\newline  {\bf else}
\newline \indent  $i_{mc} \rightarrow i_{mc} + 1$
\newline  {\bf endif}
\newline  {\bf (7)} \; $i_{mc} \rightarrow i_{mc} - 1;$ \; {\em assign}
$\hat{I}_{Z}(\{\vec{r}_{i}\})=\hat{S}^{(i_{mc})}(\{\vec{r}_{i}\})$,
$\{\vec{r}_{i}\} \in \tilde{G}$, {\em where} $\hat{S}^{(i_{mc})}(\{\vec{r}_{i}\} \in G_{s}) \equiv S_{s}$.

The main difference with the INNC case is the absence of a loop over different
classes since the initial spin discretization corresponds to the number of classes.

\section{Classification of Simulated Random Field Data}
\label{sec:class-simul}

\subsection{Simulation Study Design}
We study the  performance of the classification models on simulated
realizations of Gaussian, $Z \sim N(m=50,\sigma=10)$, and lognormal,
$\ln Z \sim N(m=4,\sigma=0.5)$ random fields~\cite{yagl87}. The
spatial correlations are imposed by means of the flexible
Whittle-Mat\'{e}rn family of covariance functions:
\begin{equation}
\label{mate} G_{\rm Z} (\|\vec{r}\|)=\sigma^{2}
\frac{{2}^{1-\nu}}{\Gamma(\nu)}(\kappa \,
\|\vec{r}\|)^{\nu}K_{\nu}(\kappa \, \|\vec{r}\|),
\end{equation}
where $\|\vec{r}\|$ is the Euclidean two-point distance,  $\sigma^2$
is the variance, $\nu$ is the smoothness parameter, $\kappa$ is the
inverse autocorrelation length, and $K_{\nu}$ is the modified Bessel
function of index $\nu$. This $G_{\rm Z} (\|\vec{r}\|)$ leads to
random field realizations that are $m$ times differentiable, where
$m$ is the largest integer smaller than $\nu.$ In addition, higher
values of $\kappa$ imply shorter correlation ranges. We generate
samples with different combinations of $\kappa = 0.2,\ 0.5$ and $\nu
= 1.5,\ 2.5$ on a square ($L_x=L_y=L$) grid $\tilde{G}$, with
$N_{G}=L^2$ nodes, where
  $L=50,\ 100,\ 200$. The samples are
generated using the spectral  method \cite{drum87,hrist02}. From the
complete data set on  $\tilde{G}$, we extract a sample $Z_{s}$ of
size $N=(1-p)\,N_{G}$ by randomly removing $P=p N_{G}$ points for
validation. For different degrees of thinning ($p=0.33$ and $0.66$),
we generate $100$ different configurations of $G_{p}$. We use two
values of class numbers ($N_c=8$ and $16$), corresponding to
different discretizations of the continuum (resolutions). The
simulated values on $G_{p}$ are classified using the spin-based
algorithms. In the classification performance evaluation, the
indicator field values $I_{Z}(G_{p})$ are compared with the
estimates $\hat{I}_{Z}(G_{p})$ that were obtained from the classification
algorithms. We  define the misclassification rate as
\begin{eqnarray}
\label{misclass} F = \frac{1}{P}\sum_{p=1}^{P}\left[
1-\delta\big(I_{Z}(\vec{r}_p),\hat{I}_{Z}(\vec{r}_p)\big) \right],
\end{eqnarray}
where $I_{Z}(\vec{r}_p)$ is the true value at the validation points,
$\hat{I}_{Z}(\vec{r}_p)$ is the classification estimate and
$\delta(I,I')=1$ if $I=I'$, $\delta(I,I')=0$ if $I \neq I'$. We also
measure the optimization CPU time, $T_{\rm cpu}$, the number of MC
sweeps required for reaching equilibrium, and the residual values of
the cost function at termination, $U^{\ast}$.

To evaluate the performance of the spin models for large $N_c$
(i.e., approaching the continuum limit), we calculate the following
``sample'' prediction errors: average absolute error AAE $=
\frac{1}{P}\sum_{p=1}^{P}|Z(\vec{r}_p) - \hat{Z}(\vec{r}_p)|$,
average relative error ARE $ =\frac{1}{P}\sum_{p=1}^{P}\frac{Z_p -
\hat{Z_p}}{Z_p}$, average absolute relative error AARE
$=\frac{1}{P}\sum_{p=1}^{P}\bigg|\frac{Z(\vec{r}_p) -
\hat{Z}(\vec{r}_p)}{Z(\vec{r}_p)}\bigg|$, root average square error
RASE $ =\sqrt{\sum_{p=1}^{P}\frac{1}{P}|Z(\vec{r}_p) -
\hat{Z}(\vec{r}_p)|^2}$, and the linear sample correlation
coefficient $R$. In the above definitions, $P$ is the number of
validation points, $Z(\vec{r}_p)$ is the true value at $\vec{r}_p$
and $\hat{Z}(\vec{r}_p)$ is the estimate of the process based on
\begin{equation}
\label{eq:Zhat}
\hat{Z}(\vec{r}_p) =
t_{\hat{I}_{Z}(\vec{r}_p)}+ \frac{t_{\hat{I}_{Z}(\vec{r}_p)+1}-t_{\hat{I}_{Z}(\vec{r}_p)}}{2}.
\end{equation}

To focus on local behavior of the classifiers we define the respective errors, in which the spatial average is replaced by a mean value calculated over a number of samples (e.g. $M=100$ realizations) at each point. Namely, at point $p$, mean absolute error = $ \langle |Z(\vec{r}_p) - \hat{Z}_{j}(\vec{r}_p)| \rangle$, mean relative error = $ \bigg\langle \frac{Z(\vec{r}_p) - \hat{Z}_{j}(\vec{r}_p)}{ Z(\vec{r}_p)} \bigg\rangle$, mean absolute relative error = $ \bigg\langle \bigg|\frac{Z(\vec{r}_p) - \hat{Z}_{j}(\vec{r}_p)}{ Z(\vec{r}_p)}\bigg| \bigg\rangle$,  and root mean squared error = $ \sqrt{\langle |Z(\vec{r}_p) - \hat{Z}_{j}(\vec{r}_p)|^{2} \rangle}$, where $\hat{Z}_{j}(\vec{r}_p)$ are predictions at point $p$ obtained from $j=1,...,M$ realizations and $ \langle . \rangle$ denotes averaging over different realizations.
The computations are performed in the Matlab\textregistered\
programming environment on a desktop computer with 1.93 GB of RAM
and an Intel\textregistered Core\texttrademark 2 CPU 6320 processor
with an 1.86 GHz clock.

\subsection{Misclassification rate} We investigate the effects of
grid size $N_{G}$, data sparseness $p$, and class number $N_c$ on
the classification performance. The results obtained by the INNC,
PNNC, and CNNC models are compared with the {\it best} results
obtained by the established KNN and FKNN methods
(see~\ref{knn_model}). Tables~\ref{tab:ka02nu25}-\ref{tab:ka05nu15}
summarize the results obtained for Gaussian random fields with
Whittle-Mat\'{e}rn parameters $(\kappa,\nu)=(0.2,2.5)$, $(0.5,2.5)$,
and $(0.5,1.5)$, respectively. Table~\ref{tab:lognka05nu25} lists
the results obtained from lognormal random fields with
Whittle-Mat\'{e}rn parameters $(\kappa,\nu)=(0.5,2.5)$.

\begin{table}[h]
\begin{small}
\caption{Classification results using the 8- and 16-level  INNC,
PNNC, and CNNC models for realizations of Gaussian random fields $\sim N(50,10)$
with Whittle-Mat\'{e}rn  covariance ($\kappa=0.2, \nu=2.5$) on a square
domain of size $L$. The
tabulated results are averages over $100$ realizations.
They include the mean value and standard
deviation of the misclassification rate $\langle F^* \rangle$  and
$\mbox{STD}_{F^*}$ ($\%$), the mean number of Monte Carlo sweeps
$\langle N_{MC} \rangle$, the mean optimization CPU time $\langle
T_{\rm cpu} \rangle$ $[sec]$, and the mean value of the cost
function at termination $\langle U^{*} \rangle$. The
$\langle F^{*}_{\rm knn} \rangle$  and $\langle F^{*}_{\rm fknn} \rangle$
represent the mean
of the lowest misclassification rates $[\%]$ achieved by KNN and FKNN algorithms respectively.  }
\label{tab:ka02nu25}
\begin{center}
\begin{tabular}{lcccccccccccccccccc}
\hline \hline
Size    & \multicolumn{6}{c}{L=50}  & \multicolumn{6}{c}{L=100} & \multicolumn{6}{c}{L=200}\\
\hline
Model          & \multicolumn{2}{c}{INNC}  & \multicolumn{2}{c}{PNNC} &  \multicolumn{2}{c}{CNNC} & \multicolumn{2}{c}{INNC}  & \multicolumn{2}{c}{PNNC}  & \multicolumn{2}{c}{CNNC} & \multicolumn{2}{c}{INNC}  & \multicolumn{2}{c}{PNNC} & \multicolumn{2}{c}{CNNC}\\
\hline
$p[\%]$        & $33$  & $66$ & $33$  & $66$ & $33$  & $66$ & $33$  & $66$ & $33$  & $66$ & $33$  & $66$ & $33$  & $66$ & $33$  & $66$ & $33$  & $66$\\
\hline
 \multicolumn{19}{c}{$N_c=8$} \\
\hline
$\langle F^{*}_{\rm knn} \rangle$ &  13.6&  20.4&  $-$&  $-$& $-$&  $-$&  10.3&  16.1&  $-$&  $-$&  $-$&  $-$& 8.2&  12.8& $-$ &  $-$& $-$&  $-$ \\
$\langle F^{*}_{\rm fknn} \rangle$ &  12.4&  18.1&  $-$&  $-$& $-$&  $-$&  9.5&  14.5&  $-$&  $-$&  $-$&  $-$& 7.6&  11.6& $-$ &  $-$& $-$&  $-$ \\
$\langle F^{*} \rangle$&  11.1&  16.1&  12.9&  19.1&  12.4&  17.5&  8.6&  12.9&  9.6&  14.9&   9.5&  14.0 & 6.9&  10.6&  7.4&  11.7& 7.3&  11.1 \\
${\rm STD}_{F^{*}}$&  1.25&  1.03&  1.34&  1.10&   1.18&  1.02& 0.52&  0.43&  0.56&  0.43&   0.52&  0.45& 0.24&  0.22&  0.26&  0.23 &  0.26&  0.21\\
$\langle N_{MC} \rangle$&  4.4&  6.2&  2.7&  6.0&  3.4&  8.5&   6.3&  6.9&  5.5&  9.1&   5.0&  12.5& 6.8&  8.0&  9.5&  13.7 &  9.9&  17.4\\
$\langle T_{\rm cpu} \rangle$&  0.11&  0.33&  0.03&  0.11&   0.05&  0.13& 0.47&  1.72&  0.13&  0.44&   0.20&  0.61& 2.02&  7.11&  0.61&  1.99&  1.06&  2.95\\
$\langle U^{*} \rangle$&  5e-4&  2e-3&  3e-5&  1e-4&   8e-9&  4e-8& 3e-4&  1e-3&  5e-6&  2e-5&  2e-9&  1e-8&  1e-4&  6e-4&  1e-6&  4e-6 &  3e-10&  2e-9 \\
\hline
 \multicolumn{19}{c}{$N_c=16$} \\
\hline
$\langle F^{*}_{\rm knn} \rangle$ &  34.2&  40.2&  $-$&  $-$& $-$&  $-$&  27.7&  34.2&  $-$&  $-$&  $-$&  $-$& 20.7&  28.0& $-$ &  $-$& $-$&  $-$ \\
$\langle F^{*}_{\rm fknn} \rangle$ &  31.9&  38.9& $-$ &  $-$&  $-$&  $-$&  25.3&  32.7&  $-$&  $-$&  $-$&  $-$& 18.9&  25.9& $-$&  $-$& $-$&  $-$ \\
$\langle F^{*} \rangle$&  21.2&  33.8&  35.1&  42.8&  23.5&  31.5&  17.2&  27.3&  27.9&  35.8&  21.2&  28.4&  13.6&  21.5&  20.5&  28.0&  18.2&  24.7\\
${\rm STD}_{F^{*}}$&  1.48&  1.60&  1.64&  1.24&  1.52&  1.56&  0.67&  0.75&  0.85&  0.52&  0.73&  0.83&  0.31&  0.36&  0.35&  0.29&  0.38&  0.74\\
$\langle N_{MC} \rangle$&  8.5&  12.9&  6.5&  17.0&  35.1&  45.1&  13.1&  15.2&  4.2&  14.7&  39.6&  43.3&  13.9&  17.1&  3.2&  2.8&  47.5&  46.7\\
$\langle T_{\rm cpu} \rangle$&  0.22&  0.62&  0.05&  0.14&  0.18&  0.32&  0.94&  3.30&  0.15&  0.55&  0.84&  1.28&  3.85&  13.92&  0.55&  1.82&  4.02&  5.32\\
$\langle U^{*} \rangle$&  7e-4&  2e-3&  1e-4&  5e-4&  2e-9&  1e-9&  4e-4&  1e-3&  2e-7&  4e-6&  3e-11&  4e-11&  2e-4&  8e-4&  6e-8&  5e-7&  1e-12&  3e-12\\
\hline \hline
\end{tabular}
\end{center}
\end{small}
\end{table}

\begin{table}[h]
\begin{small}
\caption{The same classification statistics are listed as in Table
\ref{tab:ka02nu25}, obtained from realizations of a Gaussian random field with
Whittle-Mat\'{e}rn covariance ($\kappa=0.5, \nu=2.5$).}
\label{tab:ka05nu25}
\begin{center}
\begin{tabular}{lcccccccccccccccccc}
\hline \hline
Size    & \multicolumn{6}{c}{L=50}  & \multicolumn{6}{c}{L=100} & \multicolumn{6}{c}{L=200}\\
\hline
Model          & \multicolumn{2}{c}{INNC}  & \multicolumn{2}{c}{PNNC} &  \multicolumn{2}{c}{CNNC} & \multicolumn{2}{c}{INNC}  & \multicolumn{2}{c}{PNNC}  & \multicolumn{2}{c}{CNNC} & \multicolumn{2}{c}{INNC}  & \multicolumn{2}{c}{PNNC} & \multicolumn{2}{c}{CNNC}\\
\hline
$p[\%]$        & $33$  & $66$ & $33$  & $66$ & $33$  & $66$ & $33$  & $66$ & $33$  & $66$ & $33$  & $66$ & $33$  & $66$ & $33$  & $66$ & $33$  & $66$\\
\hline
 \multicolumn{19}{c}{$N_c=8$} \\
\hline
$\langle F^{*}_{\rm knn} \rangle$ &  25.6&  32.3&  $-$&  $-$& $-$&  $-$&  25.8&  32.6&  $-$&  $-$&  $-$&  $-$& 21.8&  28.8& $-$ &  $-$& $-$&  $-$ \\
$\langle F^{*}_{\rm fknn} \rangle$ &  23.5&  30.3&  $-$&  $-$&  $-$&  $-$&  23.7&  30.6&  $-$&  $-$&  $-$&  $-$& 20.2&  26.7 & $-$&  $-$& $-$&  $-$ \\
$\langle F^{*} \rangle$&  19.7&  27.5& 25.6 & 32.3 & 22.2 & 28.7 & 19.4 & 27.0 & 24.9 & 32.5 & 22.2 & 28.8  & 17.2 & 23.7 & 21.2 & 28.1 & 19.4 & 25.5 \\
${\rm STD}_{F^{*}}$&  1.36&  1.23& 1.63 & 1.19 & 1.56 & 1.33 & 0.74 & 0.69 & 0.85 & 0.63 & 0.76 & 0.71  & 0.34 & 0.30 & 0.43 & 0.27 & 0.33 &  0.29 \\
$\langle N_{MC} \rangle$&  5.0&  6.2&  6.6 & 12.5 & 7.8 & 7.9 & 5.8 & 7.7 & 14.6 & 23.3 & 6.5 &  7.9 & 5.9 & 8.6 & 13.0 & 19.4 & 6.0 & 6.6 \\
$\langle T_{\rm cpu} \rangle$&  0.11&  0.35& 0.04 & 0.12 & 0.07 & 0.14 & 0.46 & 1.60 & 0.18 & 0.56 & 0.24 & 0.53  & 1.90 & 6.62 & 0.74 & 2.30 & 0.87 & 2.07 \\
$\langle U^{*} \rangle$&  7e-4&  2e-3& 1e-4 & 7e-4 & 9e-10 & 2e-9 & 6e-4 & 2e-3 & 3e-5 & 2e-4 & 3e-10 & 1e-9  & 4e-4 & 2e-3 & 5e-7 & 4e-6 & 1e-10 &  6e-10\\
\hline
 \multicolumn{19}{c}{$N_c=16$} \\
\hline
$\langle F^{*}_{\rm knn} \rangle$ &  51.9&  55.4&  $-$&  $-$& $-$&  $-$&  53.9&  56.5&  $-$&  $-$&  $-$&  $-$& 48.1&  51.4& $-$ &  $-$& $-$&  $-$ \\
$\langle F^{*}_{\rm fknn} \rangle$ &  49.7&  54.7&  $-$&  $-$&  $-$&  $-$&  51.5&  56.1&  $-$&  $-$&  $-$&  $-$& 45.4&  50.9& $-$&  $-$& $-$&  $-$ \\
$\langle F^{*} \rangle$&  39.0&  54.0& 52.7 & 58.5 & 37.2 & 48.7 & 38.7 & 53.9 & 52.9 & 59.3 & 36.2 & 48.1  & 33.7 & 48.1 & 46.9 & 54.1 & 32.7 &  44.1\\
${\rm STD}_{F^{*}}$&  1.77&  1.55& 1.55 & 1.00 & 2.30 & 1.69 & 0.87 & 0.76 & 0.77 & 0.54 & 1.34 & 0.96  & 0.47 & 0.42 & 0.41 & 0.24 & 0.55 &  0.44\\
$\langle N_{MC} \rangle$&  10.3&  12.7& 19.1 & 27.3 & 23.1 & 19.9 & 11.6 & 14.3 & 39.7 & 50.1 & 23.3 &  20.6 & 11.9 & 16.8 & 59.6 & 70.4 & 24.4 & 21.5 \\
$\langle T_{\rm cpu} \rangle$&  0.21&  0.65& 0.07 & 0.17 & 0.14 & 0.21 & 0.91 & 2.99 & 0.34 & 0.88 & 0.57 &  0.85 & 3.78 & 12.44 & 1.90 & 4.69 & 2.41 &  3.57\\
$\langle U^{*} \rangle$&  8e-4&  2e-3& 2e-3 & 2e-2 & 9e-11 & 3e-10 & 7e-4 & 2e-3 & 2e-3 & 2e-2 & 2e-11 & 1e-10  & 5e-4 & 2e-3 & 3e-4 & 7e-3 & 1e-11 & 7e-11 \\
\hline \hline
\end{tabular}
\end{center}
\end{small}
\end{table}

\begin{table}[h]
\begin{small}
\caption{The same classification statistics are listed as in  Table
\ref{tab:ka02nu25}, obtained from realizations of a Gaussian random field  with
Whittle-Mat\'{e}rn covariance ($\kappa=0.5, \nu=1.5$).}
\label{tab:ka05nu15}
\begin{center}
\begin{tabular}{lcccccccccccccccccc}
\hline \hline
Size    & \multicolumn{6}{c}{L=50}  & \multicolumn{6}{c}{L=100} & \multicolumn{6}{c}{L=200}\\
\hline
Model          & \multicolumn{2}{c}{INNC}  & \multicolumn{2}{c}{PNNC} &  \multicolumn{2}{c}{CNNC} & \multicolumn{2}{c}{INNC}  & \multicolumn{2}{c}{PNNC}  & \multicolumn{2}{c}{CNNC} & \multicolumn{2}{c}{INNC}  & \multicolumn{2}{c}{PNNC} & \multicolumn{2}{c}{CNNC}\\
\hline
$p[\%]$        & $33$  & $66$ & $33$  & $66$ & $33$  & $66$ & $33$  & $66$ & $33$  & $66$ & $33$  & $66$ & $33$  & $66$ & $33$  & $66$ & $33$  & $66$\\
\hline
 \multicolumn{19}{c}{$N_c=8$} \\
\hline
$\langle F^{*}_{\rm knn} \rangle$ &  38.9&  44.8&  $-$&  $-$& $-$&  $-$&  36.7&  42.4&  $-$&  $-$&  $-$&  $-$& 30.7 &  37.1 & $-$ &  $-$& $-$&  $-$ \\
$\langle F^{*}_{\rm fknn} \rangle$ &  36.7&  42.9&  $-$&  $-$&  $-$&  $-$&  34.6&  40.6&  $-$&  $-$&  $-$&  $-$& 28.9&  34.9& $-$&  $-$& $-$&  $-$ \\
$\langle F^{*} \rangle$& 31.7 & 39.5 & 38.6 & 45.0 & 35.9 & 42.8 & 29.5 & 37.2 & 36.0 & 42.4 & 33.9 & 40.2  & 25.6 & 32.0 & 29.7 & 36.1 & 29.9 & 35.4 \\
${\rm STD}_{F^{*}}$& 1.66 & 1.31 & 1.59 & 1.14 & 1.71 & 1.10 & 0.79 & 0.68 & 0.69 & 0.57 & 0.78 & 0.75  & 0.37 & 0.32 & 0.36 & 0.28 & 0.39 & 0.30 \\
$\langle N_{MC} \rangle$& 3.9 & 5.5 & 12.2 & 16.9 & 3.4 & 3.2 & 4.1 & 6.6 & 21.9 & 27.5 & 3.2 & 3.2  & 4.2 & 7.3 & 29.6 & 37.2 & 0.76 & 1.3 \\
$\langle T_{\rm cpu} \rangle$& 0.12 & 0.36 & 0.05 & 0.14 & 0.05 & 0.11 & 0.48 & 1.62 & 0.22 & 0.64 & 0.18 & 0.45  & 1.98 & 6.76 & 1.05 & 2.86 & 0.53 & 1.69 \\
$\langle U^{*} \rangle$& 9e-4 & 3e-3 & 3e-3 & 1e-2 & 4e-9 & 2e-8 & 7e-4 & 3e-3 & 1e-3 & 7e-3 & 2e-9 &  1e-8 & 4e-4 & 2e-3 & 8e-4 & 4e-3 & 1e-9 & 8e-9 \\
\hline
 \multicolumn{19}{c}{$N_c=16$} \\
\hline
$\langle F^{*}_{\rm knn} \rangle$ &  64.8&  67.3&  $-$&  $-$& $-$&  $-$&  63.5&  65.7&  $-$&  $-$&  $-$&  $-$& 58.5 &  61.0& $-$ &  $-$& $-$&  $-$ \\
$\langle F^{*}_{\rm fknn} \rangle$ &  63.2&  66.6&  $-$&  $-$&  $-$&  $-$&  61.4&  65.1&  $-$&  $-$&  $-$&  $-$& 56.3&  60.5& $-$&  $-$& $-$&  $-$ \\
$\langle F^{*} \rangle$& 54.8 & 65.8 & 65.1 & 69.4 & 56.7 & 64.5 & 53.3 & 64.8 & 62.2 & 67.3 & 54.8 & 62.9  & 47.9 & 58.4 & 57.3 & 62.7 & 50.8 & 58.6 \\
${\rm STD}_{F^{*}}$& 1.71 & 1.34 & 1.46 & 1.07 & 1.77 & 1.24 & 0.89 & 0.76 & 0.78 & 0.54 & 0.89 & 0.67  & 0.44 & 0.34 & 0.42 & 0.28 & 0.50 &  0.37 \\
$\langle N_{MC} \rangle$& 8.6 & 11.0 & 21.0 & 29.8 & 9.2 & 8.2 & 8.7 & 12.4 & 40.5 & 51.9 & 9.2 & 7.9  & 8.9 & 13.8 & 61.3 & 73.2 & 8.9 & 7.8 \\
$\langle T_{\rm cpu} \rangle$& 0.23 & 0.67 & 0.07 & 0.18 & 0.09 & 0.15 & 0.94 & 3.09 & 0.36 & 0.91 & 0.32 &  0.61 & 3.96 & 12.78 & 1.84 & 4.76 & 1.28 & 2.49 \\
$\langle U^{*} \rangle$& 8e-4 & 2e-3 & 2e-2 & 1e-1 & 9e-10 & 4e-9 & 7e-4 & 2e-3 & 1e-2 & 9e-2 & 6e-10 &  3e-9 & 5e-4 & 2e-3 & 8e-3 & 6e-2 & 3e-10 & 2e-9 \\
\hline \hline
\end{tabular}
\end{center}
\end{small}
\end{table}

\begin{table}[h]
\begin{small}
\caption{The same classification statistics are listed as in
Table~\ref{tab:ka02nu25}, obtained from realizations of a lognormal random field  with
Whittle-Mat\'{e}rn covariance  ($\kappa=0.5, \nu=2.5$).}
\label{tab:lognka05nu25}
\begin{center}
\begin{tabular}{lcccccccccccccccccc}
\hline \hline
Size    & \multicolumn{6}{c}{L=50}  & \multicolumn{6}{c}{L=100} & \multicolumn{6}{c}{L=200}\\
\hline
Model          & \multicolumn{2}{c}{INNC}  & \multicolumn{2}{c}{PNNC} &  \multicolumn{2}{c}{CNNC} & \multicolumn{2}{c}{INNC}  & \multicolumn{2}{c}{PNNC}  & \multicolumn{2}{c}{CNNC} & \multicolumn{2}{c}{INNC}  & \multicolumn{2}{c}{PNNC} & \multicolumn{2}{c}{CNNC}\\
\hline
$p[\%]$        & $33$  & $66$ & $33$  & $66$ & $33$  & $66$ & $33$  & $66$ & $33$  & $66$ & $33$  & $66$ & $33$  & $66$ & $33$  & $66$ & $33$  & $66$\\
\hline
 \multicolumn{19}{c}{$N_c=8$} \\
\hline
$\langle F^{*}_{\rm knn} \rangle$ &  18.7&  24.3&  $-$&  $-$& $-$&  $-$&  21.5&  27.5&  $-$&  $-$&  $-$&  $-$& 16.5&  22.3& $-$ &  $-$& $-$&  $-$ \\
$\langle F^{*}_{\rm fknn} \rangle$ &  17.3&  22.3&  $-$&  $-$&  $-$&  $-$&  19.9&  25.4&  $-$&  $-$&  $-$&  $-$& 15.3&  20.3& $-$&  $-$& $-$&  $-$ \\
$\langle F^{*} \rangle$& 15.2 & 20.9 & 18.4 & 23.8 & 16.3 & 21.6 & 16.8 & 23.1 & 20.6 & 26.6 & 18.2 & 24.0  & 13.7 & 18.8 & 15.6 & 20.9 & 14.9 & 19.8 \\
${\rm STD}_{F^{*}}$& 1.29 & 1.17 & 1.31 & 0.95 & 1.15 & 1.13 & 0.70 & 0.63 & 0.76 & 0.52 & 0.67 & 0.61  & 0.30 & 0.30 & 0.35 & 0.31 & 0.31 & 0.31 \\
$\langle N_{MC} \rangle$& 3.7 & 4.6 & 6.0 & 10.0 & 7.0 & 8.4 & 4.5 & 5.7 & 15.0 & 20.3 & 6.9 & 7.1  & 4.6 & 6.5 & 19.5 & 20.6 & 5.1 &  5.3\\
$\langle T_{\rm cpu} \rangle$& 0.08 & 0.19 & 0.04 & 0.12 & 0.07 & 0.14 & 0.32 & 0.87 & 0.18 & 0.57 & 0.24 & 0.53  & 1.26 & 3.27 & 0.86 & 2.34 & 0.83 & 2.28 \\
$\langle U^{*} \rangle$& 6e-4 & 1e-3 & 1e-4 & 7e-4 & 6e-9 & 7e-9 & 6e-4 & 2e-3 & 4e-5 & 2e-4 & 2e-10 & 9e-10  & 3e-4 & 1e-3 & 4e-6 & 8e-6 & 8e-11 & 6e-10 \\
\hline
 \multicolumn{19}{c}{$N_c=16$} \\
\hline
$\langle F^{*}_{\rm knn} \rangle$ &  37.1&  41.7&  $-$&  $-$& $-$&  $-$&  42.7&  47.1&  $-$&  $-$&  $-$&  $-$& 35.1&  40.2& $-$ &  $-$& $-$&  $-$ \\
$\langle F^{*}_{\rm fknn} \rangle$ &  34.9&  40.0&  $-$&  $-$&  $-$&  $-$&  40.4&  45.8&  $-$&  $-$&  $-$&  $-$& 33.1&  38.7& $-$&  $-$& $-$&  $-$ \\
$\langle F^{*} \rangle$& 28.8 & 37.9 & 37.0 & 42.4 & 29.7 & 37.1 & 32.0 & 42.2 & 41.7 & 48.5 & 31.5 &  41.4 & 26.1 & 35.1 & 34.0 & 40.8 & 27.2 & 35.6 \\
${\rm STD}_{F^{*}}$& 1.85 & 1.43 & 1.62 & 1.07 & 1.73 & 1.41 & 0.96 & 0.79 & 0.75 & 0.56 & 1.15 & 0.74  & 0.35 & 0.40 & 0.37 & 0.30 & 0.46 & 0.42 \\
$\langle N_{MC} \rangle$& 8.9 & 10.4 & 13.8 & 21.4 & 18.8 & 18.8 & 10.3 & 12.5 & 35.7 & 45.3 & 19.9 &  17.5 & 10.8 & 14.1 & 52.1 & 63.9 & 19.7 & 17.2 \\
$\langle T_{\rm cpu} \rangle$& 0.16 & 0.33 & 0.06 & 0.16 & 0.12 & 0.20 & 0.61 & 1.53 & 0.31 & 0.84 & 0.54 & 0.84  & 2.35 & 5.7 & 1.63 & 4.31 & 2.14 & 3.44 \\
$\langle U^{*} \rangle$& 7e-4 & 2e-3 & 8e-4 & 6e-3 & 2e-9 & 2e-10 & 7e-4 & 2e-3 & 7e-4 & 6e-3 & 3e-11 &  2e-10 & 5e-4 & 2e-3 & 1e-4 & 2e-3 & 9e-12 &  7e-11\\
\hline \hline
\end{tabular}
\end{center}
\end{small}
\end{table}

 As expected, the
misclassification rate overall decreases with increasing $N_G$ and
increases proportionally with $p$. Comparing the performance of
different models at fixed $N_{G}$, $p$, $N_c$ and distributional
assumptions, the INNC model performs uniformly better for $N_c=8$.
For $N_c=16$ and $(\kappa,\nu)=(0.2,2.5)$, the INNC model gives in
general the lowest $\langle F^{*} \rangle$, except one case in which
the CNNC model performs best. Generally, the CNNC model is expected
to perform better when the cross-class correlations make finite
contributions to the correlation energy. This  is likely as the
number of classes or the spatial variation of the data increase. The
case of $N_c=16$ and $(\kappa,\nu)=(0.5,2.5)$ provides an example
combining higher $N_c$ and shorter-range variations than $N_c=8$ and
$\kappa=0.2$. This example shows a clear advantage of the CNNC model
(especially at $p=0.66$) over the others. However, since the
interactions are restricted to nearest neighbors, cross-class
correlations in CNNC can increase the sensitivity to noisy or
``rough'' data. This is evidenced in the
 $N_c=16$ case for $(\kappa,\nu)=(0.5,1.5)$. The random field realizations
for $\nu=1.5$ are only once differentiable, in contrast with the
$\nu=2.5$ case, where the realizations are twice differentiable.
For $\nu=1.5$ the classification performance of the
CNNC model approaches that of the INNC model. The highest
misclassification rate among the spin models investigated is
displayed consistently by the PNNC model. This could be attributed
to the fact that it incorporates less information than its
counterparts. Namely, the INNC model sequentially uses the lower
level classification results in the higher level estimates. On the
other hand, the CNNC model differentiates between neighbors of
various classes. In contrast, the PNNC model can only distinguish if
the neighbors belong to the same or different classes. As we show
below, the Potts model leads to degenerate states, which are
undesirable for classification purposes.

\subsection{Impact of Initial State Selection} 
\begin{figure}[!t]
  \begin{center}
    \subfigure[Misclassification rate]{\label{fig:ass_frm}
    \includegraphics[scale=0.4]{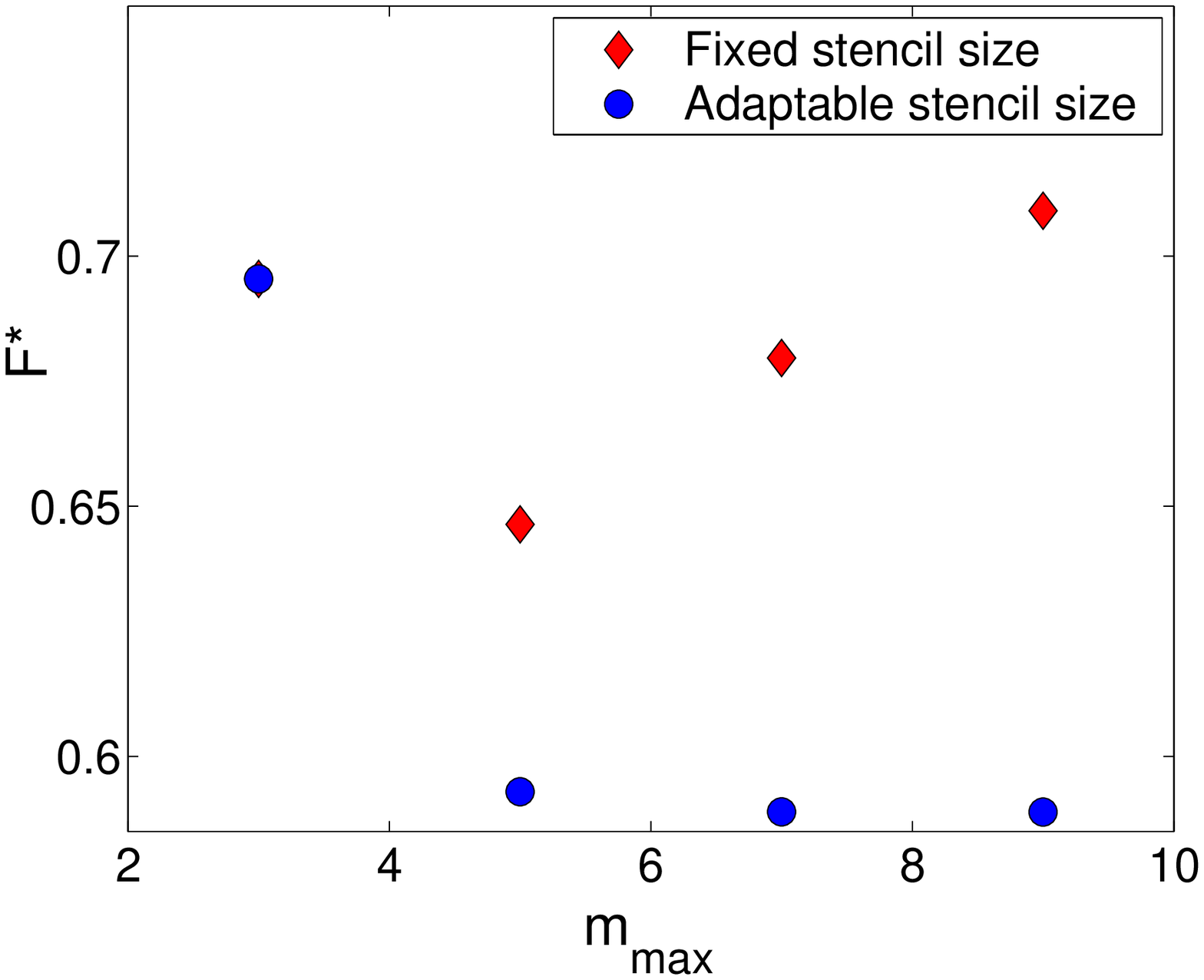}}
    \subfigure[CPU time]{\label{fig:ass_cpu}
    \includegraphics[scale=0.4]{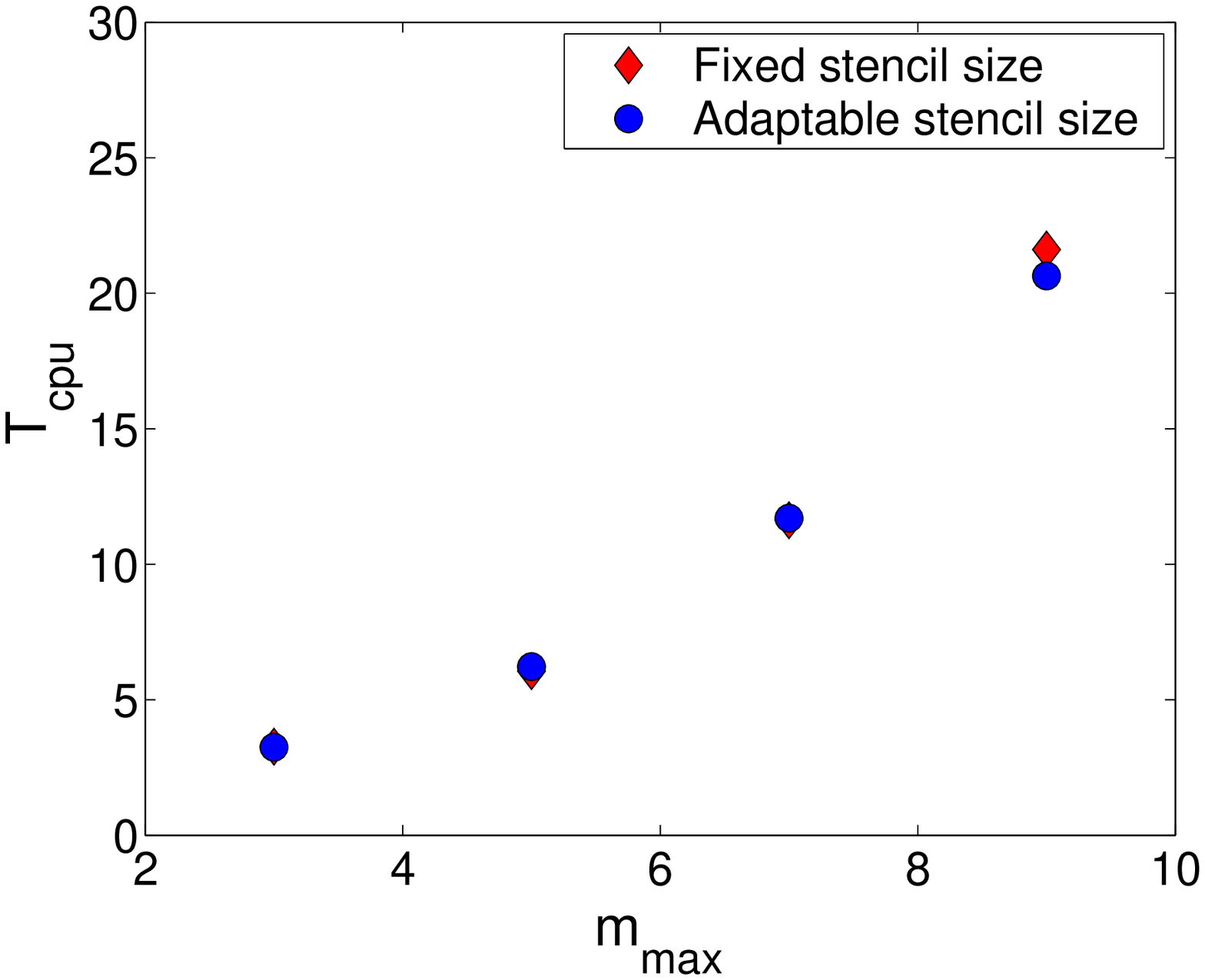}}
  \end{center}
  \caption{(Color online) Dependence of misclassification
  rate, $F^{\ast},$ on the maximum  stencil size $m_{\rm max}$ used in
   the initialization with both adaptable and fixed stencil size (a).
   Dependence of the CPU time
  on $m_{\rm max}$ (b).
  The 16-class INNC model is applied to a sample  generated from a
  realization of a Gaussian random field
  with Whittle-Mat\'{e}rn covariance ($\kappa=0.5$, $\nu=1.5$)
  on a square  grid ($L=200$)  by removing
  $66\%$ of the points.}
  \label{fig:ass}
\end{figure}

Fig.~\ref{fig:ass}
illustrates the behavior of the misclassification rates, $F^{\ast}$,
and the CPU times versus $m_{\rm max}$, considering fixed (FSS) and
adaptable (ASS) stencil size approaches. Relatively small $m_{\rm
max}=5, 7$, for FSS and ASS respectively, lead to the lowest $F^{\ast}$ values.
The computational time increases quadratically with
$m_{\rm max}$. There is no significant difference in computational
time between the two methods. However, using ASS leads to a significantly lower misclassification rate.

For the results obtained in
Tables~(\ref{tab:ka02nu25})-(\ref{tab:lognka05nu25}), the initial
state was based on majority vote with ASS. This accelerates the
relaxation process and prevents trapping in a ``poor" local minimum.
To test the sensitivity of the spin models on different initial
states, we repeat the simulations using randomly assigned initial
states for  Gaussian data with $(\kappa,\nu)=(0.5,2.5)$ on a $L=200$
grid. The results for different models and two values of $N_c$ are
given in Table~\ref{tab:ka05nu25_rand}. The greatest impact on the
classification performance is observed for the INNC model. On the
other hand, the CNNC model is the most robust with respect to
changes of the initial state, especially for lower $p$ and  higher
$N_c$. For instance, at $p=0.33$ and $N_c=16$, there is practically
no difference between the misclassification rates obtained using
initial states based on random versus ASS majority rule  selection
of the spins. Using a random initial state increases  by $73 \%$ the
number of MC sweeps needed to achieve equilibrium. However, the CPU
time is increased only by $17 \%$, because the random assignment
strategy leads to faster determination of the initial state.

\begin{table}[h]
\begin{small}
\caption{Classification results using the 8- and 16-level  INNC,
PNNC, and CNNC models for realizations of a Gaussian random field $\sim N(50,10)$
with Whittle-Mat\'{e}rn covariance $(\kappa=0.5, \nu=2.5)$ and $L=200$. The
simulations start from random initial states. The same
statistics are listed as in Table~\ref{tab:ka02nu25}. }
\label{tab:ka05nu25_rand}
\begin{center}
\begin{tabular}{lcccccccccccc}
\hline \hline
& \multicolumn{6}{c}{$N_c=8$} & \multicolumn{6}{c}{$N_c=16$}\\
\hline
Model       & \multicolumn{2}{c}{INNC}  & \multicolumn{2}{c}{PNNC} & \multicolumn{2}{c}{CNNC} & \multicolumn{2}{c}{INNC}  & \multicolumn{2}{c}{PNNC} & \multicolumn{2}{c}{CNNC}\\
\hline
$p[\%]$        & $33$  & $66$ & $33$  & $66$ & $33$  & $66$ & $33$  & $66$ & $33$  & $66$ & $33$  & $66$\\
\hline
$\langle F^{*} \rangle$&  34.3&   91.3&  25.6&  42.5&  22.0&  36.3 &  67.1&   98.8&  56.5&  69.8&  32.7&  49.4\\
${\rm STD}_{F^{*}}$&  0.50&   0.20&  0.34&  0.43&  0.37&  0.47 &  18.3&   0.07&  0.40&  0.43&  0.59&  0.64\\
$\langle N_{MC} \rangle$&  7.0&   3.0&  27.3&  74.9&  31.7&  68.9 &  13.1&   5.2&  23.3&  32.3&  42.4&  49.8\\
$\langle T_{\rm cpu} \rangle$&  1.15&   1.24&  0.63&  2.54&  2.12&  4.92 &  2.28&   2.0&  0.53&  1.16&  2.82&  3.73\\
$\langle U^{*} \rangle$&  1e-1&   1e-1&  2e-7&  8e-4&  2e-11&  3e-8 &  9e-2&   6e-2&  4e-6&  3e-6&  9e-12&  4e-11\\
\hline \hline
\end{tabular}
\end{center}
\end{small}
\end{table}

\subsection{Spatial Degeneracy} The higher sensitivity of the INNC
model to the initial state  is due  to the sequential spin
estimation, which propagates the misclassification from lower to
higher levels, thus resulting in a higher overall misclassification
rate. The misclassification at lower levels can occur due to the
presence of degenerate states which correspond to different spatial
configurations, even though their energies  are numerically identical.
Generally, the degeneracy
increases with $p$, as a result of relaxing the spatial constraints.
As the size of the prediction set is reduced at higher levels, due
to the inclusion of the classified lower levels in the sampling set,
the degeneracy is accordingly diminished. However, the lower level
misclassifications propagate to higher levels. A high level of
degeneracy is also responsible for the relatively poor
classification performance of the PNNC model; since the latter does
not include cross-class correlations, the energy contributions of
all the nearest neighbor pairs that do not belong in the same class
are zero. We believe that the robustness of the CNNC model and its
competitive classification performance is   partially  due to the
reduction of degeneracy achieved by differentiating between the
energy contributions of neighbors belonging to different classes.

\begin{table}[h]
\begin{small}
\caption{Degeneracy and residual values of the cost  function
corresponding to 100 random initial states. The sample is a
realization of a Gaussian random field $\sim N(50,10)$ with
Whittle-Mat\'{e}rn covariance
 $(\kappa=0.5, \nu=2.5)$ on a grid of size $L=50$.}
\label{tab:ka05nu25_deg}
\begin{center}
\begin{tabular}{lcccccccccccc}
\hline \hline
& \multicolumn{6}{c}{$N_c=8$} & \multicolumn{6}{c}{$N_c=16$}\\
\hline
Model       & \multicolumn{2}{c}{INNC}  & \multicolumn{2}{c}{PNNC} & \multicolumn{2}{c}{CNNC} & \multicolumn{2}{c}{INNC}  & \multicolumn{2}{c}{PNNC} & \multicolumn{2}{c}{CNNC}\\
\hline
$p[\%]$        & $33$  & $66$ & $33$  & $66$ & $33$  & $66$ & $33$  & $66$ & $33$  & $66$ & $33$  & $66$\\
\hline
degeneracy &  6 &   8&  77&  10&  64 &  7&  8&   4&  61&  53&  6 &  0\\
$\langle U^{*} \rangle$&  8e-2&   1e-1&  7e-7&  4e-4&  1e-10&  7e-9 &  8e-2&   6e-2&  4e-6&  5e-6&  5e-11&  1e-10\\
\hline \hline
\end{tabular}
\end{center}
\end{small}
\end{table}

\begin{figure}[!t]
  \begin{center}
    \subfigure[$p=0.33$]{\label{fig:ka05nu25_q8_msng33_deg}
    \includegraphics[scale=0.4]{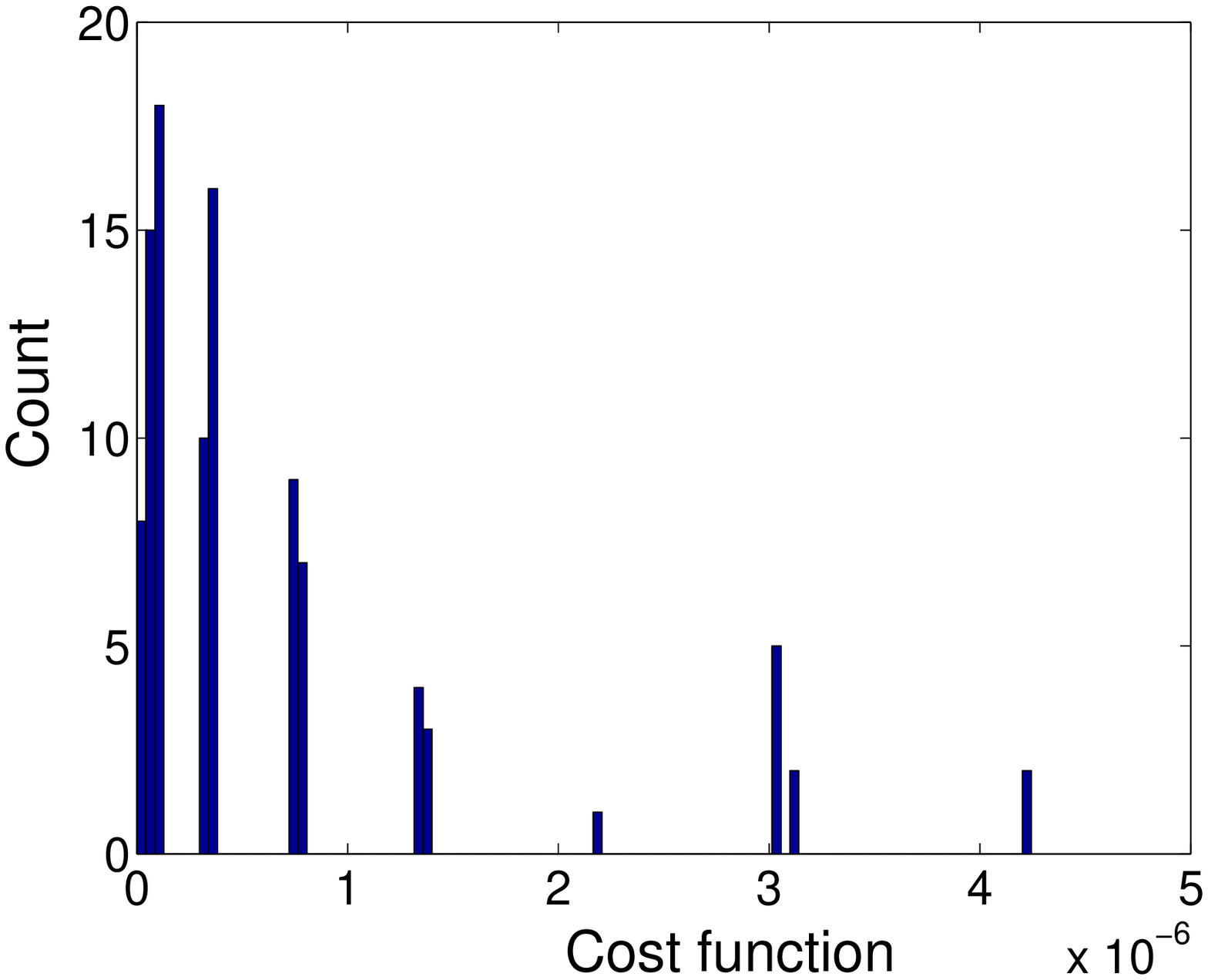}}
    \subfigure[$p=0.66$]{\label{fig:ka05nu25_q8_msng66_deg}
    \includegraphics[scale=0.4]{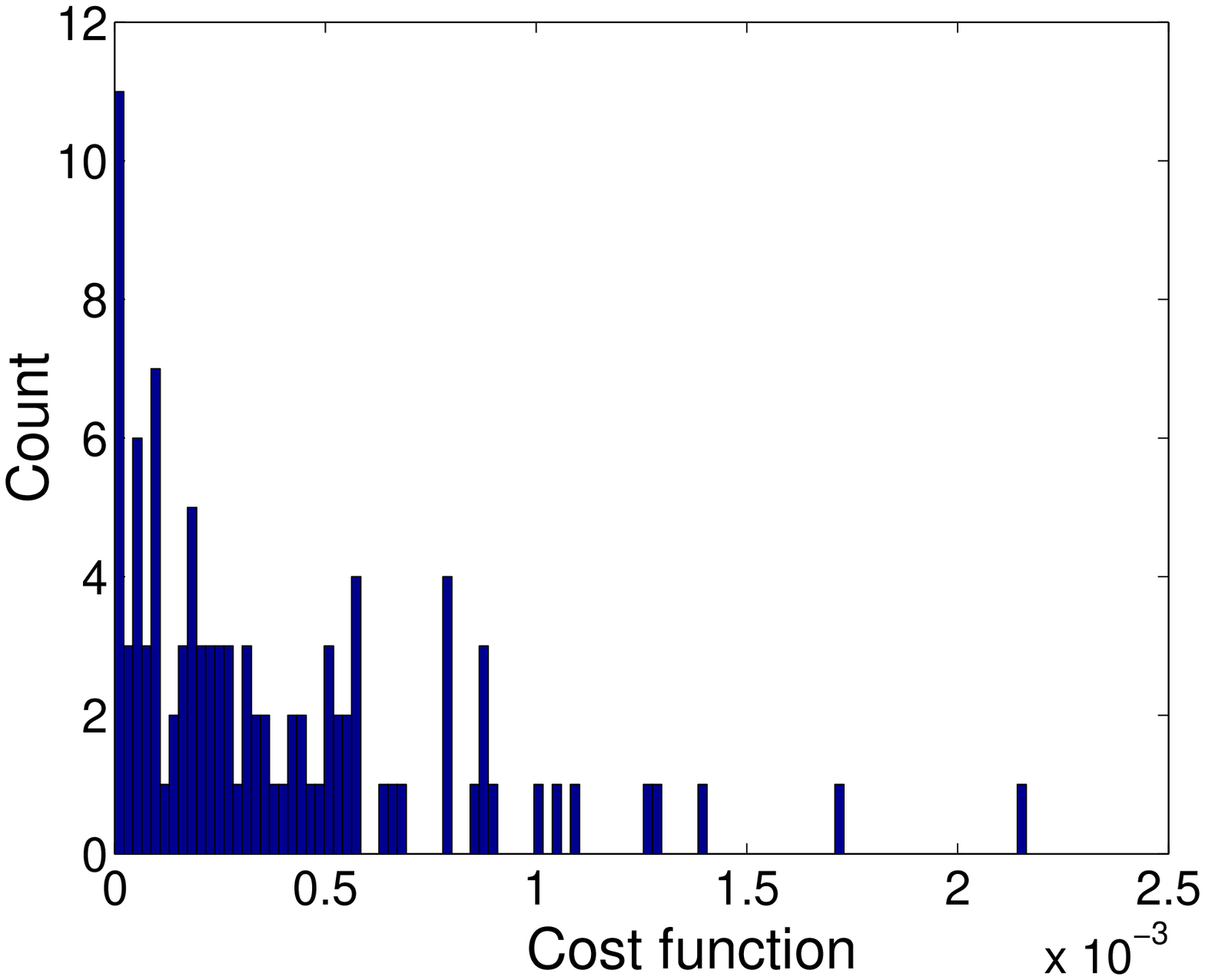}}
  \end{center}
  \caption{(Color online) Histograms of the residual values of the cost
  function obtained from the PNNC model
  for one sampling realization and 100 different random
  initial states, with $N_c=8$ and different
  values of $p$. Each bar may include states with similar but not
identical values of the cost function due to finite bar width. 
Note different scales of the abscissas.}
  \label{fig:ka05nu25_q8_deg}
\end{figure}

To investigate the degeneracy of the final class configurations
obtained by the different spin models, we use the same  sampling
configuration on a grid of size $L=50$. We then run the MC
simulation  for $100$ different (randomly assigned) sets of initial
values at the prediction points. We measure the degeneracy as the
number of final configurations (states) with the same residual
cost function value  as at least
one other state from which they differ in the
 misclassification rate. The results are summarized in
Table~\ref{tab:ka05nu25_deg}. Note that for the INNC model, the
degenerate states  only appear at the first ($q=1$) level. The
highest degeneracy is observed in the PNNC model. The histograms of
the residual cost function value obtained for the PNNC model at
$p=0.33$ and $p=0.66$ are shown in Fig.~\ref{fig:ka05nu25_q8_deg}.
Intuitively, one would expect the degeneracy to increase with $p$.
However, the opposite tendency is observed in the simulations. The
overall shift of $\langle U^{*} \rangle$ to higher values and the
shape of the histogram (Fig.~\ref{fig:ka05nu25_q8_deg}) suggest that
the reduction
 of degeneracy is due to a scattering of energy levels. This can be
viewed as a result of the cost function developing multiple local
minima (multimodality). A reduction of the degeneracy  is also
observed with increasing $L$: for example, the configurations
produced by the CNNC model for $L=200$ do not exhibit spatial
degeneracy for any of  the $N_c$ and $p$ combinations tested.

\begin{figure}[!t]
  \begin{center}
    \includegraphics[scale=0.5]{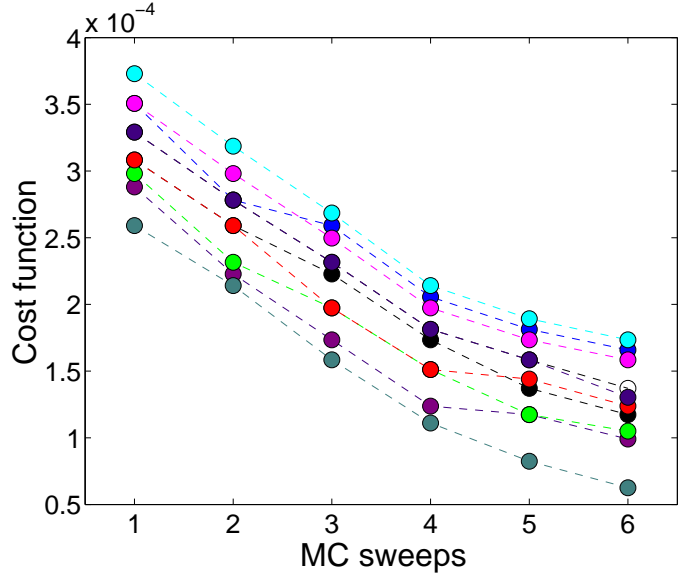}
  \end{center}
  \caption{(Color online) Cost function evolution versus the number of full (over the entire grid) MC sweeps.
  Ten initial states are generated by the majority rule with
adaptable stencil size, leading to ten $U^{\ast}$ evolution curves.
The curves are based on the 8-class PNNC model simulations performed
on a single realization of a  Gaussian random field
$(\kappa=0.5,\nu=2.5)$ on a grid of size $L=50$ from which $33\%$ of
the points are randomly removed. }
  \label{fig:one_samp_10u}
\end{figure}

\subsection{Classification Uncertainty}

\begin{figure}[!t]
  \begin{center}
    \subfigure[Cost function (PNNC, $L=50$)]
    {\label{fig:one_samp_potts_u}
    \includegraphics[scale=0.4]{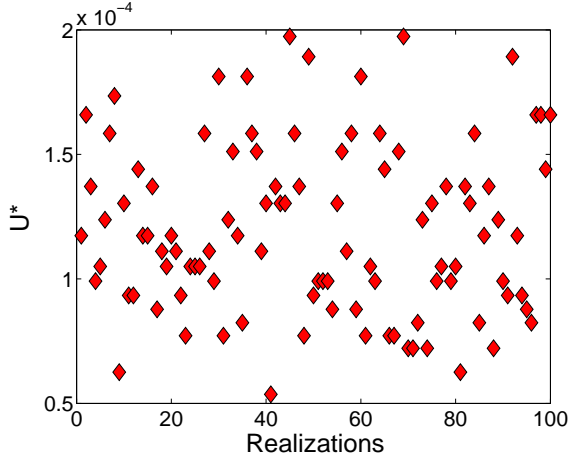}}
    \subfigure[Misclass. rate (PNNC, $L=50$)]
    {\label{fig:one_samp_potts_f}
    \includegraphics[scale=0.4]{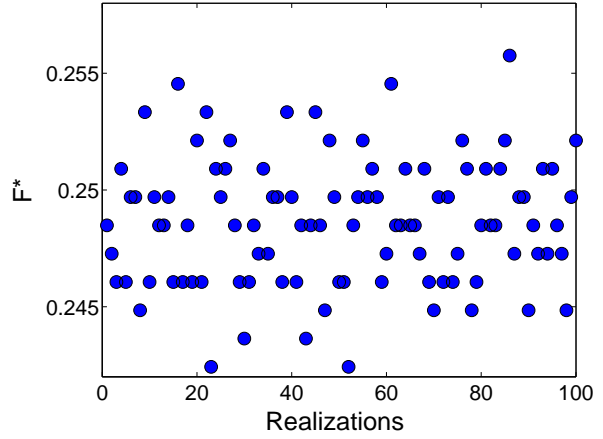}}\\
    \subfigure[Cost function (CNNC, $L=200$)]
    {\label{fig:one_samp_clock_u}
    \includegraphics[scale=0.4]{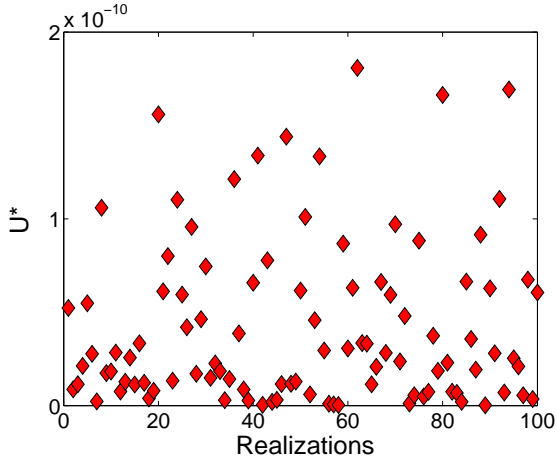}}
    \subfigure[Misclass. rate (CNNC, $L=200$)]
    {\label{fig:one_samp_clock_f}
    \includegraphics[scale=0.4]{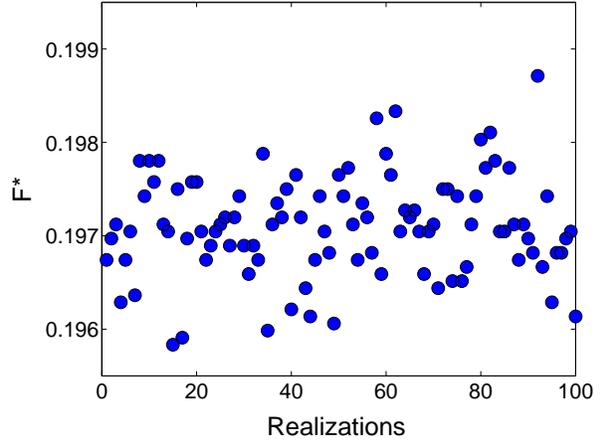}}
  \end{center}
  \caption{(Color online) Residual values of the cost function at
  termination, $U^{\ast},$
  and the misclassification rate,  $F^{\ast},$ for the 8-class PNNC model, (a), (b) and for the
  8-class CNNC model (c), (d).
  The sample is obtained from one realization of a Gaussian random field with
  Whittle-Mat\'{e}rn parameters ($\kappa=0.5, \nu=2.5$) on grids of size $L=50$ (PNNC) and
  $L=200$ (CNNC) respectively, by randomly
  removing $p=0.33$ of the points. The $U^{\ast}, F^{\ast}$ values are obtained
from 100 initial states generated by the majority rule with
adaptable stencil size.}
  \label{fig:one_samp}
\end{figure}

Multimodality and degeneracy introduce an uncertainty in the
classification outcome, starting from different initial states. The
cost function multimodality persists, even in cases where the
degeneracy vanishes. An additional source of uncertainty is
termination criterion II: if the relaxation process stops before the
(local) optimum is reached, the final configuration depends on the
initial state, thus mimicking multimodality of the cost function.
Termination criterion II is arbitrary and  aims at computational
efficiency. Hence, it does not guarantee that the resulting
configuration is in a perfect (quasi-)equilibrium state. Uncertainty
is primarily generated by the ASS random initial assignment of those
prediction points for which majority rule is not established.
Fig.~\ref{fig:one_samp_10u} demonstrates the cost function evolution
during the relaxation process. We use  the 8-class PNNC model
classification on a single realization of a Gaussian random field
($\kappa=0.5, \nu=2.5)$ on a grid of size $L=50$ (high spatial
degeneracy) thinned by $33\%$. The ten curves correspond to initial
states generated by the majority rule with ASS. Different initial
states follow different relaxation paths, leading to different local
minima (based on the current value of termination criterion II).
Figs.~\ref{fig:one_samp_potts_u} and \ref{fig:one_samp_potts_f} are
respectively the residual values of the cost function at termination
and the misclassification rates for the 8-class PNNC model, obtained
from $100$ different initial states generated by the majority rule
with ASS. Figs.~\ref{fig:one_samp_clock_u} and
\ref{fig:one_samp_clock_f} show the same quantities as Figs.
\ref{fig:one_samp_potts_u} and \ref{fig:one_samp_potts_f}, but for
$L=200$ using the 8-class CNNC model (for which no degeneracy was
observed). In both  cases,  the resulting values vary, with the
variations being more pronounced in the degenerate case.

\begin{figure}[!t]
  \begin{center}
    \subfigure[Original]{\label{fig:so_mate_ka05_nu25_L50}
    \includegraphics[scale=0.28]{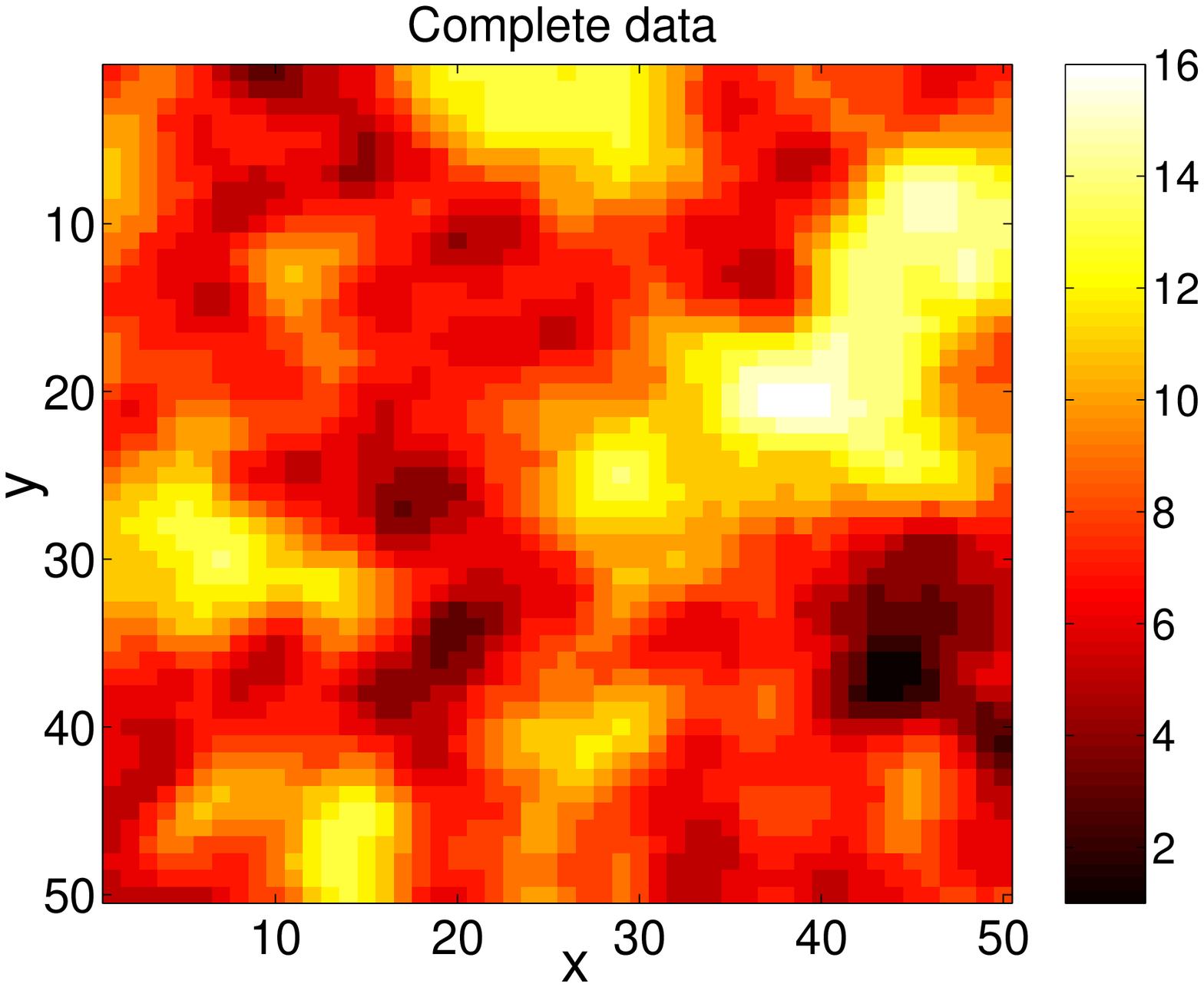}}
    \subfigure[Sample]{\label{fig:sm_mate_ka05_nu25_L50_msng33}
    \includegraphics[scale=0.28]{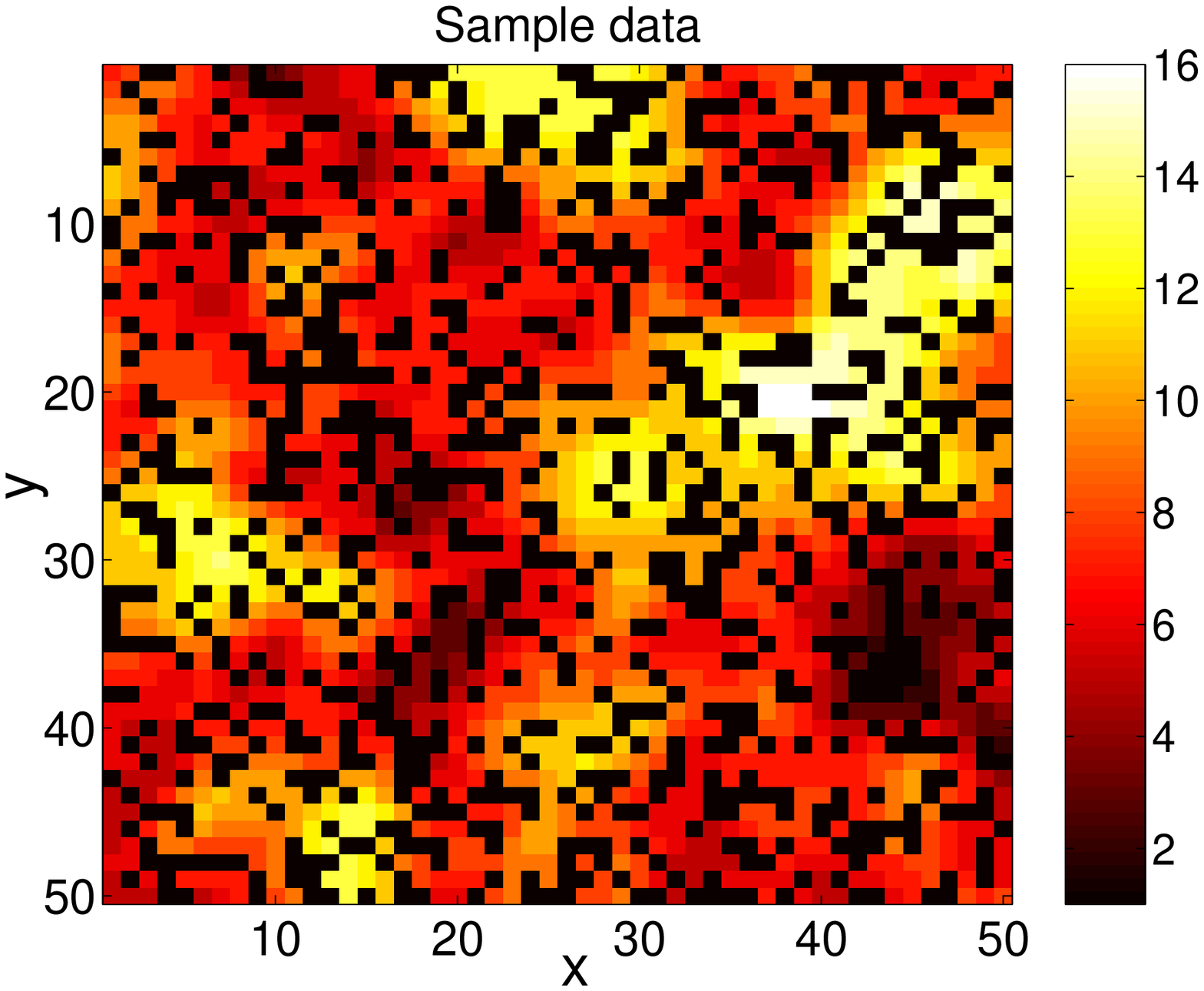}}
    \subfigure[Reconstructed]
    {\label{fig:sr_mate_ka05_nu25_L50_q16_msng33_cnnc}
    \includegraphics[scale=0.28]
    {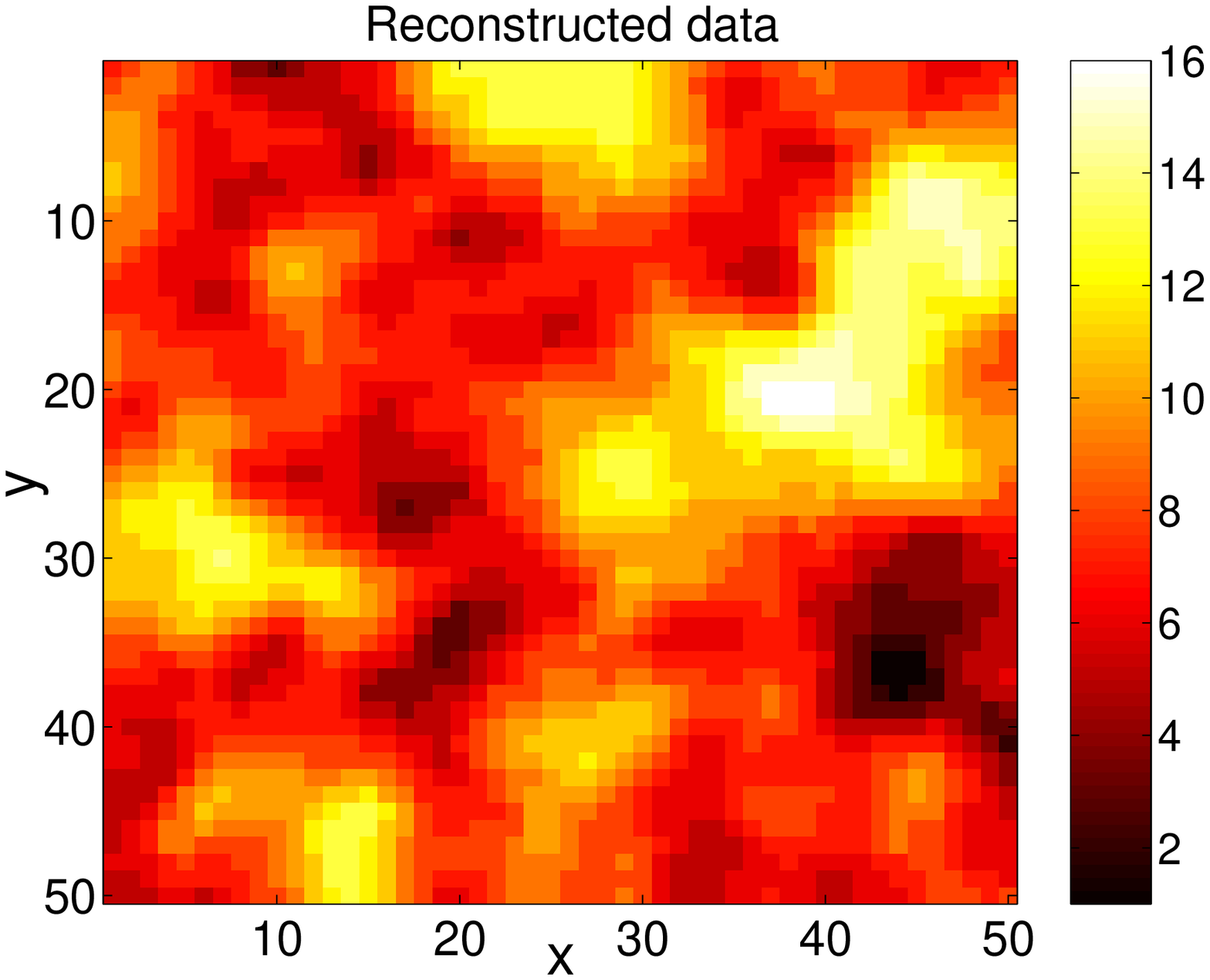}} \\
    \subfigure[Histograms]
    {\label{fig:hist_sr_med_mate_ka05_nu25_L50_q16_msng33_cnnc}
    \includegraphics[scale=0.28]
    {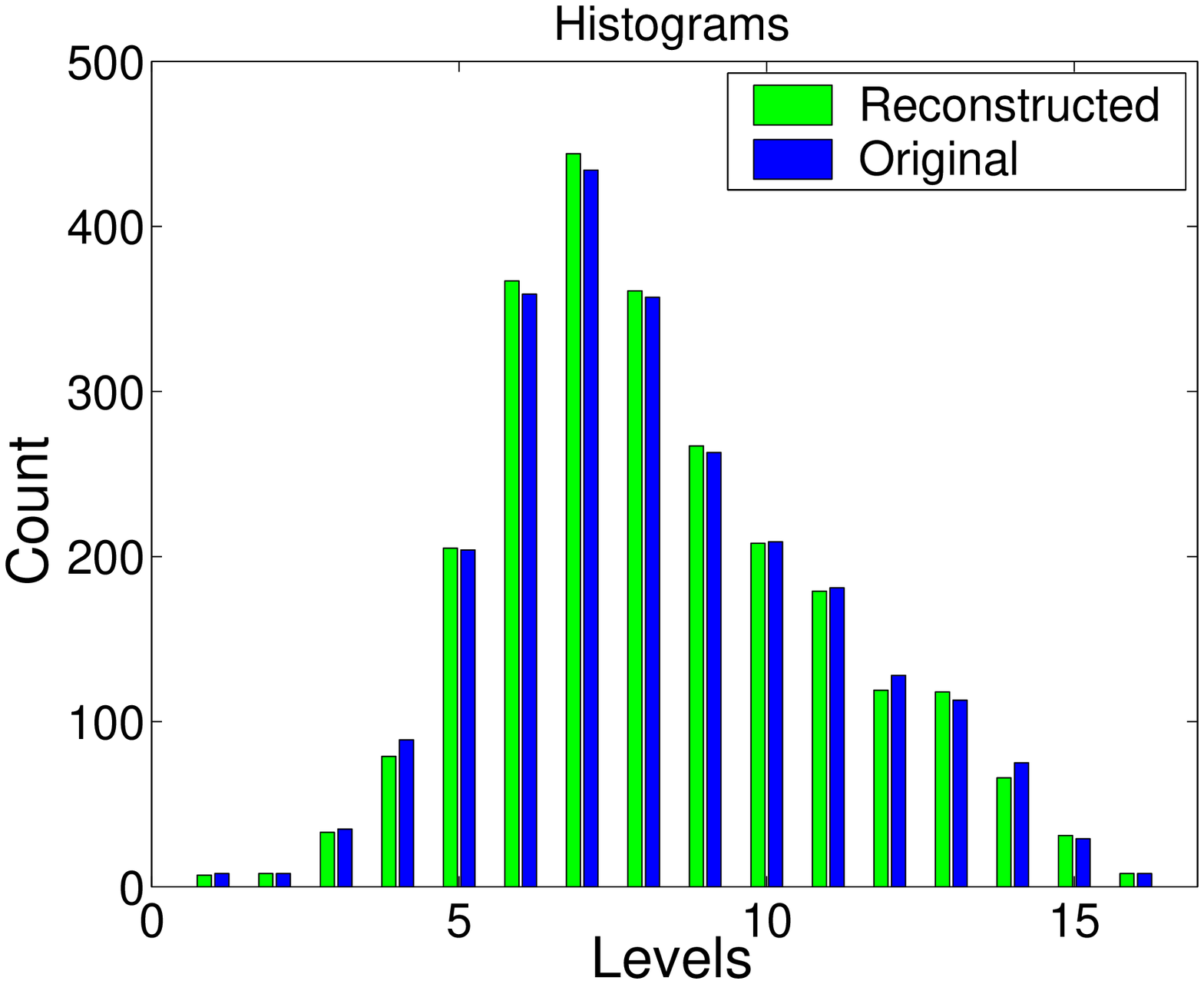}}
    \subfigure[Conf. interval]
    {\label{fig:sr_conf_int_mate_ka05_nu25_L50_q16_msng33_cnnc}
    \includegraphics[scale=0.28]
    {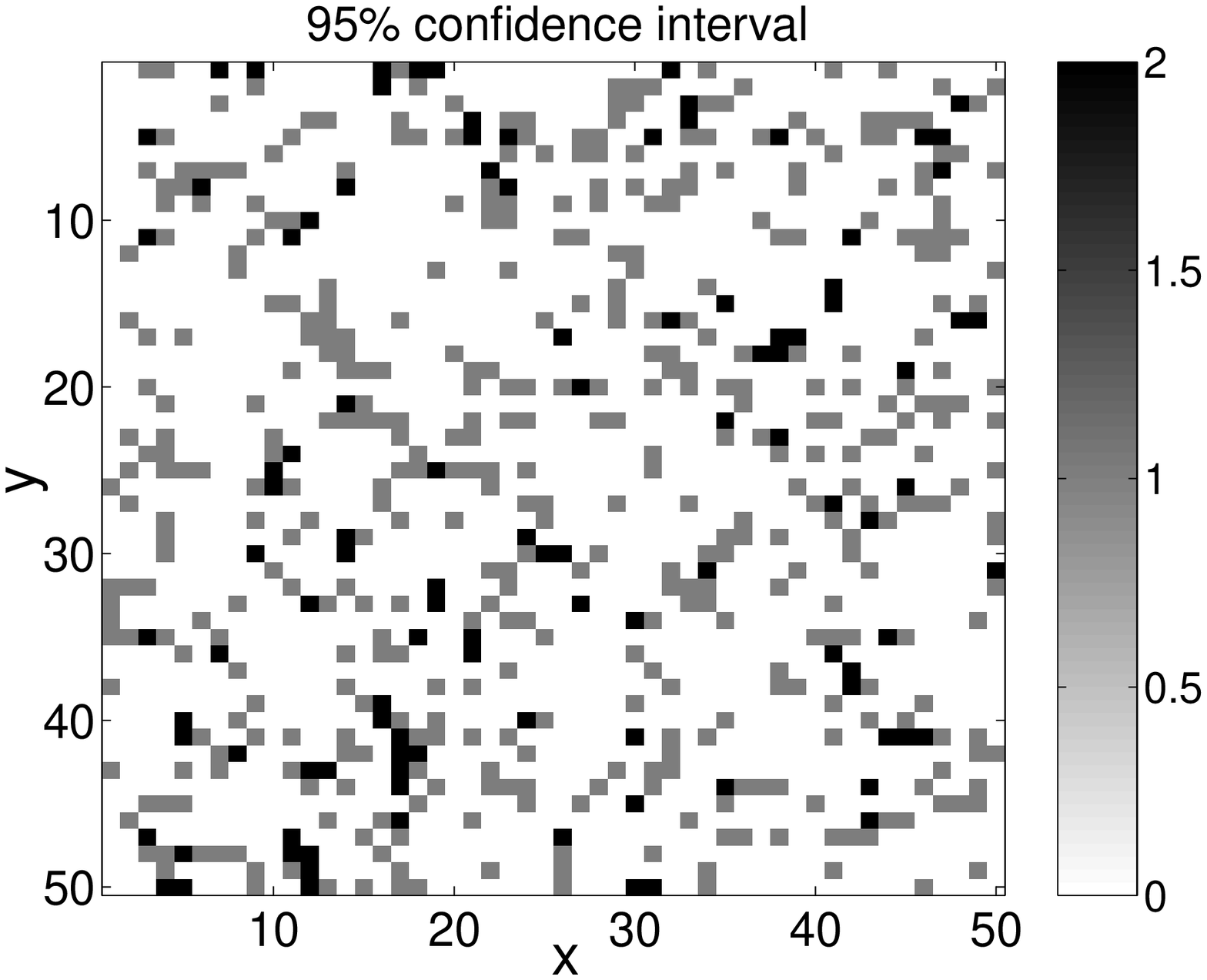}}
    \subfigure[RMSE]
    {\label{fig:sr_rmse_mate_ka05_nu25_L50_q16_msng33_cnnc}
    \includegraphics[scale=0.28]
    {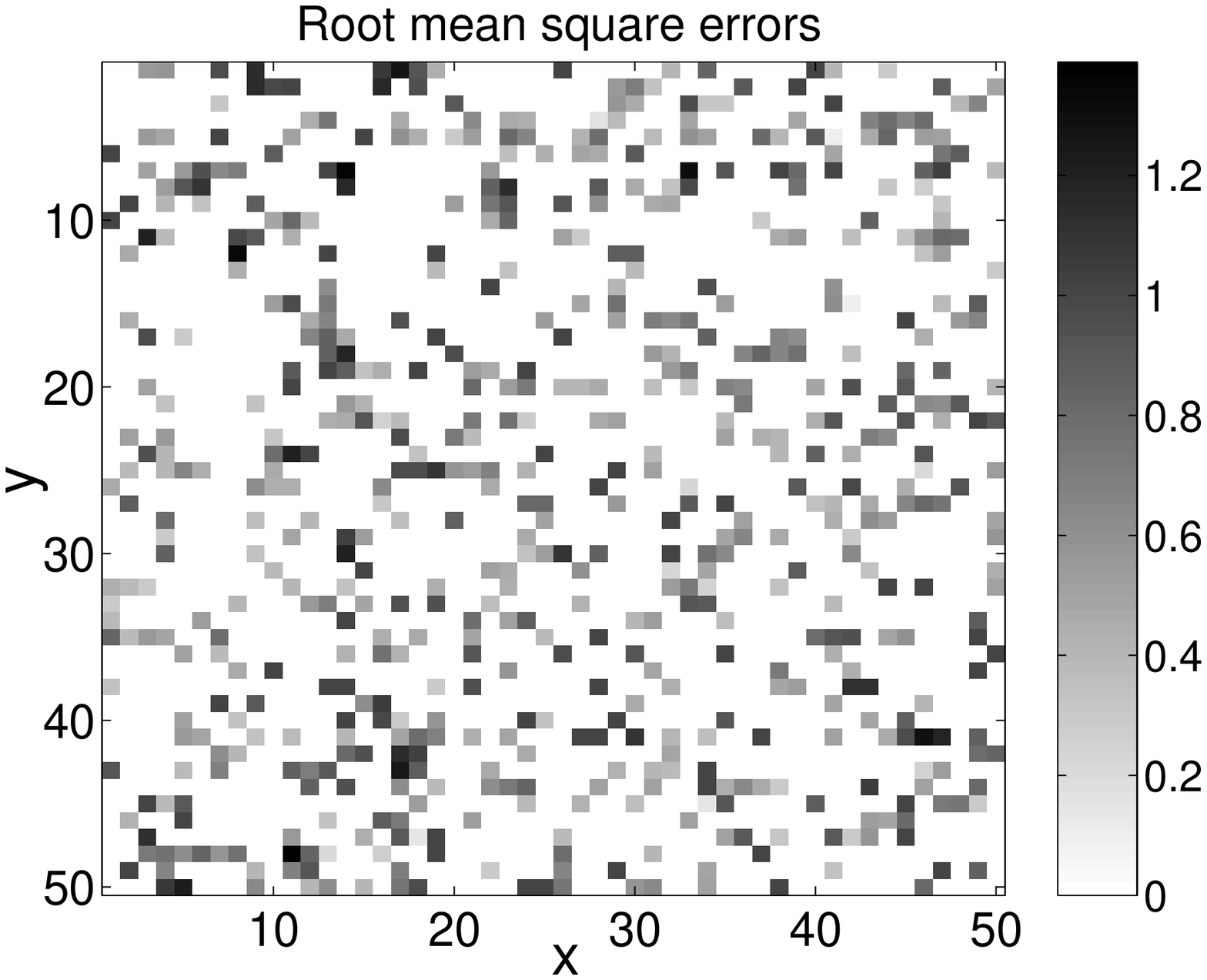}}
  \end{center}
  \caption{(Color online) Classification results obtained from the
  16-class CNNC model. The sample is obtained from one realization of a Gaussian random field with
  Whittle-Mat\'{e}rn parameters ($\kappa=0.5, \nu=2.5$) on a grid of size $L=50$  by randomly
  removing $p=0.33$ of the points.  One hundred reconstructions are generated
 starting from 100 initial states
    obtained by the majority rule with adaptable stencil size.
    Plots (a)-(c)  show
   class maps for the complete realization, the sample with the
   missing data, and the reconstructed image, respectively. The latter is based on the medians of
   the class values  obtained from the 100 reconstructions.
  Plot (d)  compares the class histograms of the
  original and reconstructed data, plot (e) shows the width of the
  $95 \%$ confidence intervals for the class predictions, and plot (f) represents
  the class root mean square error at the prediction points.   }
  \label{fig:one_samp_mate_clock}
\end{figure}

In light of the above, the spin-based classification methods permit the
estimation of the classification uncertainty by sampling over different
initial configurations. This allows us to determine points in space with
increased chance of misclassification. Therefore, multiple Monte
Carlo runs starting from different initial states permit statistical
estimates. This repetitive application can also improve estimation
results compared to a single run. For example, we consider one
sample realization of a Gaussian random field with $\kappa=0.5$,
$\nu=2.5$, on a grid of size $L=50$ reduced by $p=0.33$. A single
simulation run using the 16-class CNNC model gave a
misclassification rate of $F^{*}=36.6\%$. Repeating the simulation
$100$ times, the median misclassification rate drops to
$F^{*}=30.1\%$. The total CPU time increases linearly with the
number of runs, leading to $T_{\rm cpu}=14.8$ seconds. We note that
comparable results can be also obtained with significantly fewer
simulation runs, e.g., $10$. Fig.~\ref{fig:so_mate_ka05_nu25_L50}
shows the complete realization,
Fig.~\ref{fig:sm_mate_ka05_nu25_L50_msng33} the sample with the
missing data points, and
Fig.~\ref{fig:sr_mate_ka05_nu25_L50_q16_msng33_cnnc} the
reconstructed image based on the medians of the estimates from the
100 runs. There is good visual agreement between the original and
the reconstructed images. The class histogram of the reconstruction
also matches satisfactorily that of the data, both shown in
Fig.~\ref{fig:hist_sr_med_mate_ka05_nu25_L50_q16_msng33_cnnc}. Note
that both histograms are asymmetric, even though the random field
values are normally distributed. This is due to the relatively long
correlation length that results in relatively more areas of higher
than lower values as shown in Fig~\ref{fig:so_mate_ka05_nu25_L50}.
Finally,
Fig.~\ref{fig:sr_conf_int_mate_ka05_nu25_L50_q16_msng33_cnnc}
 displays maps representing the width of the $95 \%$ class confidence
interval and
Fig.~\ref{fig:sr_rmse_mate_ka05_nu25_L50_q16_msng33_cnnc} the class root
mean square error at each prediction point.

\subsection{Computational efficiency} Thanks to the vectorization
of the algorithm and judicious choice of the initial state, all the models perform
the classification task very efficiently. The mean CPU time for one
realization ranged from 0.03 seconds for the 8-class PNNC model,
with $L=50$ and $p=0.33,$ up to almost 14 seconds for the 16-class
INNC model, with $L=200$ and $p=0.66$. The INNC displayed the
highest CPU times, in spite of the fact that the equilibration on
the respective levels was extremely fast due to the binary nature of
the data. The major part of the total CPU time was spent on
 the initial state assignments, which is repeated at every level
 unlike the other models. This results in an
approximately linear increase of the INNC CPU time with $N_c$.
The best energy matching, marked by the lowest values of $\langle
U^{*} \rangle$, was observed in the CNNC model, which  can closely approximate
the sample energy due to the flexibility of the multi-level interactions.
In general, the poorest matching is shown by the
PNNC model at high $N_c$ and $p$, indicating a high multimodality of
the cost function and a failure to reach a ``good'' local
optimum.
In comparison with the reference models, the INNC and CNNC  models
gave systematically better classification performance than the best
results obtained by the KNN and the FKNN algorithms.
The PNNC model showed systematically worse classification
performance than the FKNN, but it gave better results than the regular KNN,
 for  relatively small $N_c$ and $p$.

\subsection{Approaching the Continuum Class Limit}
  If $N_c$ is high, the classification problem becomes  a
regression problem. It is then relevant to monitor quantities
 such as prediction errors and the
correlation coefficient between the true (not discretized) and the
predicted values back-transformed to the original data scale.
Table \ref{tab:xyka05nu25} focuses
on the performance of the CNNC model with gradual increase of number
of classes up to $N_c=128$ (extension to XYNNC model), for the
selected size and types of data. The errors (except for MARE)
show a decreasing tendency while the mean correlation
coefficient $MR$ increases.
These tendencies seem to persist up to an optimal value of $N_c$,
above which they level off or even reverse. With increasing $N_c$,
generally, one might expect a dramatic increase in the MC relaxation
time due to an exponential increase of the state space. However,
using the greedy MC algorithm, we only observe a gentle increase in
both the number of MC sweeps and the CPU time, while achieving
excellent optimization results in terms of very low residual values
of the cost function.

\begin{table}[h]
\begin{small}
\caption{Validation statistics of ``discrete-level interpolation''
obtained by the CNNC model, for samples of Gaussian and lognormal
random fields on grids of size $L=100$ with Whittle-Mat\'{e}rn
parameters $\kappa=0.5,\nu=2.5$. MAAE: Mean average absolute error.
MARE: Mean average relative error. MAARE: Mean average  absolute
relative error. MRASE: Mean root average square error. MR: Mean
Correlation coefficient. First, averages
are evaluated over the prediction points for each
realization, and then means are calculated over an
ensemble of $100$ realizations.}
\label{tab:xyka05nu25}
\begin{center}
\begin{tabular}{lcccccccccccc}
\hline \hline
Data    & \multicolumn{6}{c}{Normal}  & \multicolumn{6}{c}{Lognormal} \\
\hline
\# of classes    & \multicolumn{2}{c}{$N_c=32$}  & \multicolumn{2}{c}{$N_c=64$} &  \multicolumn{2}{c}{$N_c=128$} & \multicolumn{2}{c}{$N_c=32$}  & \multicolumn{2}{c}{$N_c=64$} &  \multicolumn{2}{c}{$N_c=128$} \\
\hline
$p[\%]$        & $33$  & $66$ & $33$  & $66$ & $33$  & $66$ & $33$  & $66$ & $33$  & $66$ & $33$  & $66$\\
\hline
MAAE& 1.32 & 1.96 & 1.24 & 1.93 & 1.21 & 1.93 & 11.88 & 16.97 & 10.97 & 16.51 & 10.73 &  16.51   \\
MARE[\%]& -0.28  & -0.49 & -0.29 & -0.53 & -0.30 & -0.54 & -0.39 & -0.12 & -0.58 & -0.39 & -0.81 & -0.84    \\
MAARE[\%]& 2.78  & 4.14 & 2.60 & 4.07 & 2.56 & 4.08 & 7.78 & 10.68 & 7.09 & 10.33 & 6.95 & 10.42    \\
MRASE& 1.71 & 2.56 & 1.61 & 2.53 & 1.59 & 2.54 & 16.12 & 23.85 & 15.10 & 23.45 & 14.86 &  23.41    \\
MR[\%]& 98.46 & 96.48 & 98.66 & 96.63 & 98.71 & 96.61 & 97.97 & 95.60 & 98.25 & 95.79 & 98.31 &  95.79   \\
$\langle N_{MC} \rangle$ & 33.6 & 28.3 & 37.1 & 31.1 & 38.7 & 32.7 & 28.8 & 24.7 & 33.0 & 27.9 & 34.8 &  29.5   \\
$\langle T_{\rm cpu} \rangle$ [s]& 0.78 & 1.13 & 1.03 & 1.50 & 1.24 & 2.04 & 0.76 & 1.14 & 0.97 & 1.35 & 1.09 &  1.87   \\
$\langle U^{*} \rangle$ & 8e-12 & 7e-11 & 7e-12 & 4e-11 & 6e-12 & 5e-11 & 1e-11 & 7e-11 & 7e-12 & 6e-11 & 7e-12 &  5e-11   \\
\hline \hline
\end{tabular}
\end{center}
\end{small}
\end{table}


\section{Applications to Real Data}
\label{sec:real}

\subsection{Remotely sensed data} 
\begin{figure}[!t]
  \begin{center}
    \subfigure[Map]{\label{fig:orig_map}
    \includegraphics[scale=0.4]{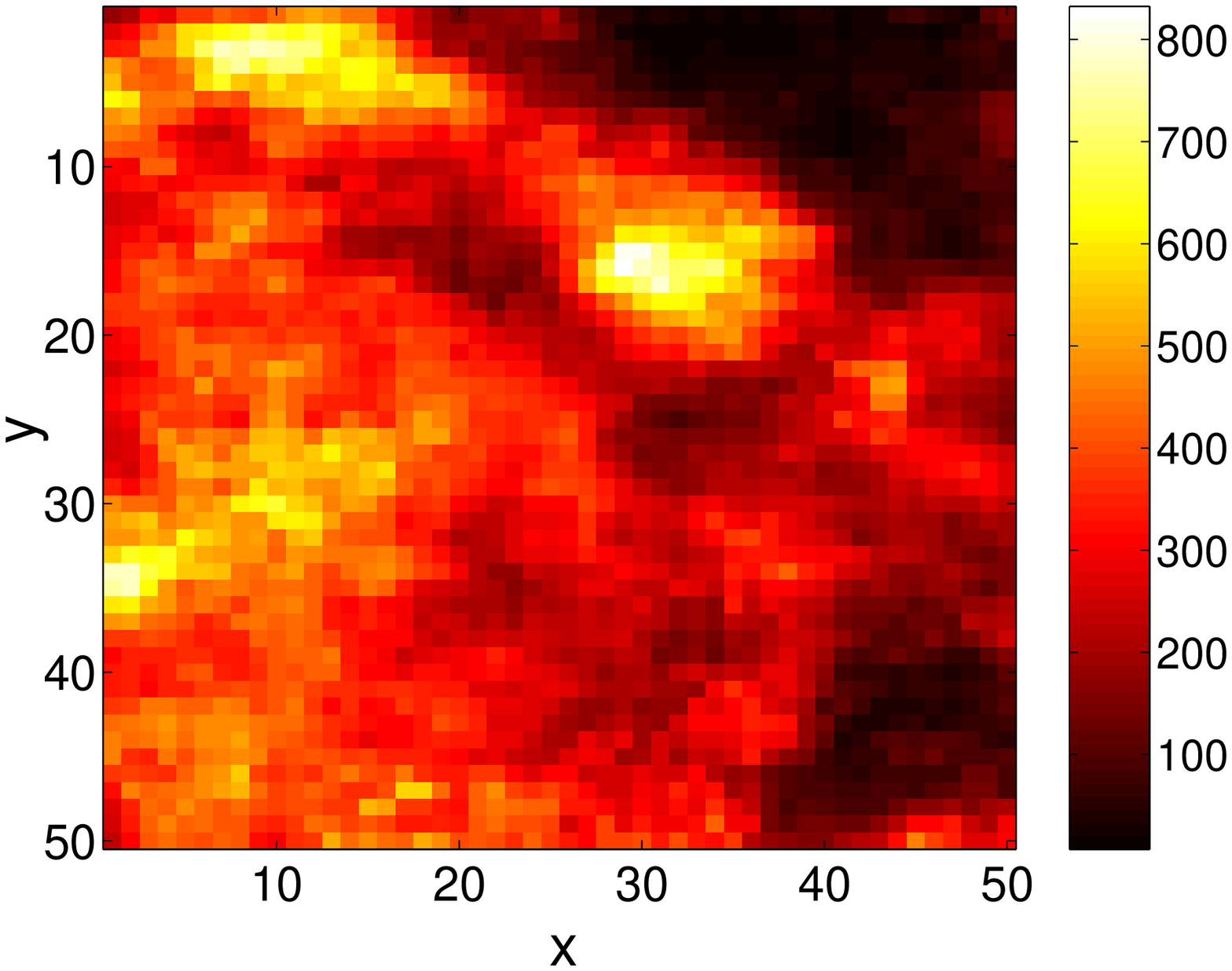}}
    \subfigure[Histogram]{\label{fig:orig_hist}
    \includegraphics[scale=0.4]{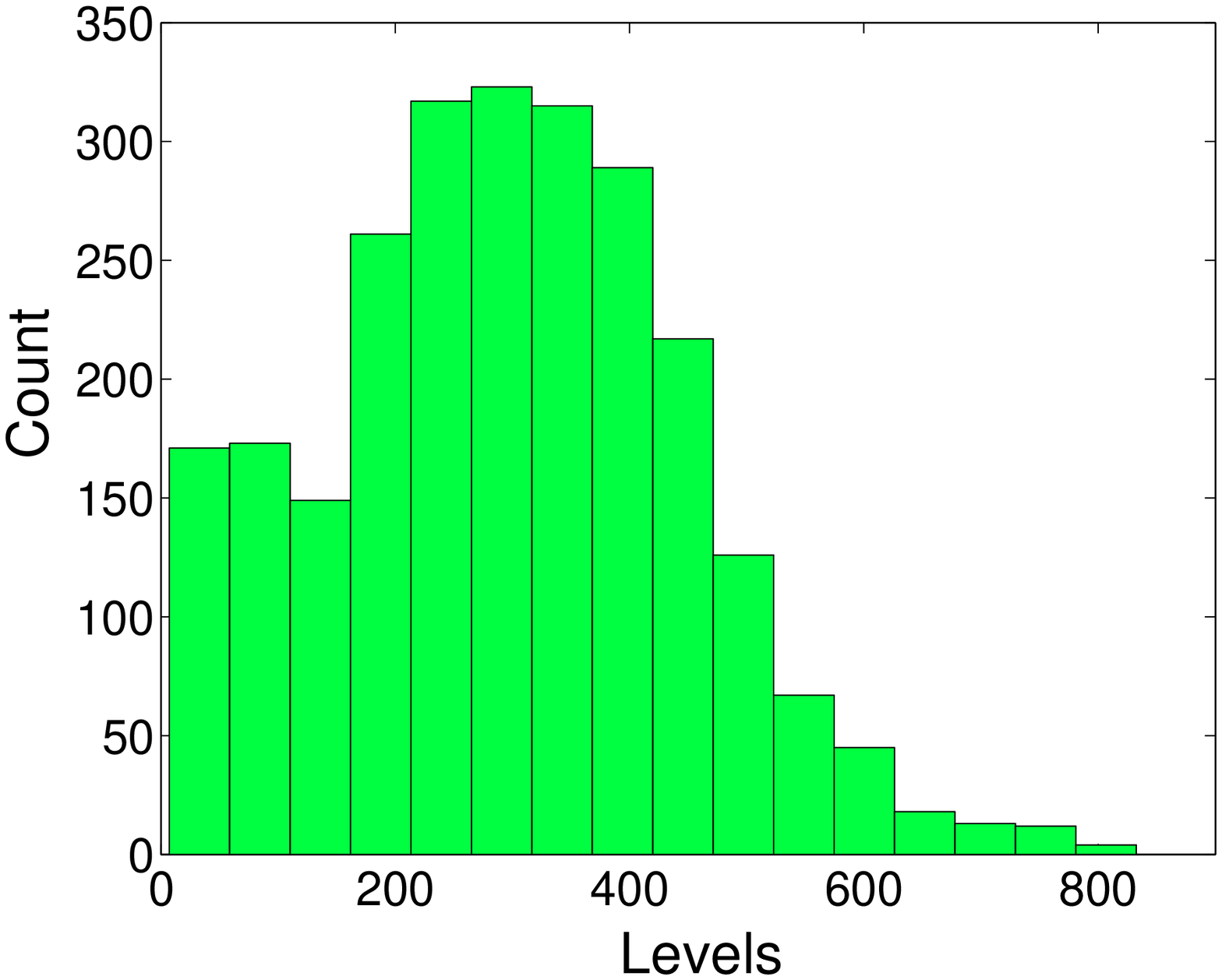}}
  \end{center}
  \caption{(Color online) Map and histogram of precipitation values for the complete rainfall data.}
  \label{fig:orig}
\end{figure}

We investigate the application of the
spin-based  models on real-world data. We compare the results with
the KNN, FKNN, and Support Vector Machine (SVM) methods. We consider
remotely sensed rainfall data on a moderately  sized  $50\times 50$
grid. The study domain covers an area of Indonesia extending from
2.5S to 10N in latitude and from 110E to 122.5E in longitude,
covered with a resolution of $0.25^\circ \times 0.25^\circ$. A map
of the precipitation distribution is shown in
Fig.~\ref{fig:orig_map}. The data represent daily accumulated
rainfall values recorded during January 2007
(\url{http://disc2.nascom.nasa.gov/Giovanni/tovas/TRMM_V6.3B42_daily.shtml})
\cite{giovanni}. The values are in millimeters and some summary
statistics are as follows: $z_{\min}=7.1$, $z_{\max}= 832.6$,
$\bar{z}=295.0$, $z_{0.50}= 293.7$, $\sigma_{z}= 151.4$. The
skewness coefficient is $0.27$ and the kurtosis coefficient $2.93$.
As evidenced in the histogram shown in Fig.~\ref{fig:orig_hist},
the precipitation p.d.f. is non-Gaussian, possibly bimodal.


The KNN and FKNN results  are obtained with an optimally tuned parameter
$k_{\rm opt}$ for each realization. In the case of SVM,  two
hyperparameters, $C$ and $\sigma_k$, need to be tuned for each of
the $N_c$ binary classifiers. For different values of $C$, we found
the bandwidth value $\sigma_{\rm kopt}$ that minimizes globally the
``testing errors'' for each of the $N_c$ classifiers (the true
values at all the  prediction points were used for testing).  We
selected a $C_{\rm opt}$ that minimizes the testing error. The
obtained misclassification rates were relatively high in this
class-adaptive approach, and further fine-tuning of the
hyperparameters did not bring noticeable improvements. Better
results were obtained by using mean values of $\sigma_{\rm kopt}$
and $C_{\rm opt}$, for all $N_c$ classifiers.  Using the same values
of the hyperparameters for all the $N_c$ classifiers  has been shown
to give satisfactory results~\cite{kane02}. Furthermore, we used the
same hyperparameters for the ten realizations, since they were all
derived from a uniform thinning of the same rainfall data set.

\begin{table}[h]
\begin{small}
\caption{Average misclassification rates for the $N_c$-level INNC,
PNNC, and CNNC models.  The averages are calculated over $10$
randomly thinned (by $p\,\%$) samples of the rainfall data. These
are compared with the best (lowest values) obtained using the KNN
and SVM techniques. The following misclassification rates are used:
$\langle F^{*}_{I} \rangle$ for the INNC, $\langle F^{*}_{P}
\rangle$ for the PNNC, $\langle F^{*}_{C} \rangle$ for the CNNC,
 $\langle F^{*}_{\rm knn}\rangle$ for the KNN,  $\langle F^{*}_{\rm fknn} \rangle$
 for the FKNN, and
 $\langle F^{*}_{\rm SVM} \rangle$ for the SVM.} \label{tab:real_rain}
\begin{center}
\begin{tabular}{lcccccccc}
\hline \hline
\# of classes    & \multicolumn{2}{c}{$N_c=2$}  & \multicolumn{2}{c}{$N_c=4$} &  \multicolumn{2}{c}{$N_c=8$} & \multicolumn{2}{c}{$N_c=16$} \\
\hline
$p[\%]$        & $33$  & $66$ & $33$  & $66$ & $33$  & $66$ & $33$  & $66$ \\
\hline
$\langle F^{*}_{I} \rangle$ & 6.81 & 8.61 & 13.41 & 16.94 & 24.51 & 30.46 & 44.95 & 53.05  \\
$\langle F^{*}_{P} \rangle$ & 6.51 & 8.43 & 14.18 & 17.53 & 27.87 & 33.01 & 51.59 & 56.25  \\
$\langle F^{*}_{C} \rangle$ & 6.86 & 8.62 & 14.68 & 17.84 & 28.63 & 32.97 & 50.32 & 55.28  \\
$\langle F^{*}_{\rm knn} \rangle$ & 6.46 & 8.47 & 13.82 & 18.01 & 28.28 & 33.85 & 51.85 & 56.12  \\
$\langle F^{*}_{\rm fknn} \rangle$ & 6.12 & 7.91 & 12.99 & 16.72 & 26.55 & 31.63 & 49.67 & 54.08  \\
$\langle F^{*}_{\rm SVM} \rangle$ & 6.78 & 8.56 & 15.61 & 17.62 & 29.70 & 33.61 & 51.27 & 55.49  \\
\hline \hline
\end{tabular}
\end{center}
\end{small}
\end{table}

\begin{figure}[!t]
  \begin{center}
    \subfigure[Best map by INNC]{\label{fig:best_recmap_Nc16_msng33_INNC}
    \includegraphics[scale=0.4]{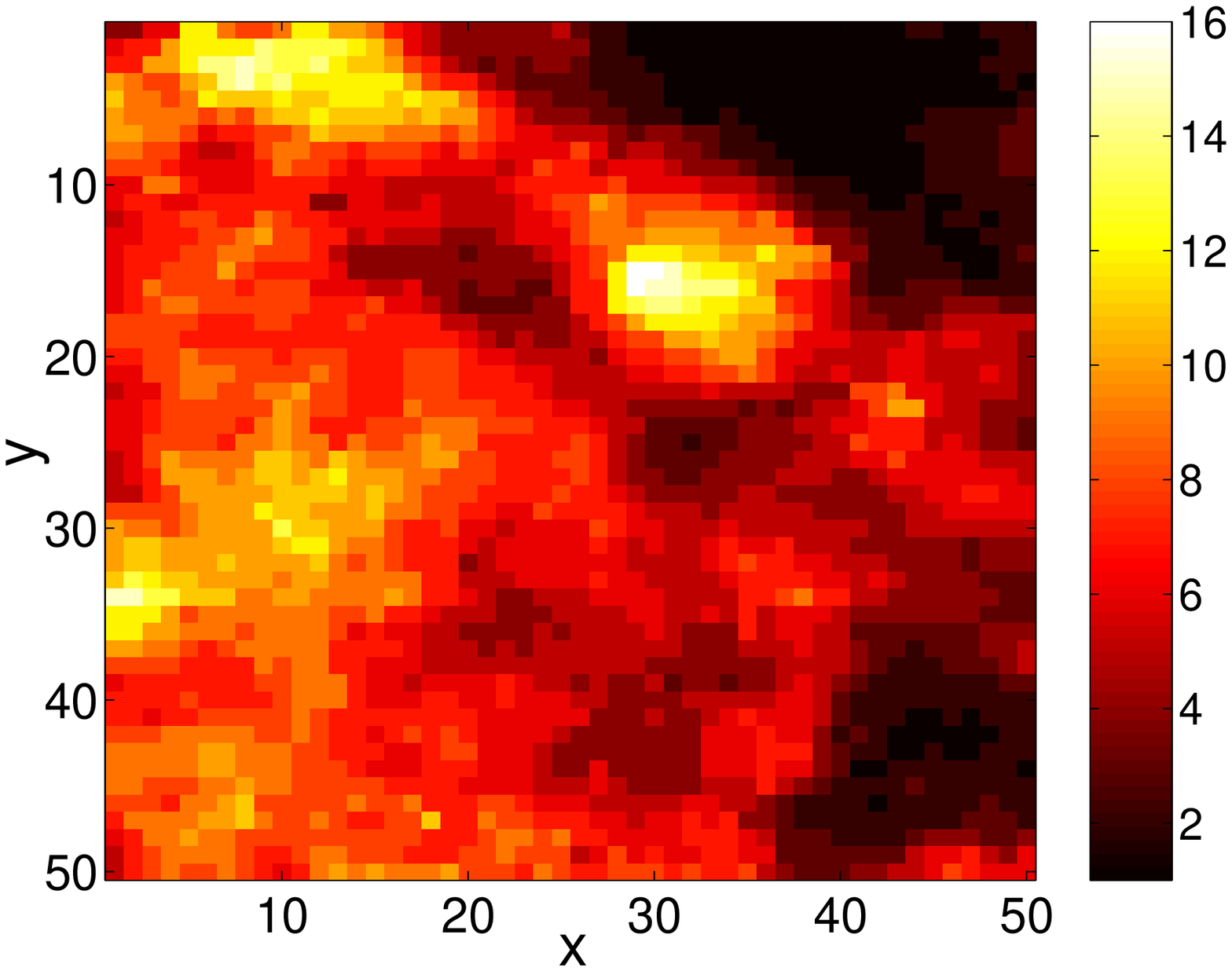}}
    \subfigure[Worst map by INNC]{\label{fig:wrst_recmap_Nc16_msng33_INNC}
    \includegraphics[scale=0.4]{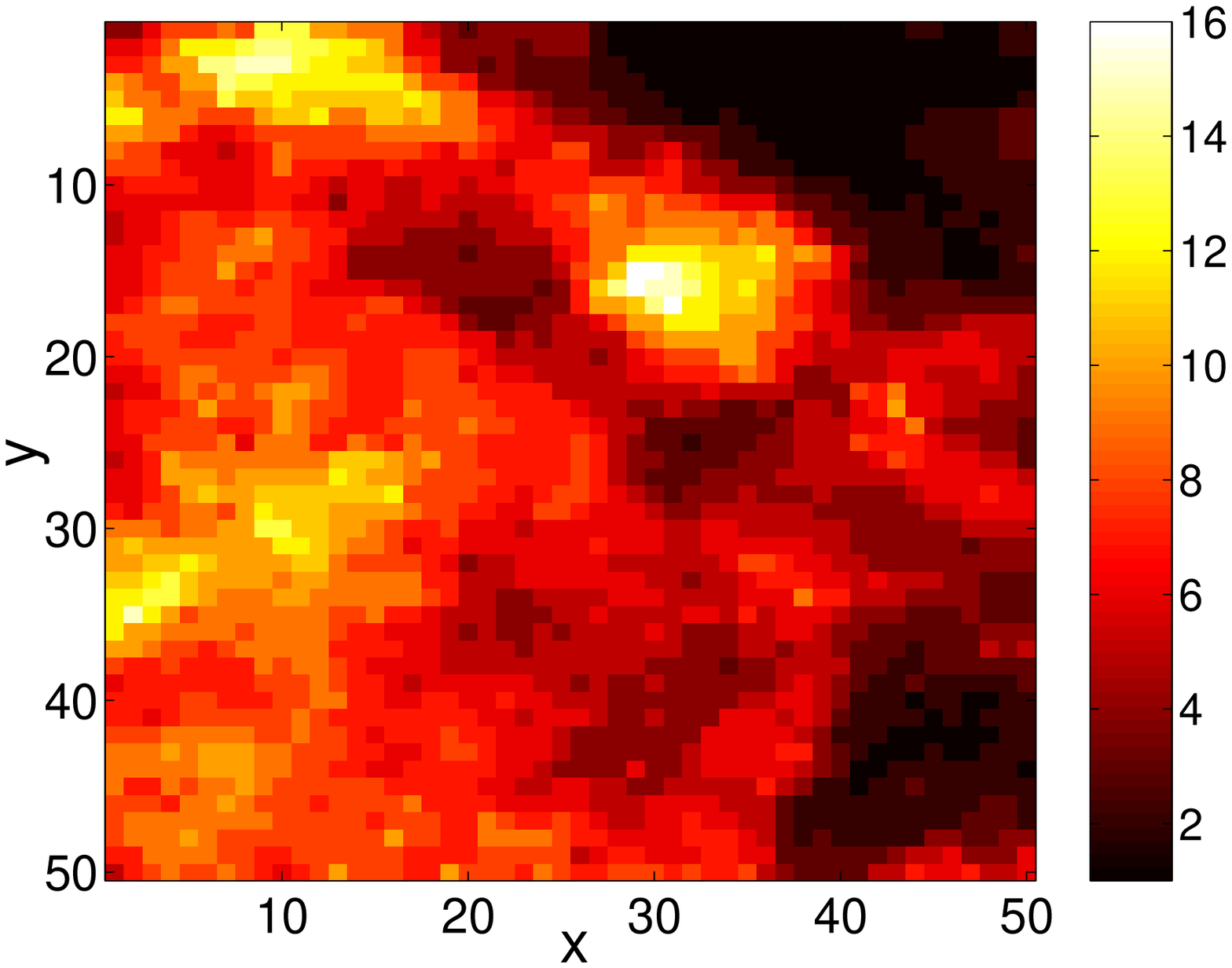}}
    \subfigure[Histograms by INNC]{\label{fig:hist_Nc16_msng33_INNC}
    \includegraphics[scale=0.4]{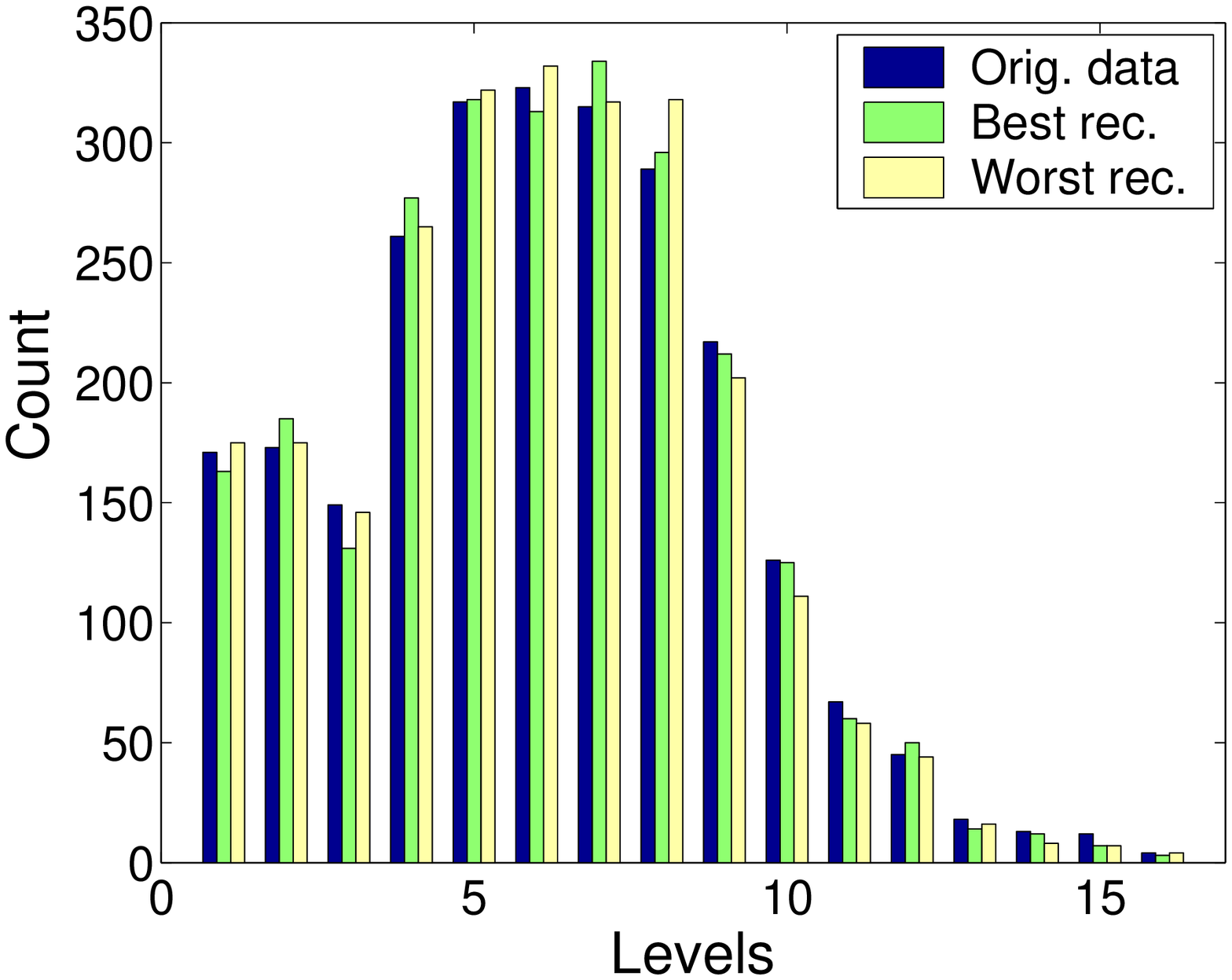}}
    \subfigure[x-variogram by INNC]{\label{fig:corrx_Nc16_msng33_INNC}
    \includegraphics[scale=0.4]{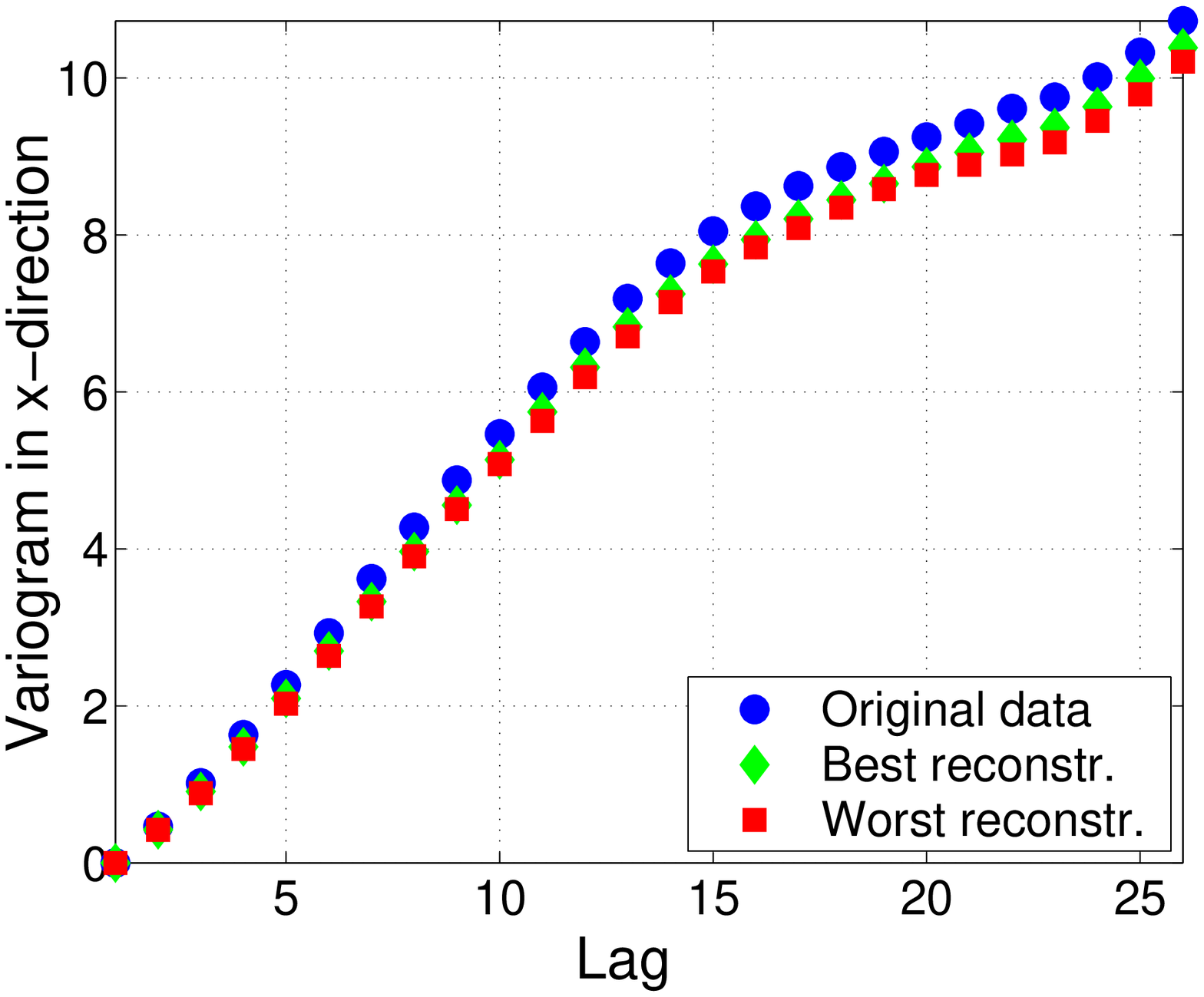}}
    \end{center}
  \caption{(Color online) 16-class classification results of the rainfall data with
    $p=33 \%$  missing values obtained by the INNC model. Using $100$ reconstructions
     derived from different  sampling configurations, we show the  class maps
  of the best  (a) and worst (b) case reconstructions,
  the class histograms of the original data as well as the best and worst reconstructions (c),
  and the empirical class variogram in the horizontal direction (d) for the original
data as well as the best and worst reconstructions.}
  \label{fig:reconstr16_innc}
\end{figure}

\begin{figure}[!t]
  \begin{center}
    \subfigure[Best map by PNNC]{\label{fig:best_recmap_Nc16_msng33_PNNC}
    \includegraphics[scale=0.4]{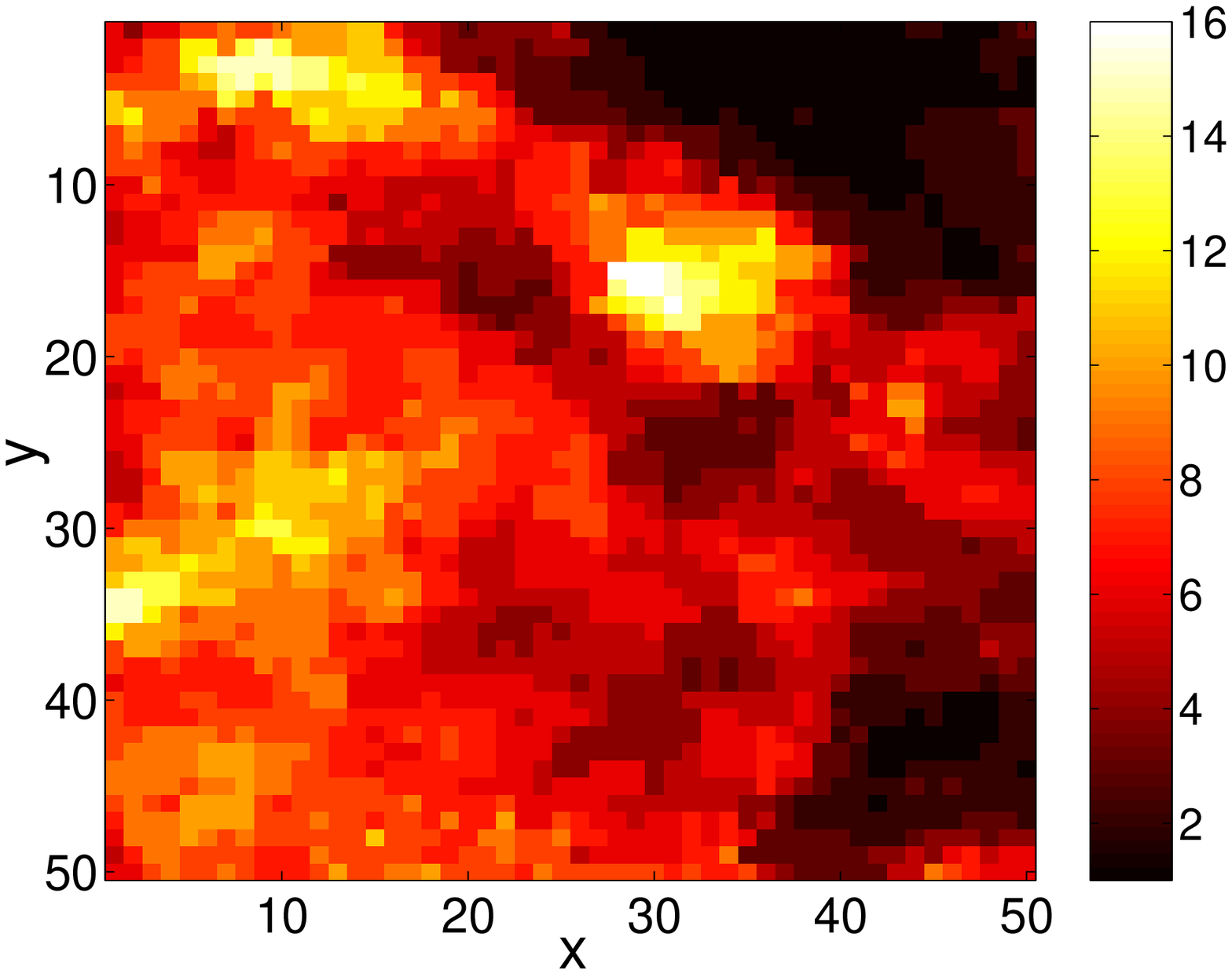}}
    \subfigure[Worst map by PNNC]{\label{fig:wrst_recmap_Nc16_msng33_PNNC}
    \includegraphics[scale=0.4]{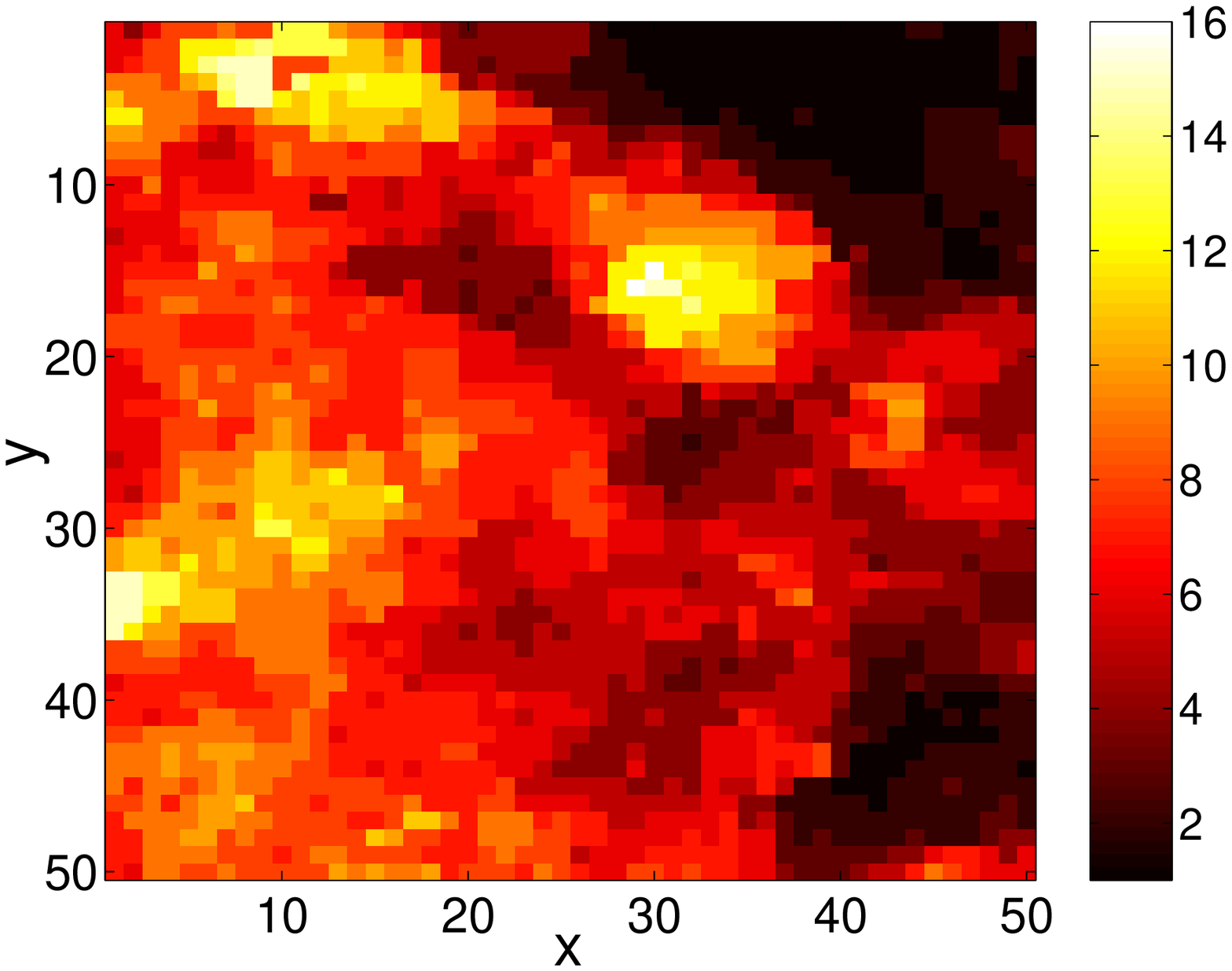}}
    \subfigure[Histograms by PNNC]{\label{fig:hist_Nc16_msng33_PNNC}
    \includegraphics[scale=0.4]{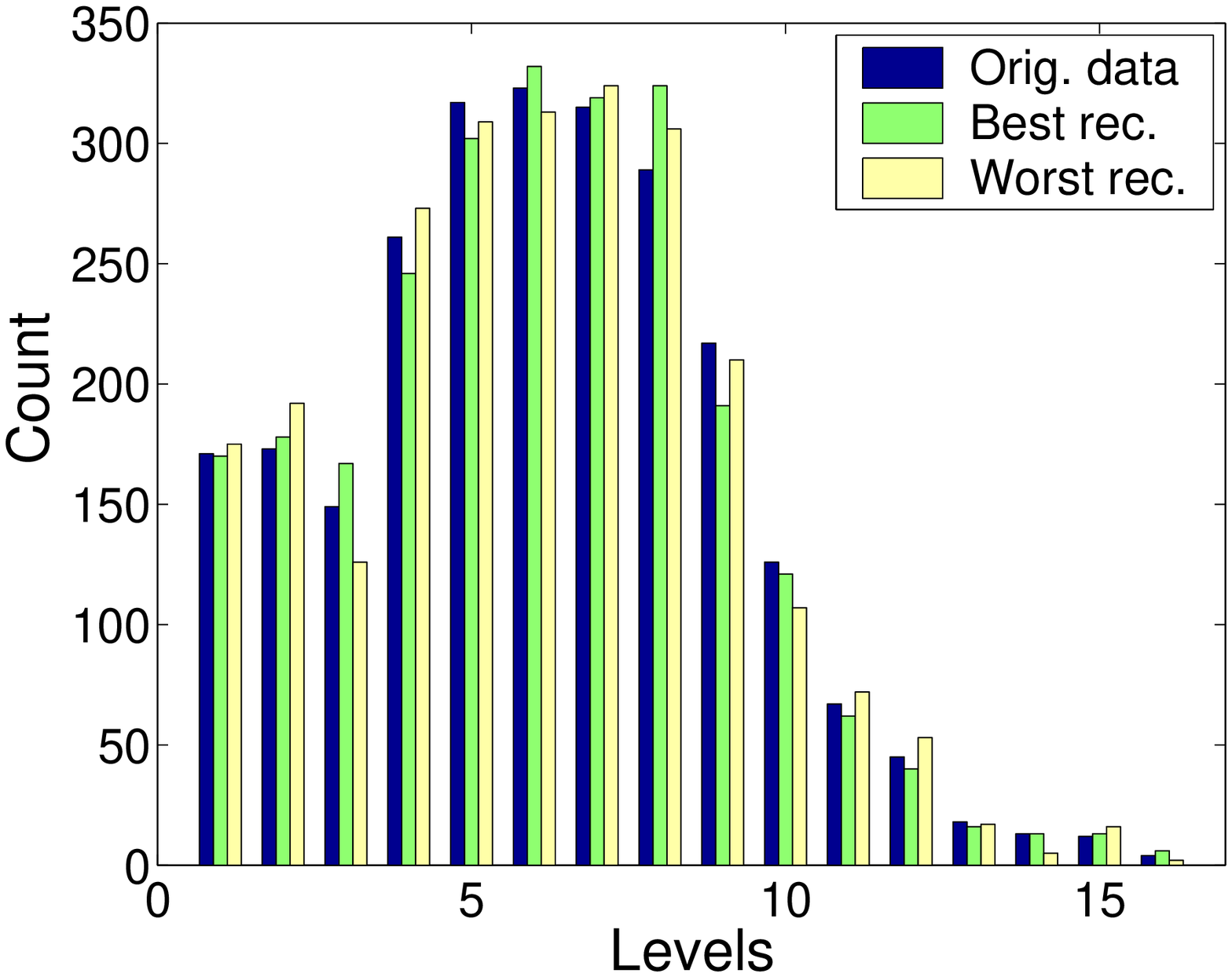}}
    \subfigure[x-variogram by PNNC]{\label{fig:corrx_Nc16_msng33_PNNC}
    \includegraphics[scale=0.4]{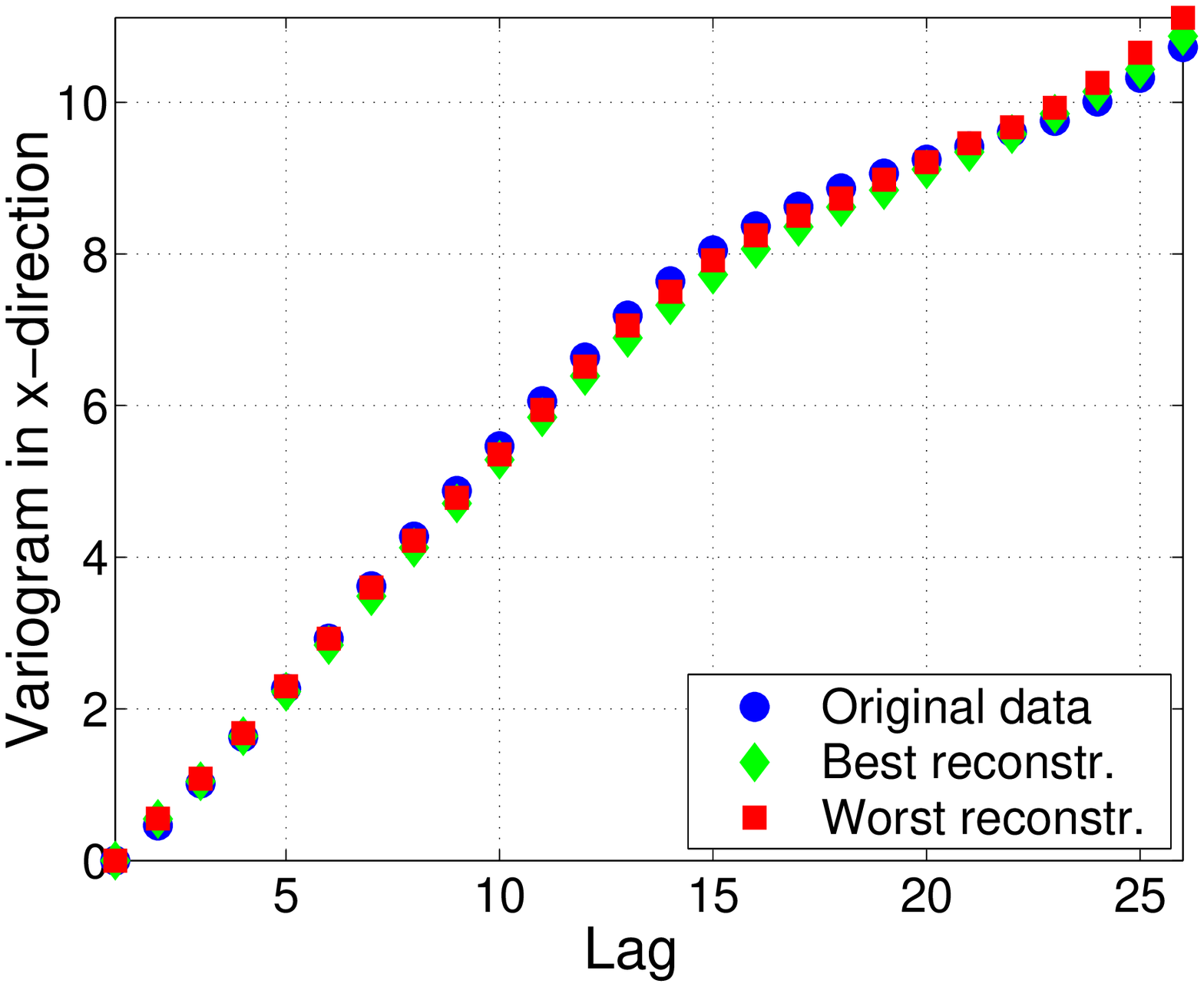}}
     \end{center}
  \caption{(Color online) 16-class classification results of the rainfall data
  obtained by the  PNNC model. The plot captions correspond to those in
   Fig.~\ref{fig:reconstr16_innc}.}
  \label{fig:reconstr16_pnnc}
\end{figure}

\begin{figure}[!t]
  \begin{center}
    \subfigure[Best map by CNNC]{\label{fig:best_recmap_Nc16_msng33_CNNC}
    \includegraphics[scale=0.4]{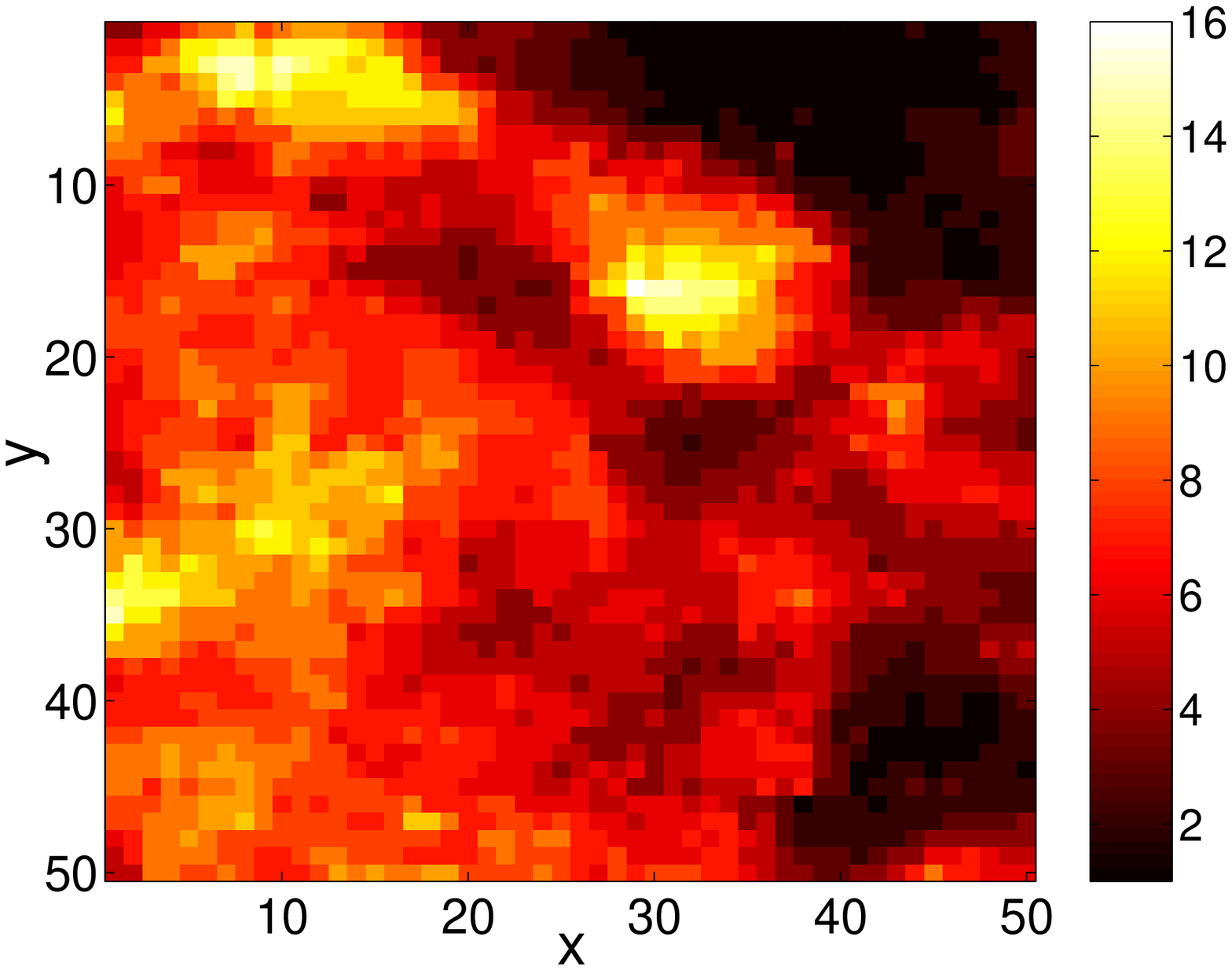}}
    \subfigure[Worst map by CNNC]{\label{fig:wrst_recmap_Nc16_msng33_CNNC}
    \includegraphics[scale=0.4]{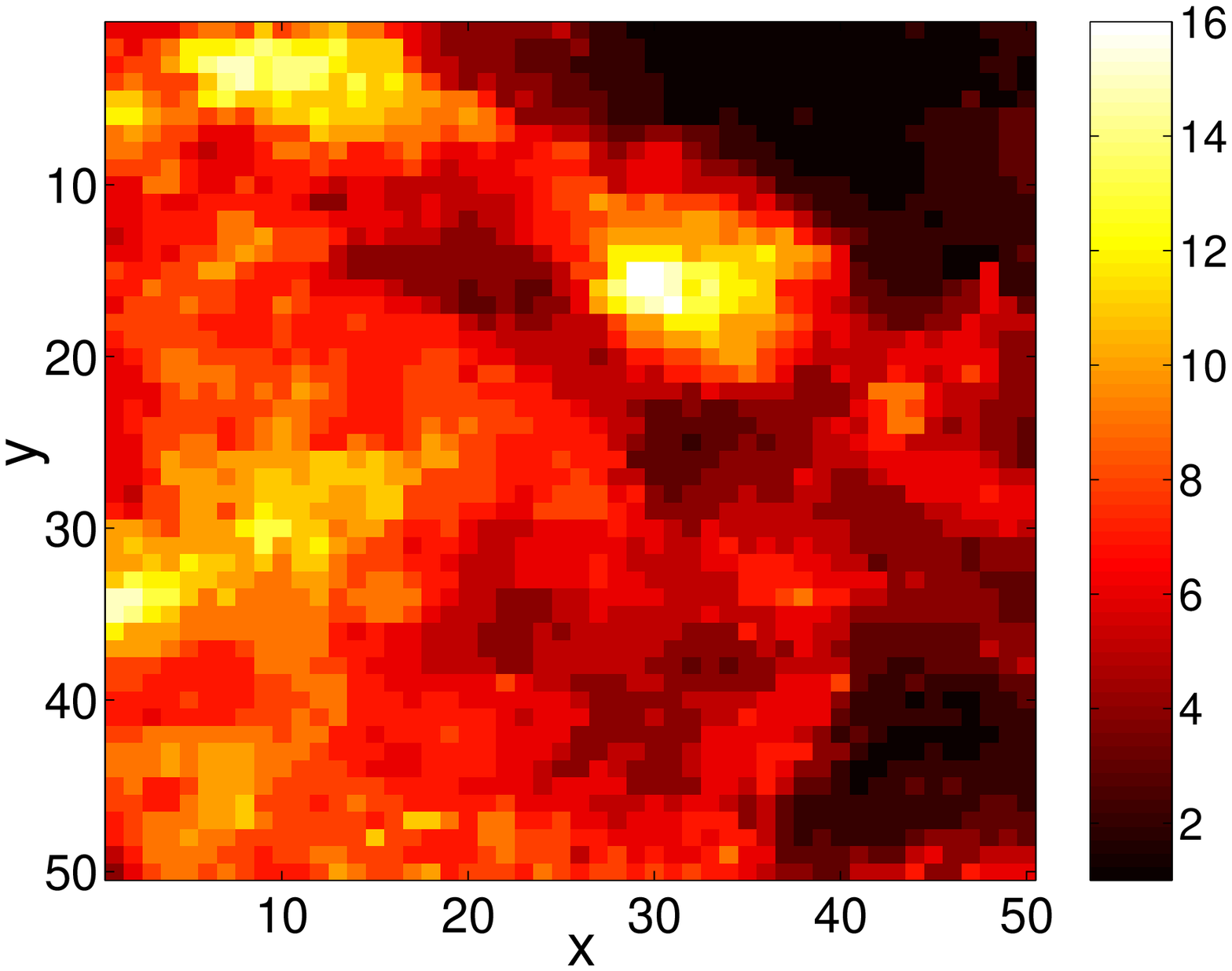}}
    \subfigure[Histograms by CNNC]{\label{fig:hist_Nc16_msng33_CNNC}
    \includegraphics[scale=0.4]{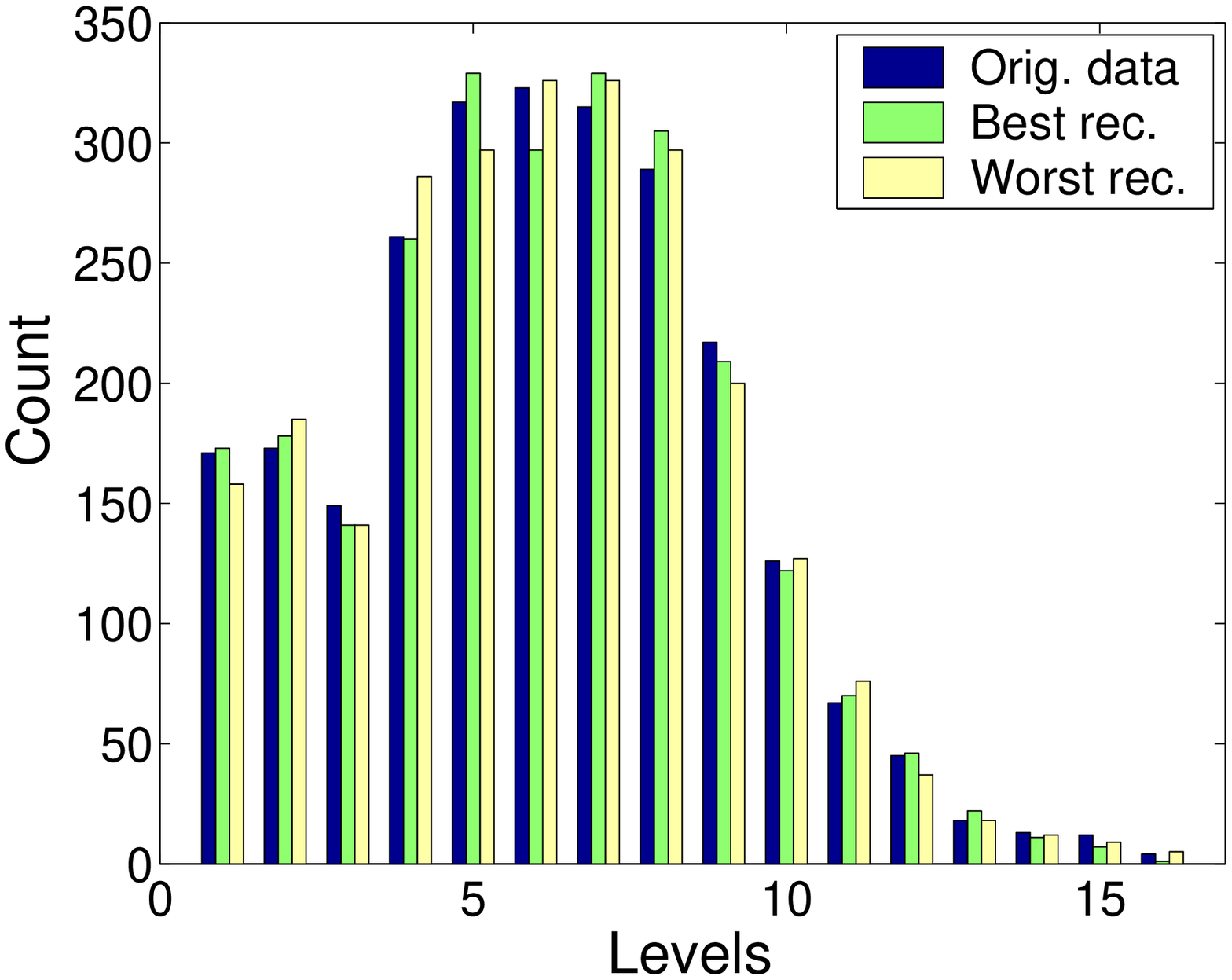}}
    \subfigure[x-variogram by CNNC]{\label{fig:corrx_Nc16_msng33_CNNC}
    \includegraphics[scale=0.4]{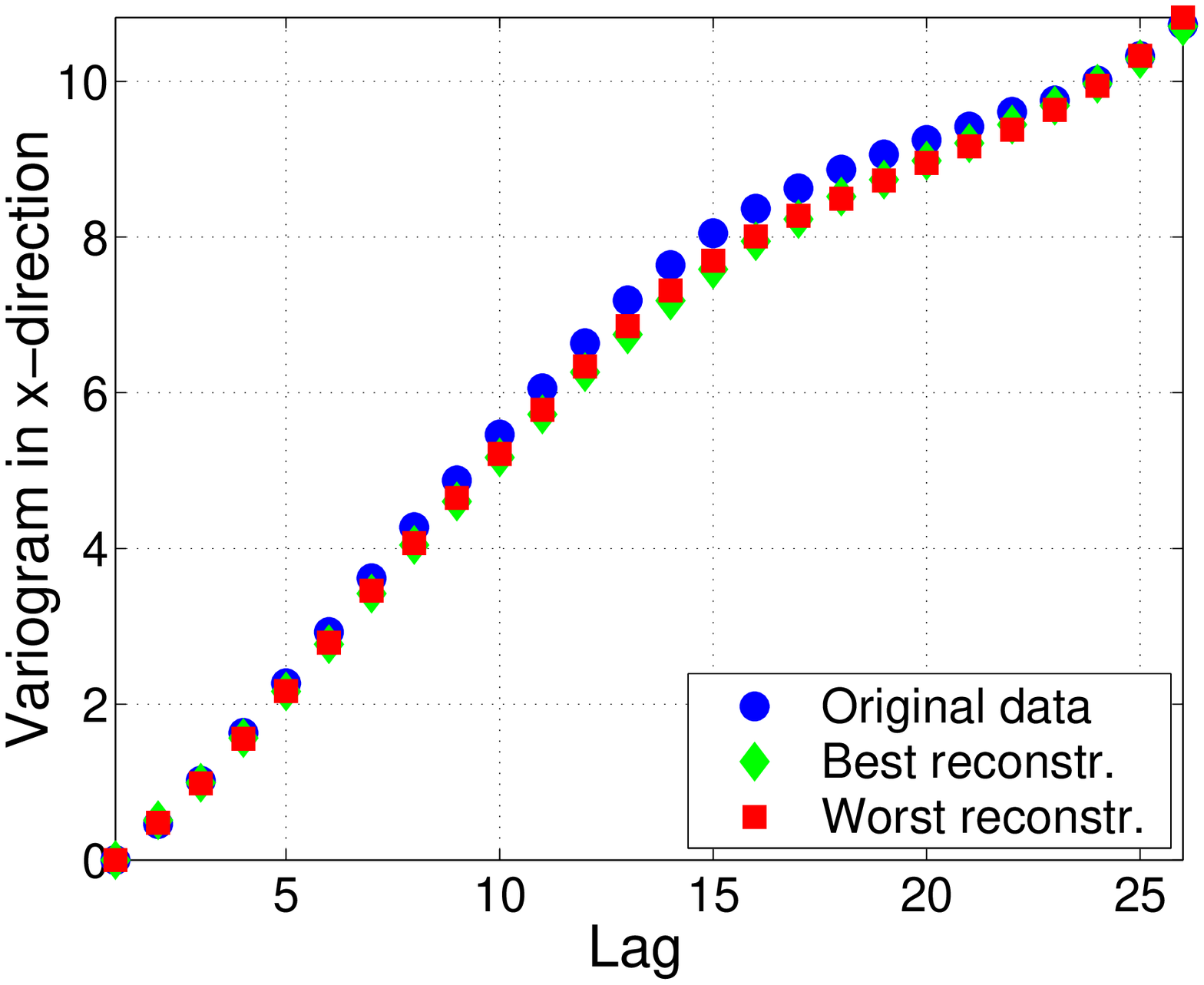}}\\
  \end{center}
  \caption{(Color online) 16-class classification results of the rainfall data
  obtained by the  CNNC model. The plot captions correspond to those in Fig.~\ref{fig:reconstr16_innc}.}
  \label{fig:reconstr16_cnnc}
\end{figure}

In Table~\ref{tab:real_rain} we compare average misclassification
rates for the spin-based models, as well as the KNN, FKNN, and SVM
classifiers based on 10 different sampling realizations. The sample
sets are derived from random thinning of the rainfall data by
$p\,\%.$ Comparing the performance of different classifiers, the
FKNN and the INNC models give the best results overall. FKKN
performs better for small number of classes $(N_c=2,4),$ while INNC
is better at larger values of $N_c$.
These results agree with the synthetic data studies, where the INNC
model also showed superior performance, especially for the
data with slower variation and moderate degree of thinning. In
contrast to the synthetic studies, the CNNC model did not perform as
satisfactorily. This could be attributed to the presence of noise
(ubiquitous in real data), as pointed out in Section
\ref{sec:class-simul}. In addition, the SVM classifier did not
perform as well as in some other comparative studies~\cite{herm99,shari03,mull97}.
This may be due to the
simplifications we adopted in the estimation of the hyperparameters.
Besides the misclassification rates, we also checked the  capacity
of the respective classifiers to reproduce the histogram and the
empirical directional variogram, $\hat{\gamma}_{Z}(r{\bf e})$ of the
data~\cite{yagl87}. The $\hat{\gamma}_{Z}(r{\bf e})$, also known as
the two-point structure function, is a measure of the spatial
continuity of the field $Z$
  in the direction ${\bf e}=\hat{\bf x}$.  On a lattice of step $\alpha$,
  the $\hat{\gamma}_{Z}(r{\bf e})$ is given by
\begin{equation}
\hat{\gamma}_{Z}(n \alpha\hat{\bf
x})=\frac{1}{2\,L(L-n)}\sum_{j=1}^{L-n} \sum_{i=1}^{L} \left[Z({\bf
s}_{i,j+n}) - Z({\bf s}_{i,j}) \right]^2, \; r=n\alpha, \;
n=1,\ldots, \frac{L}{2},
\end{equation}
where  ${\bf s}_{i,j}= \alpha \, (i \hat{\bf y} + j \hat{\bf x}).$
In Figs.~\ref{fig:reconstr16_innc}-\ref{fig:reconstr16_cnnc}, we
show the reconstructed maps, histograms, and variograms in the
direction of the $x$ axis, for the best (lowest $F^*$) and worst
(highest $F^*$) reconstructed realizations for  $N_c=16$ and
$p=0.33$, based on 100 realizations. The variogram along the $y$
axis (not shown) has similar behavior. In all the cases, the
statistics are recovered satisfactorily, and there are no
significant differences between the respective models. Even though
the PNNC model gave the highest misclassification rate, it
reproduces the histograms and the variogram reasonably well.


\begin{figure}[!t]
  \begin{center}
    \subfigure[Original]{\label{fig:so_rain}
    \includegraphics[scale=0.28]{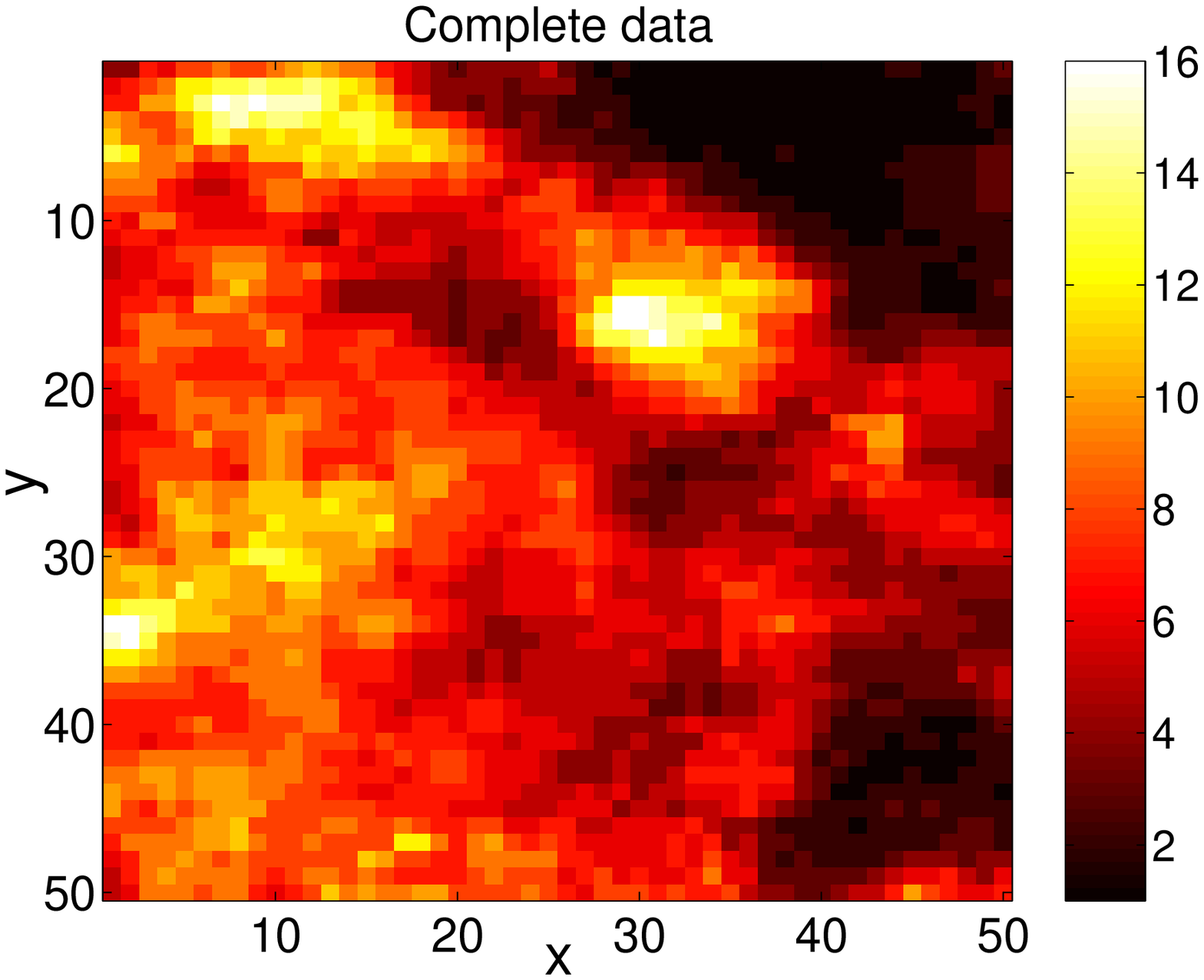}}
    \subfigure[Sample]{\label{fig:sm_rain_msng33}
    \includegraphics[scale=0.28]{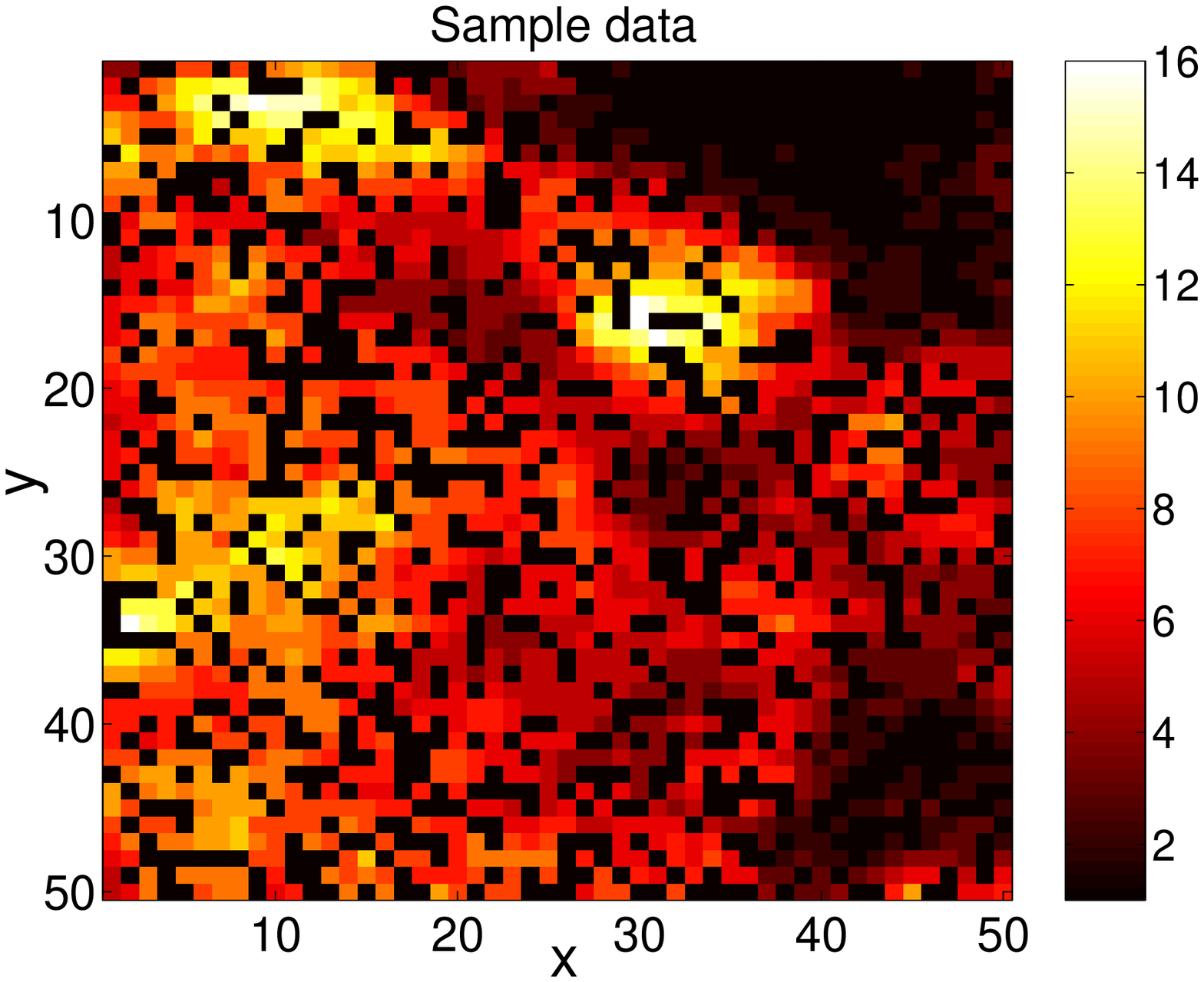}}
    \subfigure[Reconstructed]{\label{fig:sr_q16_msng33_cnnc}
    \includegraphics[scale=0.28]{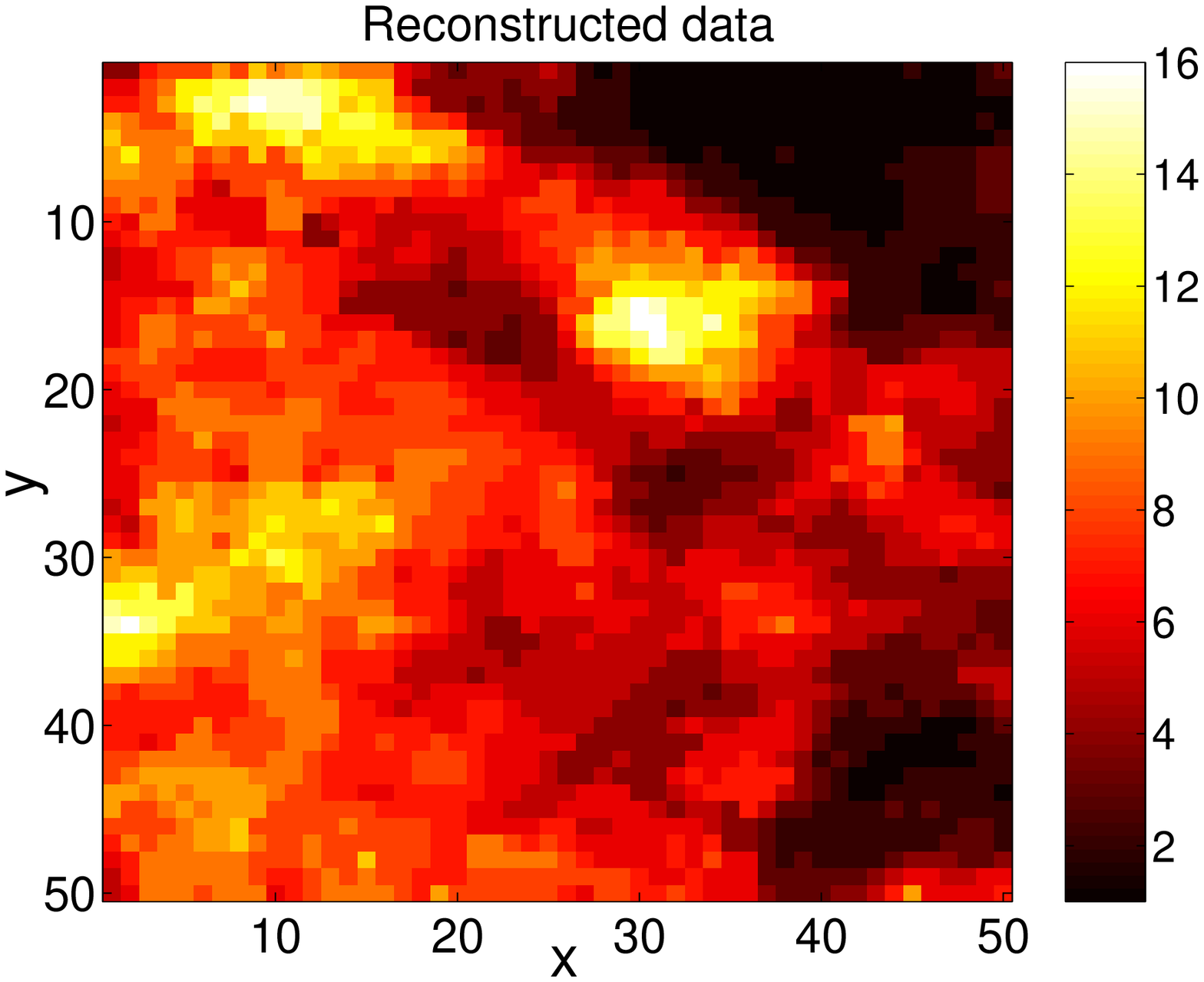}} \\
    \subfigure[Histograms]{\label{fig:hist_sr_med_rain_q16_msng33_cnnc}
    \includegraphics[scale=0.28]{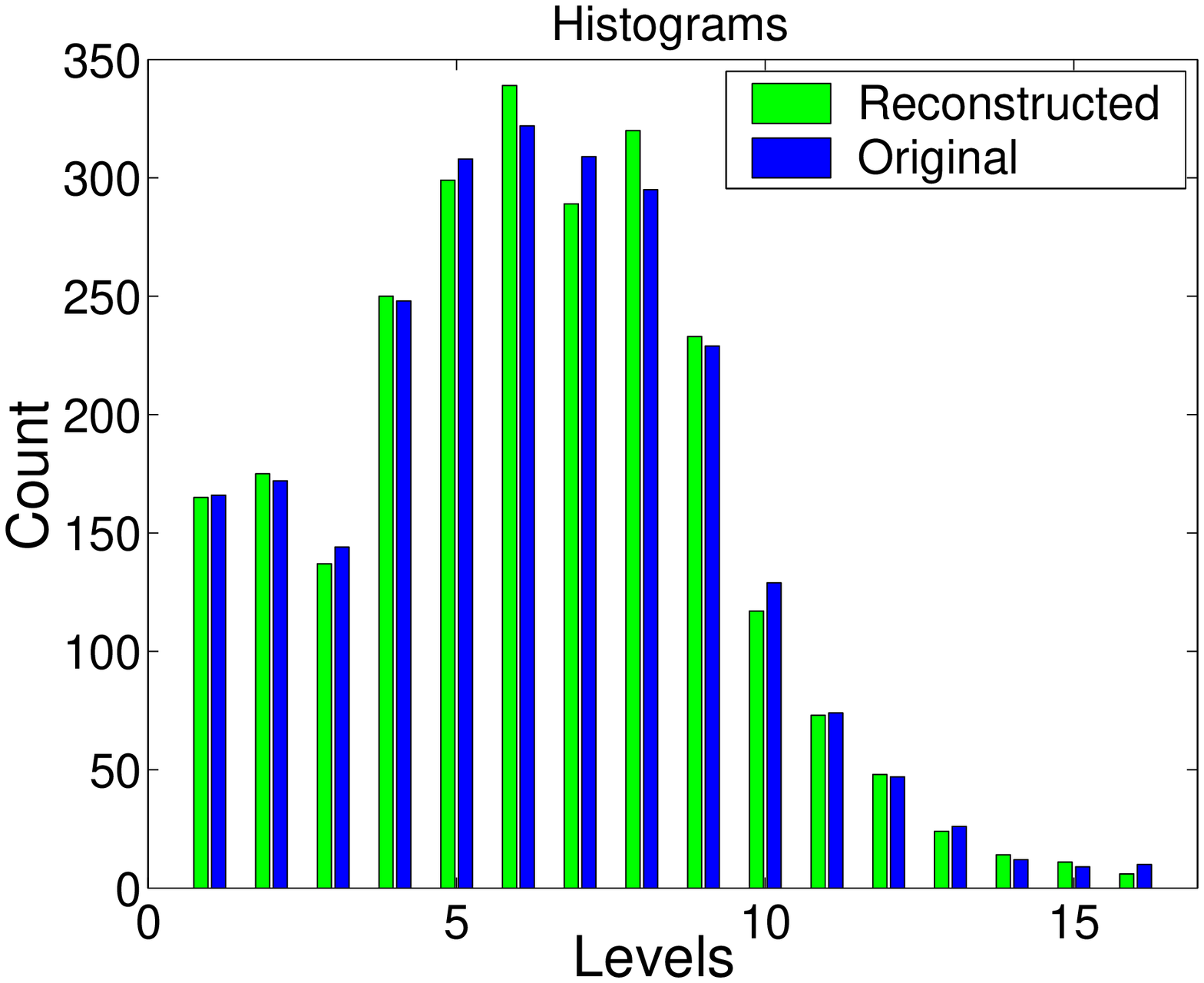}}
    \subfigure[Conf. interval]{\label{fig:sr_conf_int_rain_q16_msng33_cnnc}\includegraphics[scale=0.28]{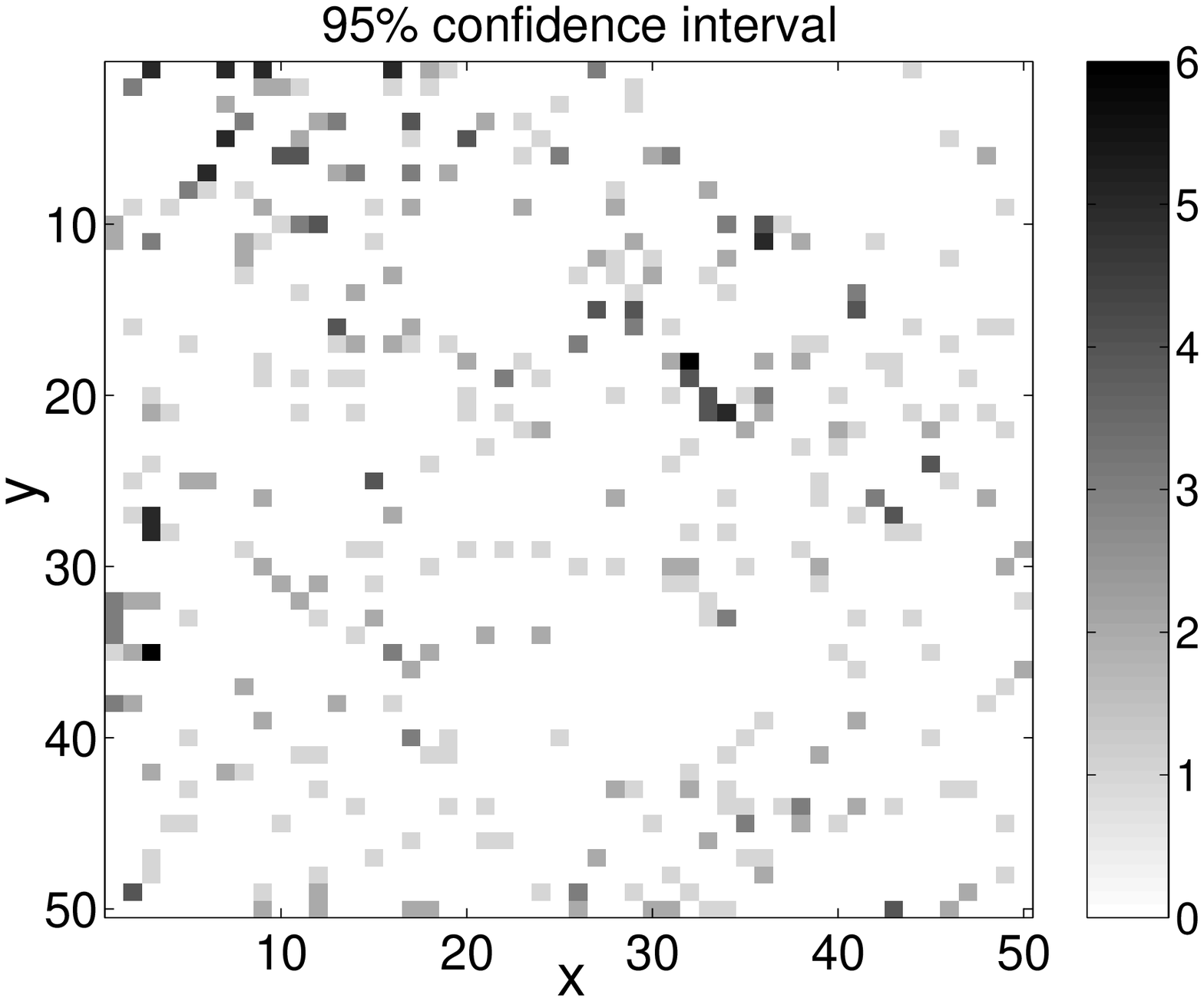}}
    \subfigure[RMSE]{\label{fig:sr_rmse_rain_q16_msng33_cnnc}
    \includegraphics[scale=0.28]{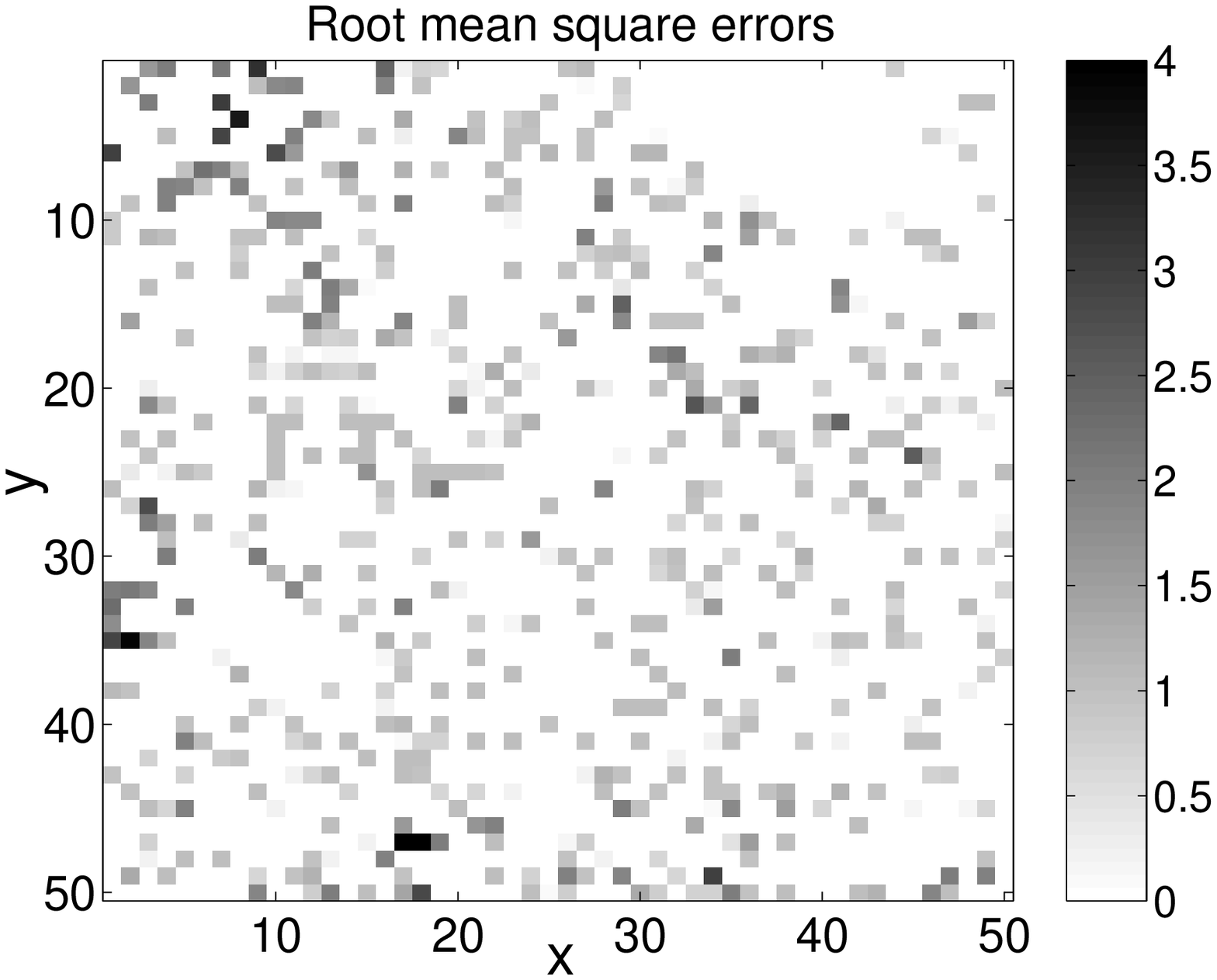}}
  \end{center}
  \caption{(Color online) Classification results obtained from the
  16-class CNNC model. The missing-values sample is obtained from the rainfall data  by randomly
  removing $p=0.33$ of the points. One hundred reconstructions are generated
 starting from 100 initial states
    obtained by the majority rule with adaptable stencil size.
  Plots (a)-(c)  show
   class maps for the complete realization, the sample with the
   missing data, and the reconstructed image, respectively. The latter is based on the medians of
   the class values  obtained from the 100 reconstructions.
  Plot (d)  compares the class histograms of the
  original and reconstructed data, plot (e) shows the width of the
  $95 \%$ confidence intervals for the class predictions, and plot (f) represents
  the class root mean square error at the prediction points.}
  \label{fig:one_samp_rain_clock}
\end{figure}

The above simulation results are based on one run for each sample
set.  As we have shown in the case of synthetic data, multiple runs
can improve the results and allow estimation of uncertainty. In the case
of the rainfall data, considering one realization (sample set)
generated with $p=0.33$ and performing 100 simulation runs using,
for example, the 16-level CNNC model gave the misclassification rate
$F^{*}=46.9\%$ requiring $T_{\rm cpu}=5.6$ seconds of total CPU time.
The multiple-run-reconstruction measures, as those shown in Fig.
\ref{fig:one_samp_mate_clock} for the synthetic data, are displayed
in Fig. \ref{fig:one_samp_rain_clock}.

\subsection{Digital image data} 
\begin{figure}[!t]
  \begin{center}
    \includegraphics[scale=0.6]{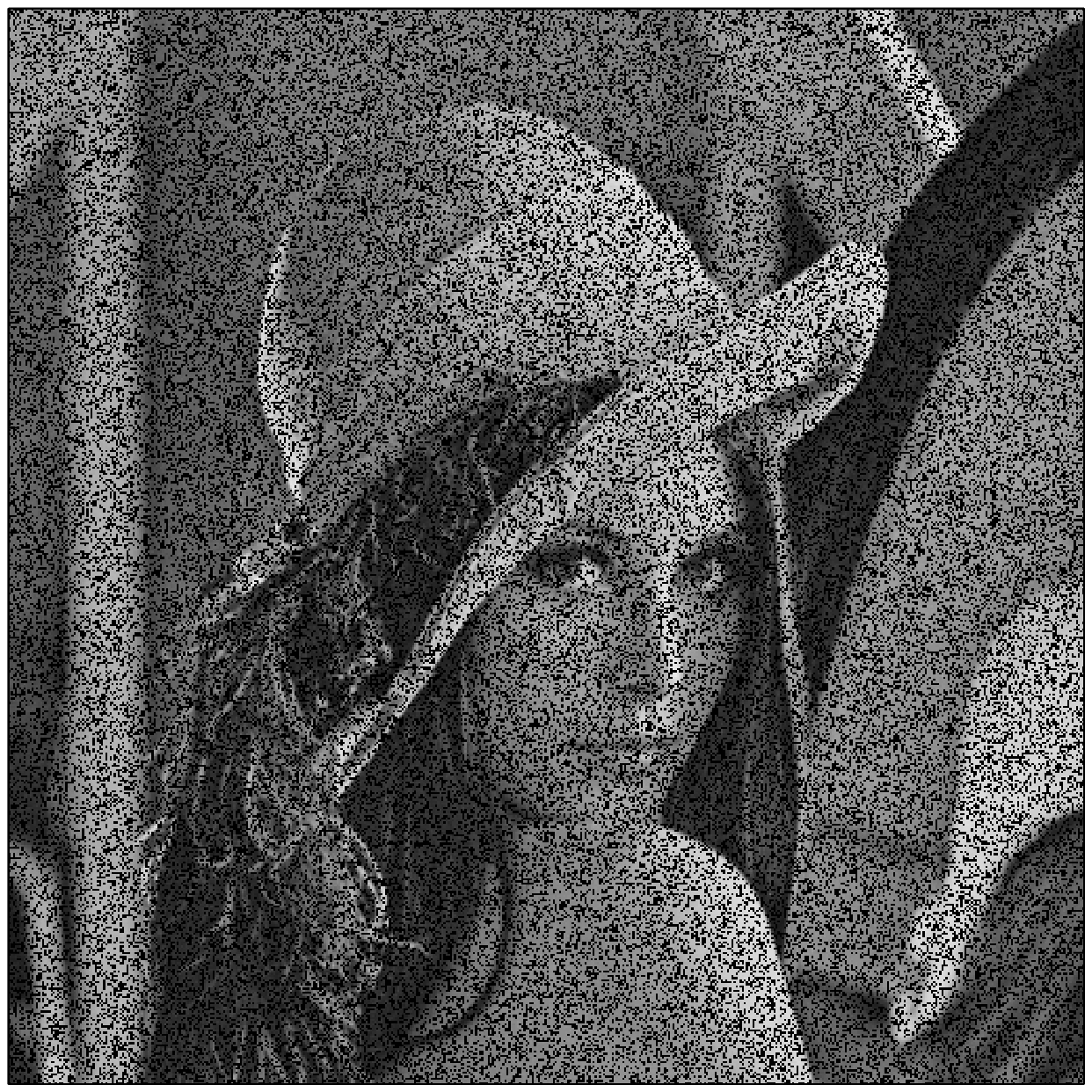}
  \end{center}
  \caption{Degraded image of Lena obtained by random removal of $33\%$ of the pixels.}
  \label{fig:lena_33}
\end{figure}

We consider the
standard 256-valued gray-scale test image of Lena on a
$512\times512$ grid. We randomly remove $p=33 \%$ of the pixels and
subsequently reconstruct the image using the spin-based models. The
degraded image is shown in Fig.~\ref{fig:lena_33}.

\begin{figure}[!t]
  \begin{center}
    \subfigure[Original]{\label{fig:lena_orig}
    \includegraphics[scale=0.4]{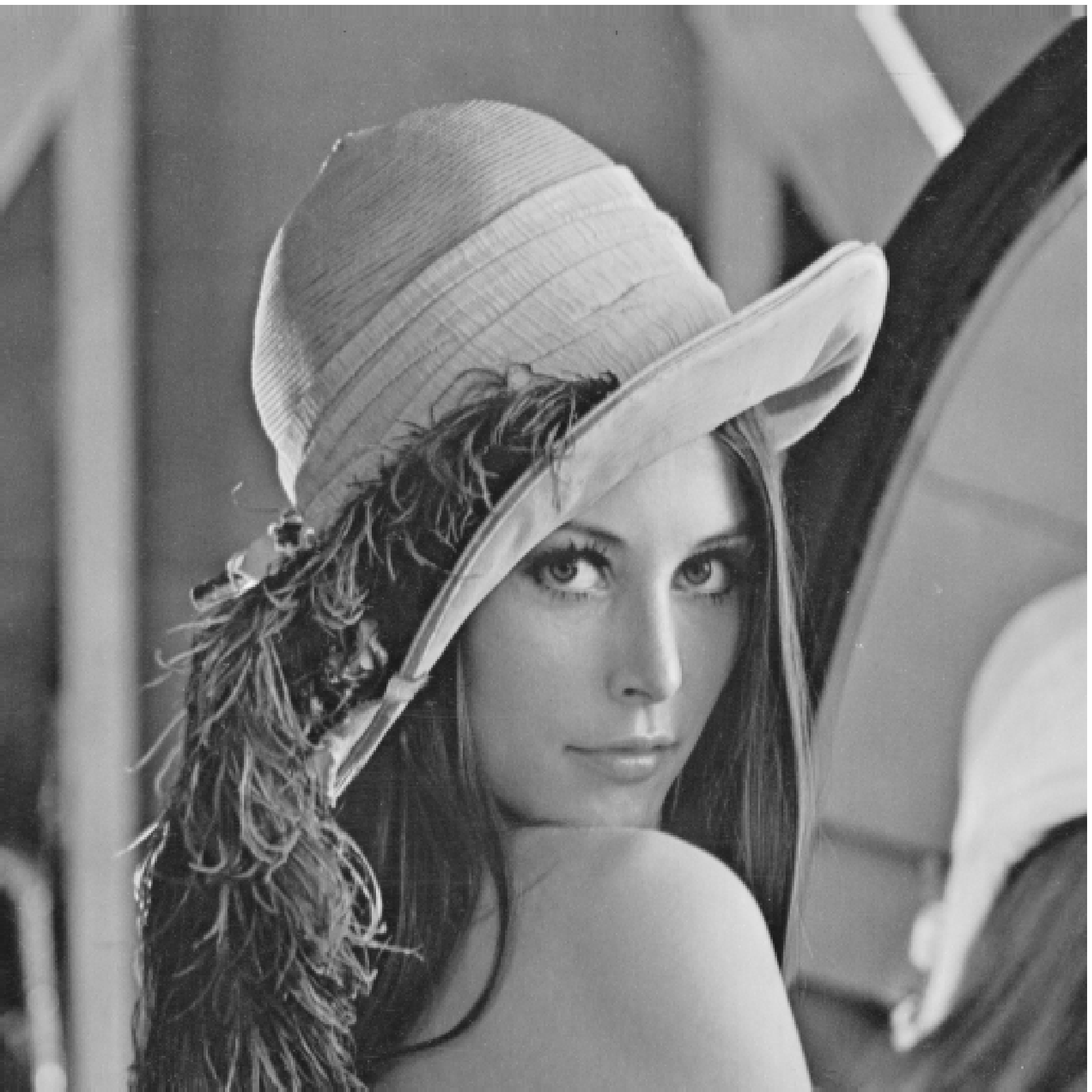}}
    \subfigure[INNC]{\label{fig:lena_INNC}
    \includegraphics[scale=0.4]{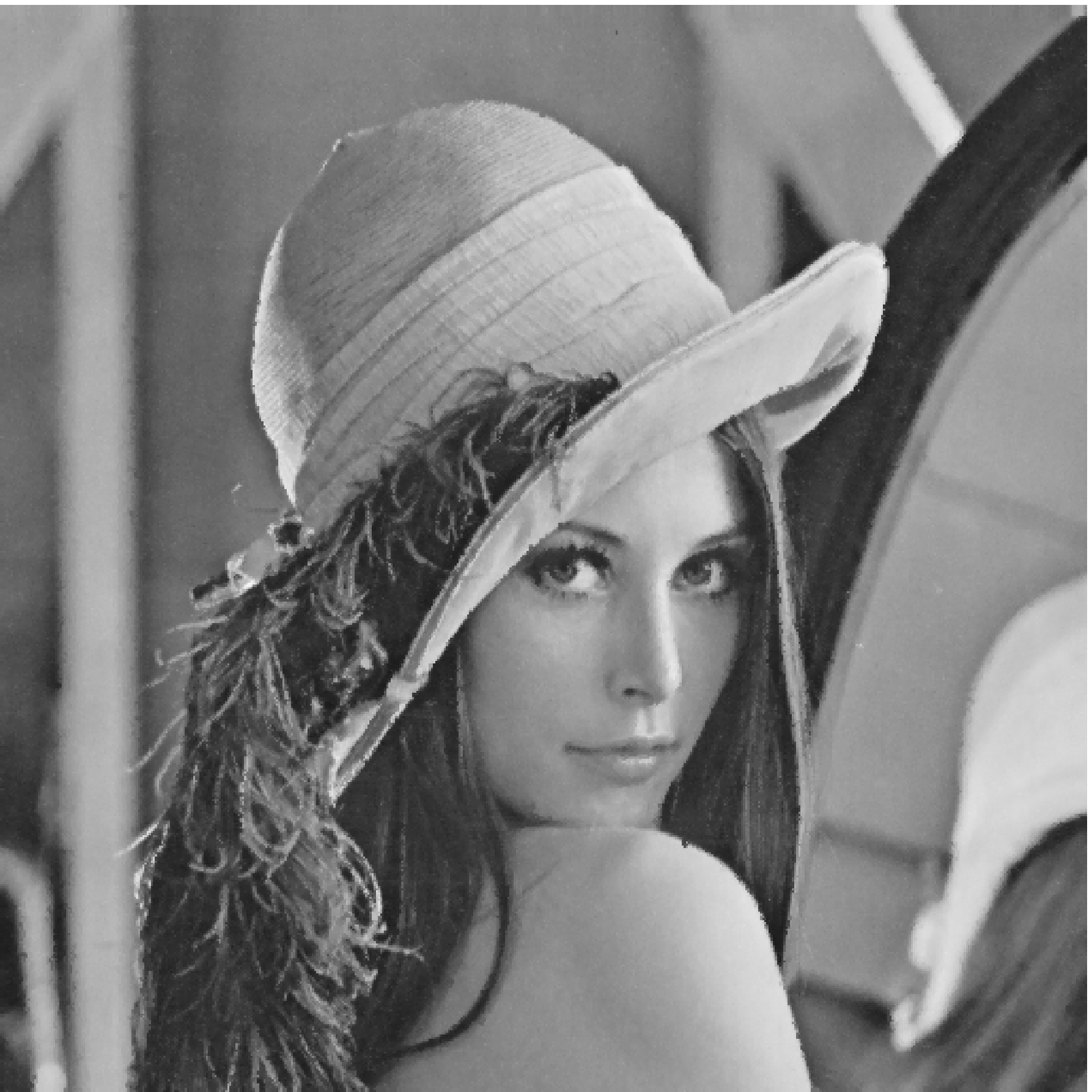}}\\
    \subfigure[PNNC]{\label{fig:lena_PNNC}
    \includegraphics[scale=0.4]{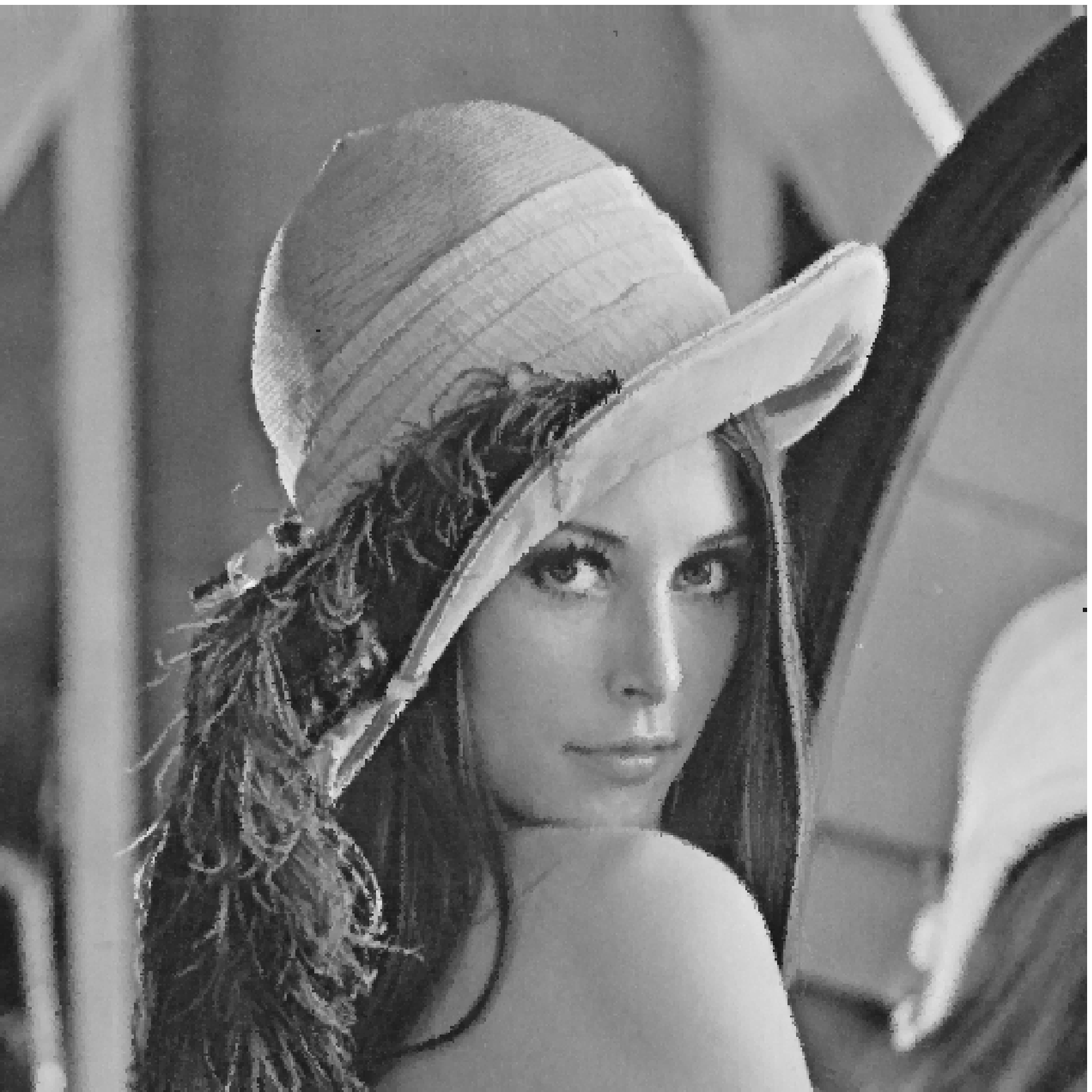}}
    \subfigure[CNNC]{\label{fig:lena_CNNC}
    \includegraphics[scale=0.4]{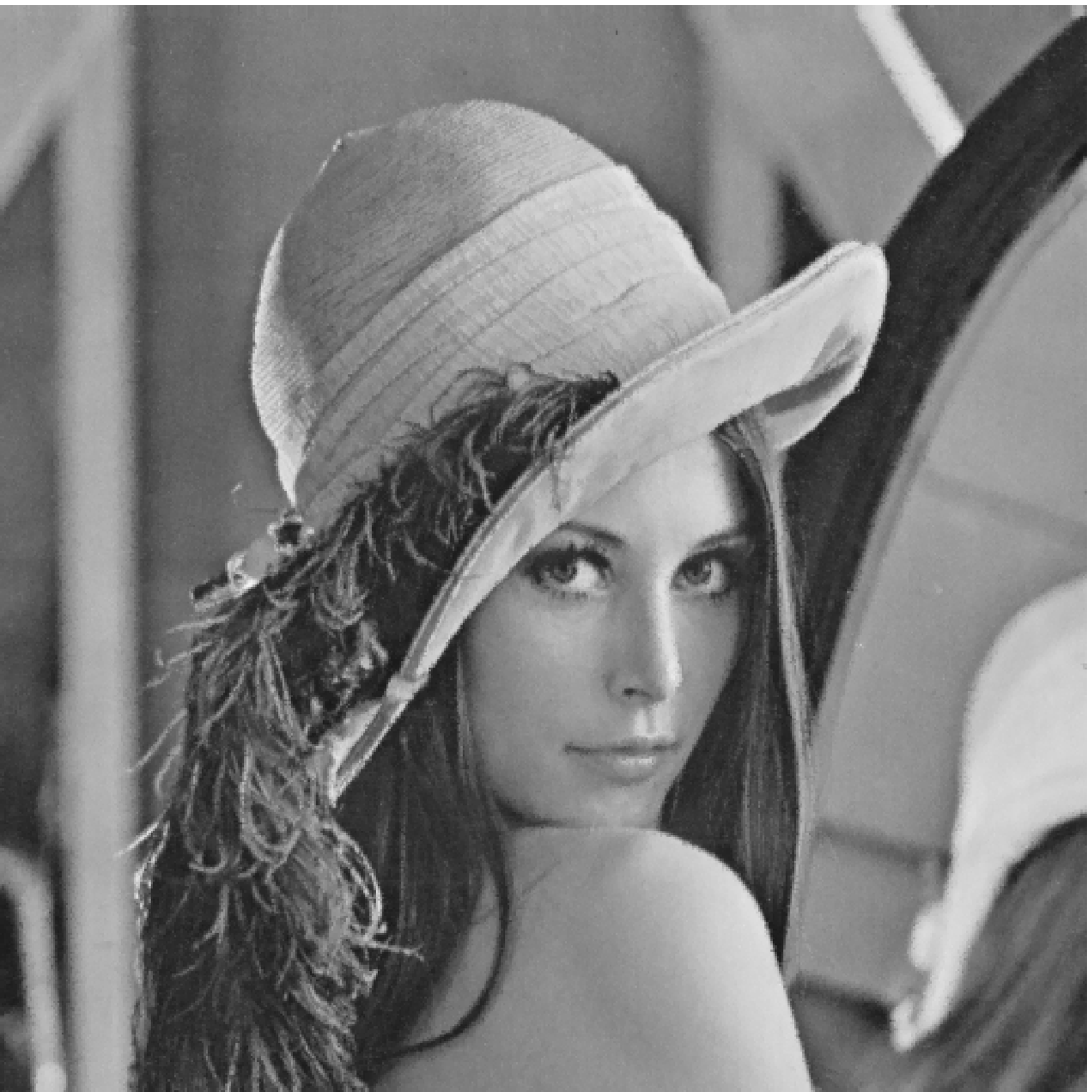}}
  \end{center}
  \caption{Original  (a) and reconstructed images of Lena using
  the INNC (b),  PNNC (c) and  CNNC (d) classification models.}
  \label{fig:lena_images}
\end{figure}

The range of the image pixel values is equal to $\max(Z)-
\min(Z)=220$, and thus we use $N_c=220$ classes of uniform width
(equal to 1). The sequence in Fig.~\ref{fig:lena_images} shows the
original image, Fig.~\ref{fig:lena_orig}, along with  the
reconstructed images obtained by the INNC, Fig.~\ref{fig:lena_INNC},
PNNC, Fig.~\ref{fig:lena_PNNC}, and CNNC, Fig.~\ref{fig:lena_CNNC}
models. Visually, all three reconstructions appear quite similar to the
original image. The INNC model misclassifies some pixels along the
edges (e.g., along the shoulder).
The numerical comparisons of univariate validation measures
shown in Table~\ref{tab:real_lena}  are
in favor  of the CNNC model. The CNNC model  is also more efficient
computationally (requiring fewer MC sweeps and less CPU time), and
it also approximates more accurately the sample energy. The worst performance in terms
of the validation measures is shown by the PNNC model.

\begin{table}[h]
\begin{small}
\caption{Prediction errors and average optimization statistics
for  reconstructions of the Lena image by
means of the INNC, PNNC, and CNNC models,
based on $10$ realizations corresponding to different sampling configurations.
MAAE: Mean average
absolute error. MARE: Mean average relative error. MAARE: Mean
average  absolute relative error. MRASE: Mean root average square
error. MR: Mean Correlation coefficient. First, averages are evaluated over
 the prediction points
for each realization, and then means are calculated with respect
to an ensemble of $10$ realizations.}
\label{tab:real_lena}
\begin{center}
\begin{tabular}{lcccccccc}
\hline \hline
& MAAE  & MARE [\%] & MAARE [\%] & MRASE & MR [\%]  & $\langle N_{MC} \rangle$ & $\langle T_{\rm cpu} \rangle$ [s]  & $\langle U^{*} \rangle$ \\
\hline
INNC & 5.04 & 2.63 & 4.97 & 9.29 & 98.35 & 106.9 & 351.1 & 3e-4  \\
PNNC & 6.20 & 0.21 & 6.48 & 11.67 & 97.08 & 571.4 & 275.7 & 0.14  \\
CNNC & 5.04 & -3e-6 & 5.33 & 8.24 & 98.52 & 8.3 & 173.7 & 2e-10  \\
\hline \hline
\end{tabular}
\end{center}
\end{small}
\end{table}

\section{Summary and conclusions}
\label{sec:conclusions}

We present  spatial classification methods for missing data on
partially sampled Cartesian grids, based on non-parametric spin
models from statistical physics (e.g., Ising, Potts, clock and XY
models). The methods are based on the idea of matching the
normalized correlation energy of the sampled spins with that of the
entire grid. The matching is performed using greedy Monte Carlo
simulation conditioned by the sample values. The non-parametric spin-based
classifiers presented
here embody isotropic nearest-neighbor correlations. Hence,
they are expected to perform well if the data exhibit a degree of
spatial continuity and structure dominated by local features. Many
geophysical and remotely sensed data as well as digital images share
these features. The models presented here are not in
principle suitable for capturing long-range correlations, such as
those characterizing the transport properties of geological media.
Nonetheless, it should be pointed out that the spatial features of
the ``reconstructed'' field are not determined exclusively by the
properties of the classification model but also by the features
(e.g., anisotropic correlations) present in the sample.


The relative performance of the spin models in the case studies investigated
varied, depending on the type of data, the sampling density, and the
number of classes considered. Overall, the INNC model, which is
based on a sequential classification algorithm, gave the most
accurate classification rates in most of the cases studied. For all
the simulated data, the PNNC model gave the highest
misclassification rates. We believe that this is mainly due to the
higher spatial degeneracy of the PNNC model.
For noise-free data with
short-range differentiable variations,
 the CNNC model that incorporates cross-class correlations
 performed best, especially for the higher class numbers. As
the number of classes increases, the CNNC model tends to the
continuous XY model, and the classification emulates spatial
interpolation. Up to a threshold, increasing the number of classes
gradually lowers spatial prediction errors at the cost of a moderate
increase in computing time. The classification performance of the
spin-based methods can be further improved by executing multiple
runs starting from different initial states. This strategy also
permits an estimate of the classification uncertainty. The
classification performance of the spin-based models was compared
with the $k$-nearest neighbor (KNN), and the fuzzy $k$-nearest neighbor
(FKNN) algorithms, and (for the rainfall data) with the Support
Vector Machine (SVM) classifier. For the synthetic
data the INNC and CNNC models gave uniformly lower misclassification
rates than the KNN and FKNN algorithms. For classification of real
data into a small number of classes, the FKNN algorithm with optimized $k$
was the most accurate of the classifiers tested.

All the spin-based models are computationally efficient.  For the
PNNC, CNNC, and XYNNC models, the mean CPU time ranged from $0.03$
seconds (PNNC model, $L=50$, $p=0.33$, $N_c=8$) to $5$ seconds (CNNC
model, $L=200$, $p=0.66$, and $N_c=16$). For the INNC model, the CPU
time is generally higher due to the cost of determining the initial
state at each level. In contrast, the time needed for the Monte
Carlo relaxation is very short. Therefore, the resulting INNC CPU
time varies almost linearly with the number of classes, and in our
study it ranged from $0.08$ to almost $14$ seconds.  We do not report
CPU times for the KNN, FKNN, and SVM computations, since in order to
optimize the classification accuracy significant computational time
was devoted to fine-tuning the hyperparameters.


An advantage of the spin-based models with respect to the
other classifiers tested is the lack of hyperparameters that need
tuning by the user. Hence the classification procedure can be
automated, and it provides competitive accuracy as well as
computational efficiency. Compared to linear spatial interpolation
algorithms (e.g., kriging), the spin-based classification methods
present the advantages of computational efficiency and ability to
handle non-Gaussian probability distributions at the expense of
introducing discrete intervals in place of continuous values.
A comparative study of the two approaches in the future could
help to quantify their relative merits.

Currently, the spin-based models are formulated on a regular grid.
Hence, potential areas of application involve the compression of
large images and the reconstruction  of missing values (e.g., image
pixels). Note that in light of the comments in
Section~\ref{sec:spin_clas}, the energy matching principle is not
suitable for the refinement (resampling) of a regular grid, e.g., by doubling
the spatial resolution. Extension to irregular sampling patterns is
possible by defining a distance-dependent interaction strength
$J_{i,j}=J_{0} \, K(\vec{r}_{ij})$, where $J_0$ is arbitrary and
$K(\vec{r}_{ij})$ is a normalized function of $\vec{r}_{ij}$, over a
specified interaction neighborhood.  Other potential
extensions include extended-range interactions and/or ``multi-point
spin" correlations in the respective Hamiltonians. This could also
help to eliminate the spatial degeneracy evident in the present
models, and provide more flexible measures of spatial dependence at
the expense of concomitant increases in computational time and
parametrization.

\begin{acknowledgments}
This research project has been supported by a Marie Curie Transfer
of Knowledge Fellowship of the European Community's Sixth Framework
Programme under contract number MTKD-CT-2004-014135.
We are grateful
to Prof. M. Kanevski (Universit\'{e} de Lausanne, Switzerland)
for providing us with the full version of the GeoSVM software.
\end{acknowledgments}

\end{document}